\RequirePackage[T1]{fontenc}
\documentclass[aps,pra,reprint,superscriptaddress,nofootinbib,floatfix]{revtex4-2}
\usepackage{graphicx}
\usepackage{amssymb}
\usepackage{mathtools}
\usepackage{braket}
\usepackage{bbold}
\usepackage{relsize}
\usepackage[normalem]{ulem}

\usepackage{import}
\usepackage[table,xcdraw,dvipsnames]{xcolor}
\usepackage{float}
\usepackage[shortcuts]{extdash}

\usepackage{siunitx}

\usepackage{tikz}
\usetikzlibrary{arrows.meta,calc,decorations.pathmorphing,fit,shadings,shapes,positioning,external}
\usepackage{pgfplots}
\pgfplotsset{compat=1.16}
\pgfplotsset{scaled x ticks=false,
tick scale binop=\times,
enlargelimits=false}
\usepgfplotslibrary{groupplots}

\usepackage{booktabs}
\usepackage{multirow}

\definecolor{HOMcolour}{RGB}{66,245,236} 
\definecolor{MZ2scolour}{RGB}{255,106,0} 
\definecolor{MZ2dcolour}{RGB}{153, 0, 204} 
\definecolor{MZ1colour}{RGB}{182, 255, 0} 

\definecolor{P1colour}{RGB}{0, 169, 255} 
\definecolor{P2colour}{RGB}{255, 25, 0} 
\definecolor{Pccolour}{RGB}{255, 142, 236} 
\definecolor{FIcolour}{RGB}{16, 255, 12} 

\usepackage{acro}
\acsetup{single}

\DeclareAcronym{HOM}{
	short = HOM,
	long = Hong-Ou-Mandel,
}
\DeclareAcronym{MZ}{
	short = MZ,
	long = Mach-Zehnder,
}
\DeclareAcronym{SPDC}{
	short = SPDC,
	long = spontaneous parametric down-conversion,
}
\DeclareAcronym{MZI}{
	short = MZI,
	long = Mach-Zehnder interferometer,
}
\DeclareAcronym{POVM}{
	short = POVM,
	long = positive-operator valued measure,
}

\usepackage{hyperref}
\hypersetup{
    colorlinks=true,
    citecolor=blue,
    filecolor=blue,
    linkcolor=blue,
    urlcolor=blue
}

\newcommand{\omegap}{\omega_{\mathrm{p}}}
\newcommand{\etamag}[1]{\eta_{\tilde{#1}}}

\NewDocumentCommand{\DB}{s+m}{\textcolor{purple}{\IfBooleanT{#1}{\bfseries} #2}}
\NewDocumentCommand{\HS}{s+m}{\textcolor{cyan}{\IfBooleanT{#1}{\bfseries} #2}}
\NewDocumentCommand{\EG}{s+m}{\textcolor{blue}{\IfBooleanT{#1}{\bfseries} #2}}
\NewDocumentCommand{\NW}{s+m}{\textcolor{magenta}{\IfBooleanT{#1}{\bfseries} #2}}

\begin{document}
\title{Noise limits on two-photon interferometric sensing}

\author{Hamish Scott}
\affiliation{SUPA, Institute of Photonics and Quantum Sciences, Heriot-Watt University, Edinburgh, EH14 4AS, UK}
\author{Dominic Branford}
\affiliation{SUPA, Institute of Photonics and Quantum Sciences, Heriot-Watt University, Edinburgh, EH14 4AS, UK}
\author{Niclas Westerberg}
\affiliation{School of Physics and Astronomy, University of Glasgow, Glasgow, G12 8QQ, United Kingdom}
\author{Jonathan Leach}
\affiliation{SUPA, Institute of Photonics and Quantum Sciences, Heriot-Watt University, Edinburgh, EH14 4AS, UK}
\author{Erik M. Gauger}
\affiliation{SUPA, Institute of Photonics and Quantum Sciences, Heriot-Watt University, Edinburgh, EH14 4AS, UK}

\date{\today}

\begin{abstract}
When a photon interferes with itself while traversing a Mach-Zehnder inteferometer, the output port where it emerges is influenced by the phase difference between the interferometer arms.
This allows for highly precise estimation of the path length difference (delay) but is extremely sensitive to phase noise.
By contrast, a delay between the arms of the two-photon Hong-Ou-Mandel interferometer directly affects the relative indistinguishability of the photon pair, affecting the rate of recorded coincidences.
This likewise allows for delay estimation; notably less precise but with the advantage of being less sensitive to perturbations of the photons' phase.
Focusing on two-photon input states, we here investigate to what degree of noise Mach-Zehnder interferometry retains its edge over Hong-Ou-Mandel interferometry.
We also explore the competing benefits of different two-photon inputs for a Mach-Zehnder interferometer, and under what parameter regimes each input performs best.
\end{abstract}

\maketitle

\section{Introduction}

Interference lies at the heart of optical metrology: observing changes in the recorded interference patterns allows for precise measurements of sample and environmental parameters.
This is commonly realised with Michelson, Mach-Zehnder, Fabry-Perot, Sagnac, and Hong-Ou-Mandel interferometers~\cite{Ley-1987,Loudon-2000,Demkowicz-Dobrzanski-2015,Bouchard-2021}.
The quintessential task for interferometry is estimating optical delays which manifest as phase shifts in the traditional \ac{MZ} interferometer or distinguishability in \ac{HOM} interferometry.
For example, laser-interferometric gravitational wave detectors use sensitivity to physical displacements to achieve extraordinarily precise measurements of mechanical displacements across a broad range of frequencies~\cite{Tse-2019,Virgo-2019}.

The operating principle of the \ac{MZ} interferometer derives from interference fringes with a period determined by the optical frequency~\cite{Loudon-2000};
these fringes shift according to the relative phase in the interferometer.
By contrast, \ac{HOM}~\cite{Hong-1987} interferometers only possess a single interference dip with a width determined by the spectral distribution of the input photons; this dip is displaced according to the relative delay between the two input paths.

In both cases, path delay can be estimated from the readout of detectors placed at the two interferometer output ports.
For single-photon \ac{MZ} this is through the ratio of clicks between detectors~\cite{Kok-2004,Ben-Aryeh-2012,Demkowicz-Dobrzanski-2015};
in the two-photon \ac{HOM} case the delay influences the rate at which photons bunch
~\cite{Olindo-2006,Lyons-2018,Chen-2019,Yang-2019,Restuccia-2019,Scott-2020,Fabre-2021}.

\ac{MZ} analyses can be extended to include multi-photon inputs with non-monochromatic~\cite{Shih-1994,Rarity-1990,Kim-2003} and monochromatic light~\cite{Lang-2014,Demkowicz-Dobrzanski-2015,Polino-2020}.
This is expected to give rise to significant benefits through the application of non-classical light which can obtain more favourable scalings in the number of particles used~\cite{Lang-2014,Demkowicz-Dobrzanski-2015,Polino-2020}.
However reaching the high-photon regime where such non-classical light becomes beneficial compared to classical light is experimentally demanding~\cite{Polino-2020}.
Instead, one of the foremost concerns is in the probing of so-called ``delicate'' samples which are sensitive to high-photon numbers and can genuinely benefit from the application of optimum few-photon probe states~\cite{Wolfgramm-2013, Taylor-2016, Casacio-2021,Triginer-Garces-2020,Xavier-2021}.

The narrow wavelength-order fringes of the Mach-Zehnder interferometry may intuitively be expected to result in higher sensitivitiy than the much broader \ac{HOM} dip, which varies on the order of the inverse spectral width rather than inverse frequency.
In principle MZ allows for arbitrarily high precision with sufficiently high frequency,
however, it suffers from the ``phase-wrapping'' problem~\cite{Itoh-1982,von-Toussaint-2015,Hayashi-2018} where multiple phases can produce the same output signal.
In practice identifying the true phase requires prior information, or the use of adaptive techniques.
By comparison the \ac{HOM} effect remains technically distinct across half the dip, affording a much larger dynamic range at the cost of generally reduced estimation precision.

This gap between phase-insensitive and phase-sensitive interferometry has recently been narrowed with experiments that have reported much improved sensitivities to HOM-based sensing~\cite{Lyons-2018,Chen-2019}.
This raises the question of whether there are settings where \ac{HOM} interferometry---which may already be practically desirable due to the wider dynamic range and relatively simpler optics---can compete with or even surpass phase-sensitive interferometry.

In this work, we thus focus on interferometry with photon pairs as input states to Mach-Zehnder and Hong-Ou-Mandel interferometers
as the archetypal phase-sensitive and phase-insensitive interferometers.
Rather than limiting to monochromatic inputs we consider Gaussian two-photon spectral distributions such as readily produced by generic down-conversion sources.
We shall be interested in the comparative performance between those different approaches, including a comparison of different two-photon MZ input states. We later additonally compare to the results for a single-photon MZ protocol.

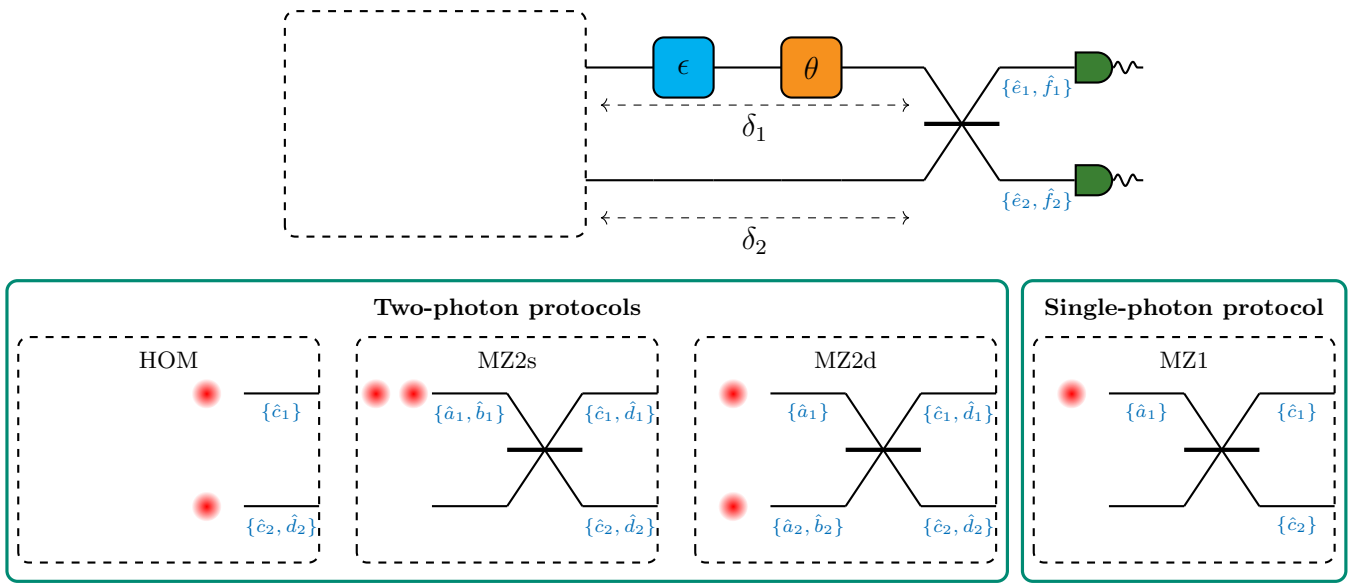
\begin{figure*}[htbp]
	\centering
	\tikzsetnextfilename{schematic}
	\begin{tikzpicture}[mode-label/.style={text=NavyBlue,font=\scriptsize,execute at begin node=\mathstrut}]
	\foreach \j [count=\n] in {0,1.5}
	{
		\draw [thick] (0.5,\j) -- (1.4,\j); 
		\draw [thick] (2.2,\j) -- (3.1,\j); 
		\draw [thick] (3.9,\j) -- (5,\j); 
		\draw [thick] (5,\j) -- (6,1.5-\j); 
		\draw [thick] (6,\j) -- (7,\j); 
		\node (det-\n) [draw,fill=OliveGreen,thick,rounded rectangle, rounded rectangle left arc=none,minimum width = 0.7cm,minimum height = 0.4cm,anchor=west] at (7,\j) {}; 
		\draw [thick,decorate,decoration={snake,segment length=0.2cm,amplitude=0.075cm}] (det-\n.east) --++ (0.40,0); 
	}
	\draw [<->,dashed] (0.7,1) -- (4.8,1); 
	\draw [<->,dashed] (0.7,-0.5) -- (4.8,-0.5); 
	\node at (2.75,0.7) {\large $\delta_1$}; 
	\node at (2.75,-0.8) {\large $\delta_2$}; 
	\draw [thick, rounded corners, fill=ProcessBlue] (1.4,1.5-0.4) rectangle (2.2,1.5+0.4); 
	\draw [thick] (1.4,0) -- (2.2,0); 
	\draw [thick, rounded corners, fill=BurntOrange] (3.1,1.5-0.4) rectangle (3.9,1.5+0.4); 
	\draw [thick] (3.1,0) -- (3.9,0); 
	\draw [thick, rounded corners, dashed] (-3.5,-0.75) rectangle (0.5,2.25); 
	\draw [ultra thick] (5,0.75) -- (6,0.75); 
	\node [mode-label] at (6.50,1.25) {$\{\hat{e}_1,\hat{f}_1\}$}; 
	\node [mode-label] at (6.50,-0.25) {$\{\hat{e}_2,\hat{f}_2\}$}; 
	\node at (1.8,1.5) {\large $\epsilon$}; 
	\node at (3.5,1.5) {\large $\theta$}; 

\begin{scope}[shift={($(current bounding box.south)+(0,-2.5)$)}]
	\draw [very thick, PineGreen, rounded corners] (-9.4,-1.75) rectangle (3.9,2.25);
	\node at (-2.75,1.875) {\bf{Two-photon protocols}};
	\draw [very thick, PineGreen, rounded corners] (4.1,-1.75) rectangle (8.4,2.25);
	\node at (6.25,1.875) {\bf{Single-photon protocol}};
	\begin{scope}[xshift=-6.75cm]
	\draw [thick, rounded corners, dashed] (-2.5,-1.5) rectangle (1.5,1.5);
	\foreach \j [count=\n] in {0,1.5}
	{
		\draw [thick] (0.5,-0.75+\j) -- (1.5,-0.75+\j);
		\shade [inner color=red] (0.0,-0.75+\j) circle (2mm);
	}
	\node at (-0.5,1.2) {HOM};
	\node [mode-label] at (1,0.5) {$\{\hat{c}_1\}$}; 
	\node [mode-label] at (1,-1) {$\{\hat{c}_2,\hat{d}_2\}$}; 
	\end{scope}

	\begin{scope}[xshift=-2.25cm]
	\draw [thick, rounded corners, dashed] (-2.5,-1.5) rectangle (1.5,1.5);
	\draw [ultra thick] (-0.5,0.0) -- (0.5,0.0);
	\foreach \j [count=\n] in {0,1.5}
	{
		\draw [thick] (-0.5,-0.75+\j) -- (0.5,0.75-\j);
		\draw [thick] (-0.5,-0.75+\j) -- (-1.5,-0.75+\j);
		\draw [thick] (0.5,-0.75+\j) -- (1.5,-0.75+\j);
	}
	\shade [inner color=red] (-1.75,0.75) circle (2mm);
	\shade [inner color=red] (-2.25,0.75) circle (2mm);
	\node at (-0.5,1.2) {MZ2s};
	\node [mode-label] at (-1,0.5) {$\{\hat{a}_1,\hat{b}_1\}$}; 
	\node [mode-label] at (1,0.5) {$\{\hat{c}_1,\hat{d}_1\}$}; 
	\node [mode-label] at (1,-1) {$\{\hat{c}_2,\hat{d}_2\}$}; 
	\end{scope}
	
	\begin{scope}[xshift=2.25cm]
	\draw [thick, rounded corners, dashed] (-2.5,-1.5) rectangle (1.5,1.5);
	\draw [ultra thick] (-0.5,0.0) -- (0.5,0.0);
	\foreach \j [count=\n] in {0,1.5}
	{
		\draw [thick] (-0.5,-0.75+\j) -- (0.5,0.75-\j);
		\draw [thick] (-0.5,-0.75+\j) -- (-1.5,-0.75+\j);
		\draw [thick] (0.5,-0.75+\j) -- (1.5,-0.75+\j);
		\shade [inner color=red] (-2.0,-0.75+\j) circle (2mm);
	}
	\node at (-0.5,1.2) {MZ2d};
	\node [mode-label] at (-1,0.5) {$\{\hat{a}_1\}$}; 
	\node [mode-label] at (-1,-1) {$\{\hat{a}_2,\hat{b}_2\}$}; 
	\node [mode-label] at (1,0.5) {$\{\hat{c}_1,\hat{d}_1\}$}; 
	\node [mode-label] at (1,-1) {$\{\hat{c}_2,\hat{d}_2\}$}; 
	\end{scope}
	
	\begin{scope}[xshift=6.75cm]
	\draw [thick, rounded corners, dashed] (-2.5,-1.5) rectangle (1.5,1.5);
	\draw [ultra thick] (-0.5,0.0) -- (0.5,0.0);
	\foreach \j [count=\n] in {0,1.5}
	{
		\draw [thick] (-0.5,-0.75+\j) -- (0.5,0.75-\j);
		\draw [thick] (-0.5,-0.75+\j) -- (-1.5,-0.75+\j);
		\draw [thick] (0.5,-0.75+\j) -- (1.5,-0.75+\j);
	}
	\shade [inner color=red] (-2.0,0.75) circle (2mm);
	\node at (-0.5,1.2) {MZ1};
	\node [mode-label] at (-1,0.5) {$\{\hat{a}_1\}$}; 
	\node [mode-label] at (1,0.5) {$\{\hat{c}_1\}$}; 
	\node [mode-label] at (1,-1) {$\{\hat{c}_2\}$}; 
	\end{scope}
\end{scope}
\end{tikzpicture}
\caption{Generalised schematic for our four protocol configurations.
The upper arm contains two phase shifts: a frequency-dependent shift $\epsilon$ and a frequency-independent shift $\theta$.
Noise is subsequently modelled as some variation in these shifts.
Photon modes are labelled at the different stages of the schematic;
at each stage we work with a combination of two orthogonal modes to model the initial indistinguishability within the photon pair.
}
\label{fig_schematic}
\end{figure*}

With our focus on photon pair inputs, we compare the performance of different protocols at varying degrees of phase noise, including the high-noise limit.
Our noise model is similar to that of Refs.~\cite{Genoni-2011,Escher-2012,Genoni-2012}, which explored the ultimate quantum limit of phase estimation\footnote{This is equivalent to delay estimation with monochromatic photons.} in the presence of phase diffusion.
Much as the best probe states for Mach-Zehnder interferometry are undermined by noise~\cite{Escher-2012,Dorner-2009,Lee-2009,Datta-2011}, we may anticipate that HOM is capable of delivering valuable precision (relative to a two-photon Mach-Zehnder) in high-noise regimes.

This Article is organised as follows: In Sec.~\ref{sec_protocols} we describe our protocols with fixed frequency-dependent and frequency-independent phase shifts.
We then move to our noisy model in Sec.~\ref{sec_noise} and derive the relevant probabilities and expressions for the Fisher information.
In Sec.~\ref{sec_alternatives} we outline how our model can be tweaked to account for frequency-independent input photons and noise that is uncorrelated between photon modes.
Sec.~\ref{sec_results} presents our results, where we compare the resilience of our various protocols and models to increased noise, and also identify and discuss a number of interesting emerging features.
Finally, we summarise these results and present some concluding comments in Sec.~\ref{sec_conclusion}.

\section{Protocols}\label{sec_protocols}
We consider and compare four different protocols, illustrated in Fig.~\ref{fig_schematic} with the common parameter encoding and measurements along the different state preparations.
First, a standard \acf{HOM} protocol: wherein two photons interfere at a beamsplitter and detectors are placed at the two output ports.
Then, three protocols using a \acf{MZI}: consisting of two beamsplitters such that the output ports of the first beamsplitter are directed towards the input ports of the second beamsplitter, with detectors at the outputs of the second beamsplitter.
Between the two beamsplitters the upper and lower interferometer arms have path lengths $\delta_1$ and $\delta_2$, respectively.
We consider a pair of two-photon \ac{MZ} protocols: both photons enter via the same initial input port (MZ2s) and both entering via different input ports (MZ2d); and in addition a conventional single-photon \ac{MZ} protocol (MZ1).

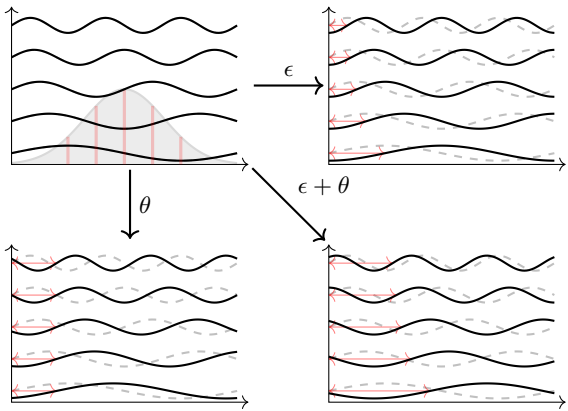
\begin{figure}
\centering
\tikzsetnextfilename{shifts}
\begin{tikzpicture}[xscale=1.5]
\gdef\fixedoff{90}
\gdef\freqoff{0.4}
\gdef\freqcount{5}

\foreach \xplt in {0,1}{
\foreach \yplt in {0,1}{
\begin{scope}[xshift={80*\xplt},yshift={-90*\yplt}]
		\newcount\xypos
		\xypos=\numexpr\xplt+\yplt\relax
	\draw [<->] (0,2.1) -- (0,0)  -- (2.1,0);
	\ifnum\xypos=0
			\draw [thick,smooth,samples=50,domain=0:2,fill=gray,opacity=0.15] plot ({\x},{exp(-((\x-1)/0.5)^2)});
	\fi
	\foreach \j [count=\k] in {0.5,0.75,1,1.25,1.5} {
	\ifnum\xypos=0
		\draw [very thick,red,opacity=0.2] (\j,0) -- (\j,{exp(-((\j-1)/0.5)^2)});
		\fi
		\draw [smooth,thick,dashed,opacity=0.25,samples=50,domain=0:2] plot ({\x},{0.15 + 1.7*(\k-1)/(\freqcount-1) + 0.1*sin(\j*\x*360)});
		\ifnum\xypos>0
			\draw [opacity=0.5,red,<->] (0,{0.15+1.7*(\k-1)/(\freqcount-1)}) -- ({\yplt*\freqoff+\xplt*\fixedoff/(360*\j)},{0.15+1.7*(\k-1)/(\freqcount-1)});
		\fi
		\draw [smooth,thick,samples=50,domain=0:2] plot ({\x},{0.15 + 1.7*(\k-1)/(\freqcount-1) + 0.1*sin(\j*(\x-\yplt*\freqoff)*360-\xplt*\fixedoff)});
	}
	\coordinate (box-north-\xplt-\yplt) at (1.05,2.1);
	\coordinate (box-east-\xplt-\yplt) at (2.1,1.05);
	\coordinate (box-south-\xplt-\yplt) at (1.05,0);
	\coordinate (box-west-\xplt-\yplt) at (0,1.05);
	\coordinate (box-nw-\xplt-\yplt) at (0,2.1);
	\coordinate (box-se-\xplt-\yplt) at (2.1,0);
\end{scope}
}}

\draw [->,thick,shorten <=2,shorten >=2] (box-south-0-0) -- (box-north-0-1) node [midway,right] {$\theta$};
\draw [->,thick,shorten <=2,shorten >=2] (box-east-0-0) -- (box-west-1-0) node [midway,above] {$\epsilon$};
\draw [->,thick,shorten <=2,shorten >=2] (box-se-0-0) -- (box-nw-1-1) node [midway,above right] {$\epsilon+\theta$};
\end{tikzpicture}
\caption{For a Gaussian wavepacket, the effect of our fixed shifts $\epsilon$ and $\theta$ on the different frequency modes is shown. $\epsilon$ is frequency-dependent, and can be thought of as some modification to the actual delay $\delta$. Upper right shows the differing intensity of an $\epsilon$ shift at differing frequencies. $\theta$ is frequency-independent, and bottom left shows all modes experience the same effect from a $\theta$ shift. Bottom right shows a combined $\epsilon$ and $\theta$ shift. Our noise model is obtained in Sec.~\ref{sec_noise} by averaging over these shifts.}\label{fig_shifts}
\end{figure}

The predominant type of noise affecting the photons can be expressed as an unknown fluctuating phase shift in one or both of the arms. We can write this as $e^{-i\phi(\omega)}$, where $\phi(\omega)$ has some unspecified frequency-dependence.
Taylor-expanding around $\omega_0$, we then write ${e^{-i\phi(\omega)}\approx e^{-i \left(\phi(\omega_0)+\frac{\partial \phi(\omega_0)}{\partial \omega}(\omega-\omega_0)+\cdots\right)}}$.
Truncating terms beyond linear order in \( \omega-\omega_0 \) we can write the total phase shift from noise as
\begin{equation}
\exp \big [-i \big\{\overbrace{\phi(\omega_0)-\frac{\partial \phi(\omega_0)}{\partial \omega}\omega_0}^\mathlarger{\mathlarger{\theta}}+\overbrace{\frac{\partial \phi(\omega_0)}{\partial \omega}}^\mathlarger{\mathlarger{\epsilon}} \omega \big\} \big].
\label{eq_noise_taylor}
\end{equation}
Here, we identify a frequency-dependent component \( \epsilon \), and a frequency-independent component $\theta$ to the phase shift.
The latter, $\epsilon$, can be thought of as representing fluctuations in the path length, such as that might arise from vibrations or heating in the system.
By contrast, $\theta$ is a ``pure'' phase shift that leaves the path length unaffected.
The effects of these shifts are illustrated for different frequency modes in Fig.~\ref{fig_shifts}.
For the remainder of these sections we will work with unknown but fixed $\epsilon$ and $\theta$; we will later average over them in Sec.~\ref{sec_noise} to capture fluctuations in time.

For simplicity, we limit these shifts to the upper arm of the interferometer: essentially assuming noise to be localised noise entirely within the upper arm.
Though in reality we expect noise to be present in both arms, this localisation gives rise to equivalent detection probabilities with phase-insensitive measurements.
For justification and further discussion, see Appendix~\ref{app_both_arms}.

\subsection{Optical modes}
A biphoton state generated by \ac{SPDC} will, in practice, exhibit some degree of non-spatial distinguishability (such as, e.g., a mismatch of polarisations) between the two photons.
Thus we write one photon in the initial superposition $\sqrt{\alpha}\>\hat{a}^\dagger(\omega)+\sqrt{1-\alpha}\>\hat{b}^\dagger(\omega)$, with $\hat{b}$ some orthogonal photonic mode.
The visibility $\alpha\in[0,1]$ therefore encodes the relative indistinguishability of the photon pair upon generation.
The modes labelled in Fig.~\ref{fig_schematic}: $\hat{a}_j$/$\hat{b}_j$, $\hat{c}_j$/$\hat{d}_j$, and $\hat{e}_j$/$\hat{f}_j$; are then the pairs of orthogonal photonic modes in a given arm at each stage of the protocol.
The subscripts $j=\{1,2\}$ denote distinct spatial modes, corresponding to the two arms of the interferometer.

\subsection{Common optics}

We describe the initial preparation illustrated by the lower part of Fig.~\ref{fig_schematic} individually in the following subsections, alongside the resulting detection probabilities.
For the MZ protocols the initial states are given in terms of the \( \{ \hat{a}_{1,2} , \hat{b}_{1,2} \} \) modes which pass through a beam splitter with transforms
\begin{equation}
\begin{gathered}
	\hat{a}_1^\dagger(\omega) \to \frac{1}{\sqrt{2}}[i\hat{c}_1^\dagger(\omega)+\hat{c}_2^\dagger(\omega)], \\*
	\hat{a}_2^\dagger(\omega) \to \frac{1}{\sqrt{2}}[\hat{c}_1^\dagger(\omega)+i\hat{c}_2^\dagger(\omega)], \\*
	\hat{b}_1^\dagger(\omega) \to \frac{1}{\sqrt{2}}[i\hat{d}_1^\dagger(\omega)+\hat{d}_2^\dagger(\omega)], \\*
	\hat{b}_2^\dagger(\omega) \to \frac{1}{\sqrt{2}}[\hat{d}_1^\dagger(\omega)+i\hat{d}_2^\dagger(\omega)].
\end{gathered}
\label{eq:input_bs}
\end{equation}
The HOM protocol inputs photons directly into the delay stage and so the input state is written in terms of the modes \( \{ \hat{c}_{1,2} , \hat{d}_2 \} \).

After the initial preparation stage all photons pass through a common set of linear optics to reach the detectors where the mode transformations are described by
\begin{equation}
\begin{gathered}
	\hat{c}_1^\dagger(\omega) \to e^{-i\omega(\delta_1+\epsilon)}e^{-i\theta}\hat{c}_1^\dagger(\omega), \\
	\hat{c}_2^\dagger(\omega) \to e^{-i\omega\delta_2}\hat{c}_2^\dagger(\omega), \\
	\hat{d}_1^\dagger(\omega) \to e^{-i\omega(\delta_1+\epsilon)}e^{-i\theta}\hat{d}_1^\dagger(\omega), \\
	\hat{d}_2^\dagger(\omega) \to e^{-i\omega\delta_2}\hat{d}_2^\dagger(\omega),
\end{gathered}
\label{eq:delay_encoding}
\end{equation}
which encode the local delays alongside the phase shifts $\epsilon$ and $\theta$.
The detection probabilities depend only on the path length difference $\delta = \delta_2 - \delta_1$.
The photons then interfere at the (final) beamsplitter:
\begin{equation}
\begin{gathered}
	\hat{c}_1^\dagger(\omega) \to \frac{1}{\sqrt{2}}[i\hat{e}_1^\dagger(\omega)+\hat{e}_2^\dagger(\omega)],\\
	\hat{c}_2^\dagger(\omega) \to \frac{1}{\sqrt{2}}[\hat{e}_1^\dagger(\omega)+i\hat{e}_2^\dagger(\omega)],\\
	\hat{d}_1^\dagger(\omega) \to \frac{1}{\sqrt{2}}[i\hat{f}_1^\dagger(\omega)+\hat{f}_2^\dagger(\omega)],\\
	\hat{d}_2^\dagger(\omega) \to \frac{1}{\sqrt{2}}[\hat{f}_1^\dagger(\omega)+i\hat{f}_2^\dagger(\omega)],
\end{gathered}
\label{eq:detector_bs}
\end{equation}
after which detectors measure whether a photon is found in one of the arms.

In Appendix~\ref{app_general_probs} we derive the general detection probabilities in terms of the form of the output state.

\subsection{HOM}
For the \ac{HOM} protocol, we take the biphoton state~\cite{Hong-1987,Branczyk-2017}
\begin{equation}
\ket{\psi_\mathrm{HOM}^\mathrm{in}} = \int d\omega \> \phi(\omega) \hat{c}_1^\dagger(\omegap-\omega) [\sqrt{\alpha}\>\hat{c}_2^\dagger(\omega)+\sqrt{1-\alpha}\>\hat{d}_2^\dagger(\omega)] \ket{0}
\label{eq_HOM_input}
\end{equation}
as input, with
\begin{equation}
\phi(\omega)=(2\pi\sigma^2)^{-1/4}e^{-\frac{(\omega-\omegap/2)^2}{4\sigma^2}}.
\label{eq_JSA}
\end{equation}
$\omegap$ is the pump frequency, and $\sigma$ the spectral width~\cite{Branczyk-2017}. 

Evolving $\ket{\psi_{\mathrm{HOM}}^{\mathrm{in}}}$ according to Eqs.~\eqref{eq:delay_encoding} and~\eqref{eq:detector_bs}, we obtain the output state:
\begin{align}
	\ket{\psi_\mathrm{HOM}^\mathrm{out}} &= \frac{1}{2} \int d\omega \> \phi (\omega) e^{-i [({\delta_1}+\epsilon ) (\omegap-\omega )+{\delta_2} \omega +\theta ]} \nonumber \\*
&\times[i \hat{e}_1^\dagger(\omegap-\omega)+\hat{e}_2^\dagger(\omegap-\omega)]\nonumber \\*
&\times[\sqrt{\alpha}\{\hat{e}_1^\dagger(\omega)+i\hat{e}_2^\dagger(\omega)\}\nonumber \\*
&\mkern32mu+\sqrt{1-\alpha}\{\hat{f}_1^\dagger(\omega)+i\hat{f}_2^\dagger(\omega)\}]\ket{0}.
\label{eq_HOM_output}
\end{align}

Eqs.~(\ref{eq_general_prob_j}, \ref{eq_general_prob_c}) together with Eq.~\eqref{eq_HOM_output} then give the probabilities of detection at one or both detectors.
We find
\begin{equation}
P_{1,\mathrm{HOM}}=P_{2,\mathrm{HOM}}=\frac{1}{4} \left(1+\alpha  e^{-2 \sigma^2 (\delta -\epsilon )^2}\right),
\end{equation}
the probability of detection at detector 1 only and detector 2 only,
\begin{equation}
P_{\mathrm{c},\mathrm{HOM}} = \frac{1}{2} \left( 1 - \alpha e^{-2\sigma^2 (\delta - \epsilon)^2} \right),
\end{equation}
the probability of coincidence at both detectors.
These probabilities depend
concur with the probabilities seen in standard HOM analyses~\cite{Branczyk-2017}.

One thing we immediately notice is that $\theta$ drops out---\ac{HOM} is not affected by arbitrary frequency-independent phase shifts and therefore is immune to frequency-independent noise\footnote{Because the \ac{HOM} state is still separable at the stage where phase shifts are applied, the frequency-independent shift $\theta$ can be thought of as a global phase for the \ac{HOM} case.}.
In traditional \ac{HOM} analyses ones often treats $P_{1,\mathrm{HOM}} + P_{2,\mathrm{HOM}}$ as the ``bunching probability'', but we here keep them separate for consistency with the \ac{MZ} analysis.

\subsection{Two-photon MZ: same input port (MZ2s)}
We now consider our MZ2s protocol, where both photons enter via the same port (specifically, we choose the upper left port).
Our initial state is thus a modified version of our biphoton state from Eq.~\eqref{eq_HOM_input}:
\begin{align}
	\ket{\psi_\mathrm{MZ2s}^\mathrm{in}} =& \frac{1}{\sqrt{1+\alpha}} \int d\omega \> \phi(\omega) [\sqrt{\alpha}\>\hat{a}_1^\dagger(\omega)+\sqrt{1-\alpha}\>\hat{b}_1^\dagger(\omega)]\nonumber \\*
\times& \hat{a}_1^\dagger(\omega_p-\omega)\ket{0}.
\end{align}
The normalisation of $\frac{1}{\sqrt{1+\alpha}}$ is required as both photons entering via the same port results in a visibility-dependent overlap of their initial modes.

Up to an irrelevant global phase, we apply
Eqs.~\eqref{eq:input_bs} to~\eqref{eq:detector_bs}
to obtain the output state:
\begin{align}
	\ket{\psi_\mathrm{MZ2s}^\mathrm{out}} =& \frac{1}{\sqrt{1+\alpha}} \int d\omega \> \phi (\omega)\nonumber \\*
\times&[\sin(\Omega^-)\hat{e}_1^\dagger(\omegap-\omega)+\cos(\Omega^-)\hat{e}_2^\dagger(\omegap-\omega)]\nonumber \\*
\times&[\sqrt{\alpha}\{\sin(\Omega^+)\hat{e}_1^\dagger(\omega)+\cos(\Omega^+)\hat{e}_2^\dagger(\omega)\}\nonumber \\*
+&\sqrt{1-\alpha}\{\sin(\Omega^+)\hat{f}_1^\dagger(\omega)+\cos(\Omega^+)\hat{f}_2^\dagger(\omega)\}]\ket{0},
\label{eq_MZ2s_output}
\end{align}
with
\begin{align}
\Omega^- &= \frac{1}{2} \{\theta -(\delta -\epsilon) (\omegap-\omega) \}
\label{eq_trig_arg_1} \\*
\Omega^+ &= \frac{1}{2} \{\theta -(\delta -\epsilon) \omega \}.
\label{eq_trig_arg_2}
\end{align}

Taking this output state with Eqs.~(\ref{eq_general_prob_j}, \ref{eq_general_prob_c}), we obtain the detection probabilities
\begin{align}
	P_{1,\mathrm{MZ2s}} &= \frac{1}{8} \left[ 2 + e^{-2 \sigma^2 (\delta -\epsilon )^2} + \cos (2 \theta -\omegap (\delta -\epsilon ))\right. \nonumber \\
	&\mkern32mu- \left. 4 e^{-\frac{1}{2} \sigma^2 (\delta -\epsilon )^2} \cos \left(\theta -\frac{\omegap}{2} (\delta -\epsilon) \right) \right],\\
	P_{2,\mathrm{MZ2s}} &= \frac{1}{8} \left[ 2 + e^{-2 \sigma^2 (\delta -\epsilon )^2} + \cos (2 \theta -\omegap (\delta -\epsilon ))\right.\nonumber \\*
	&\mkern32mu+\left.4 e^{-\frac{1}{2} \sigma^2 (\delta -\epsilon )^2} \cos \left(\theta -\frac{\omegap}{2} (\delta -\epsilon )\right) \right],\\
P_{c,\mathrm{MZ2s}} &= \frac{1}{4} \left[ 2 - \cos (2 \theta -\omegap (\delta -\epsilon ))
- e^{-2 \sigma^2 (\delta-\epsilon)^2} \right].
\end{align}
As both photons enter the same port, the state at every point has form \( \hat{a}_{\zeta}(\omega) [ \sqrt{1-\alpha} \hat{a}_{\zeta}(\omega') + \sqrt{\alpha} \hat{b}_{\zeta'}(\omega') ] \) where \( \hat{a}_{\zeta} \) and \( \hat{b}_{\zeta} \) are orthogonal photonic modes and \( \hat{a}_{\zeta} \) (\( \hat{b}_{\zeta'} \)) can be a superposition of the spatial modes \( \hat{a}_1 \) and \( \hat{a}_2 \) (\( \hat{b}_1 \) and \( \hat{b}_2 \)).
This prohibits any HOM-like interference which only occurs when the photons enter a beam splitter in non-identical spatial (superposition) modes.
Moreover, as the photonic modes experience identical optical transforms the non-zero distinguishability (\( \alpha < 1 \)) does not affect the measured outputs, hence visibility dependence vanishes from the MZ2s output probabilities.

\subsection{Two-photon MZ: different input ports (MZ2d)}
We now consider the MZ2d protocol, where both photons enter the \ac{MZI} via different ports. This is similar to the two-photon case of Ref.~\cite{Holland-1993}, though our protocol is generalised to non-monochromatic inputs.

Our initial state is the same as in Eq.~\eqref{eq_HOM_input}, but with $\hat{c}$ and $\hat{d}$ relabelled as $\hat{a}$ and $\hat{b}$ to reflect that we now have an additonal beamsplitter.
Again up to an irrelevant global phase, we apply Eqs.~\eqref{eq:input_bs} to~\eqref{eq:detector_bs} to obtain the output state:
\begin{align}
	\ket{\psi_\mathrm{MZ2d}^\mathrm{out}} &= \int d\omega \> \phi (\omega) \nonumber \\*
	&\times [\sin(\Omega^-)\hat{e}_1^\dagger(\omegap-\omega)+\cos(\Omega^-)\hat{e}_2^\dagger(\omegap-\omega)]\nonumber \\*
	&\times [\sqrt{\alpha}\{\cos(\Omega^+)\hat{e}_1^\dagger(\omega)+\sin(\Omega^+)\hat{e}_2^\dagger(\omega)\}\nonumber \\*
	+&\sqrt{1-\alpha}\{\cos(\Omega^+)\hat{f}_1^\dagger(\omega)+\sin(\Omega^+)\hat{f}_2^\dagger(\omega)\}]\ket{0},
\label{eq_MZ2d_output}
\end{align}
with $\Omega^-$ and $\Omega^+$ as defined in Eqs.~(\ref{eq_trig_arg_1}, \ref{eq_trig_arg_2}). 

Together with Eqs.~(\ref{eq_general_prob_j}, \ref{eq_general_prob_c}), we obtain the detection probabilities
\begin{align}
	P_{1,\mathrm{MZ2d}} &= P_{2,\mathrm{MZ2d}} = \frac{1}{8} \left[ 2 - (1-\alpha) e^{-2 \sigma^2 (\delta -\epsilon )^2} \right. \nonumber \\*
&\mkern64mu-\left. (1+\alpha) \cos (\omegap (\epsilon -\delta )+2 \theta ) \right],
\\
	P_{\mathrm{c},\mathrm{MZ2d}} &= \frac{1}{4} \left[ 2 + (1-\alpha) e^{-2 \sigma^2 (\delta -\epsilon )^2} \right. \nonumber\\*
	&\mkern64mu+ \left. (1+\alpha) \cos (\omegap (\epsilon -\delta )+2 \theta )\right].
\label{eq_MZ2d_Pc}
\end{align}
Visibility dependence is seen for MZ2d, while it was absent from MZ2s.
We also note that, like the HOM case, ${P_1 = P_2}$.
This initially seems a major deviation from the conventional MZ1 scenario, where varying the phase can bias the photon towards one detector or the other; a feature that remains present in MZ2s.
The crucial difference in this configuration is that after the initial beamsplitter, both photons have opposite phase.
Setting a phase in the upper arm that biases one photon towards a certain detector must equally bias the other photon towards the other detector: no combination of delay and fixed shifts can result in $P_1$ or $P_2$ more likely than the other, and hence $P_1=P_2$.

In the limiting case of $\delta=\epsilon=0$, our MZ2d coincidence probability in Eq.~(\ref{eq_MZ2d_Pc}) concurs with the coincidence probability in Ref.~\cite{Jachura-2016}, which examines two photons entering opposite input ports of an \ac{MZI} with equal path lengths in both arms, but a fixed frequency-independent phase shift in one arm.

\subsection{Single-photon MZ (MZ1)}
For our single-photon \ac{MZ} protocol, we have the input state
\begin{equation}
	\ket{\psi_\mathrm{MZ1}^\mathrm{in}} = \int d\omega \> \phi(\omega) \hat{a}^\dagger(\omega)\ket{0},
	\label{eq_MZ1_input}
\end{equation}
with $\phi(\omega)$ as defined in Eq.~(\ref{eq_JSA}).

Applying the \ac{MZ} mode evolutions described by Eqs.~\eqref{eq:input_bs} to~\eqref{eq:detector_bs}
we obtain the output state
\begin{align}
	\ket{\psi_\mathrm{MZ1}^\mathrm{out}} =& \frac{1}{2} \int d\omega \> \phi(\omega) \left[(e^{-i \delta_2 \omega }-e^{-i(\theta+ (\delta_1+\epsilon) \omega) })\hat{c}_1^\dagger(\omega)\right. \nonumber \\*
+&\left.(i e^{-i(\theta+ (\delta_1+\epsilon) \omega) }+i e^{-i \delta_2 \omega })\hat{c}_2^\dagger(\omega)\right]\ket{0}.
\end{align}

The resulting detection probabilities are then
\begin{align}
	P_{1,\mathrm{MZ1}} &= \frac{1}{4}\int d\omega \> |\phi(\omega)|^2 \> |e^{-i \delta_2 \omega }-e^{-i(\theta+ (\delta_1+\epsilon) \omega) }|^2,\nonumber \\*
&= \frac{1}{2} - \frac{1}{2} e^{-\frac{1}{2} \sigma^2 (\delta -\epsilon )^2} \cos \left(\theta - \frac{\omegap}{2}(\delta -\epsilon ) \right) ,
\label{eq_MZ1_P1}
\end{align}
the probability of detection at detector 1 and
\begin{align}
	P_{2,\mathrm{MZ1}} &= \frac{1}{4}\int d\omega \> |\phi(\omega)|^2 \> |i e^{-i(\theta+ (\delta_1+\epsilon) \omega) }+i e^{-i \delta_2 \omega }|^2,\nonumber \\*
	&= \frac{1}{2} + \frac{1}{2} e^{-\frac{1}{2} \sigma^2 (\delta -\epsilon )^2} \cos \left(\theta -\frac{\omegap}{2} (\delta -\epsilon )\right),
\label{eq_MZ1_P2}
\end{align}
the probability of detection at detector 2.

\section{Modelling noise}\label{sec_noise}
To move from our probabilities in Sec.~\ref{sec_protocols} to those that model noise, we now assume some uncertainty in our previously fixed phase shifts.
Assuming both $\epsilon$ and $\theta$ fluctuate around zero, we average over them with the Gaussian weighting factors
\begin{equation}
\begin{aligned}
	J_\epsilon(\varepsilon) &= \frac{ e^{-\frac{\varepsilon^2}{2\eta_\epsilon^2}} }{ \sqrt{2\pi} \eta_\epsilon },
&
	J_\theta(\vartheta) &= \frac{ e^{-\frac{\vartheta^2}{2\eta_\theta^2}} }{ \sqrt{2\pi} \eta_\theta }.
\end{aligned}
\label{eq_dists}
\end{equation}
$\eta_\epsilon$ and $\eta_\theta$ control the width of the Gaussian weighting factors---the strength of these noise processes.
These terms give rise to shot-to-shot variations on top of the path length \( \delta \) on the order of \( \eta_{\theta} / \omegap \) and \( \eta_{\epsilon} \).

The probabilities for such noise distributions are given by
\begin{equation}
	P_{j}^\eta 
	= \int_{-\infty}^\infty \int_{-\infty}^\infty d\vartheta d\varepsilon \> J_\epsilon(\varepsilon) J_\theta(\vartheta) P_{j}.
	\label{eq_dist_int}
\end{equation}

For simplicity, we have chosen to use a Gaussian noise distribution with support $(-\infty,\infty)$ for frequency-independent ($\theta$) noise. Note however this is, for the probabilities under consideration, equivalent to a more conventional wrapped $[-\pi,\pi]$ Gaussian distribution. This equivalence is shown in Appendix~\ref{app_dist}, and our core results are contrasted against results where a von Mises distribution is chosen for frequency-independent noise.
\footnote{The frequency-dependent ($\epsilon$) noise, like $\delta$, does not have such a symmetry as the different frequency modes accumulate a phase due to $\epsilon$ noise (or change in $\delta$).}

It is important to emphasise that although $\eta_\epsilon$ and $\eta_\theta$ represent the degree of uncertainty in the pair of phase shifts, and these shifts may fluctuate over time as multiple photon pairs traverse the interferometer, in a single run of the protocol both photons will experience the same constant (but unknown) $\epsilon$ and $\theta$ shifts simultaneously.
This fact leads to some interesting consequences that we will observe in Sec.~\ref{sec_results} and explore in more detail in Appendix~\ref{app_cc}.

From these probabilities we calculate the Fisher information, which quantifies the information obtained about a parameter of interest (the path length difference $\delta$) from an average measurement, given a set of measurement outcomes and their associated probabilities.
For a parameter $\delta$ and measurement outcomes $m\in\mathcal{M}$, with $P(m|\delta)$ the probability of outcome $m$ given $\delta$, the Fisher information can be written~\cite[Chap.~3]{Kay-1993}
\begin{equation}
F(\delta) = \sum_{m\in\mathcal{M}} \frac{1}{P(m|\delta)} \left(\frac{\partial}{\partial \delta} P(m|\delta)\right)^2.
\label{eq_FI}
\end{equation}

The single-parameter Fisher information can then be used to bound the variance of an unbiased estimator for that parameter. With $\tilde{\delta}$ an unbiased estimator for $\delta$ and $N$ the number of independent measurements, the Cram\'er-Rao bound is given by
\begin{equation}
\mathrm{var}(\tilde{\delta})\geq \frac{1}{N F(\delta)},
\end{equation}
and represents the ultimate limit on the precision of an unbiased estimator $\tilde{\delta}$.

\subsection{HOM}
For our \ac{HOM} protocol the probabilities are
\begin{align}
	P_{1,\mathrm{HOM}}^\eta = P_{2,\mathrm{HOM}}^\eta &= \frac{1}{4} \left(1+\frac{\alpha  e^{-\frac{2 \delta^2 \sigma^2}{4 \eta_\epsilon^2 \sigma^2+1}} }{ \sqrt{4 \eta_\epsilon^2 \sigma^2+1} } \right),
	\\
	P_{c,\mathrm{HOM}}^\eta &= \frac{1}{2} \left(1-\frac{\alpha  e^{-\frac{2 \delta^2 \sigma^2}{4 \eta_\epsilon^2 \sigma^2+1}}}{\sqrt{4 \eta_\epsilon^2 \sigma^2+1}}\right).
\end{align}
If we let $\eta_\epsilon=0$, the scenario with no noise, this reduces to $P_{j,\mathrm{HOM}}$ with $\epsilon=0$, as we would expect.

From these probabilities, we can calculate the Fisher information:
\begin{equation}
	F_\mathrm{HOM}^\eta = \frac{16 \alpha^2 \delta^2 \sigma ^4}{\left(4 \eta_\epsilon^2 \sigma^2+1\right)^2 \left(\left(4 \eta_\epsilon^2 \sigma^2+1\right) e^{\frac{4 \delta^2 \sigma^2}{4 \eta_\epsilon^2 \sigma^2+1}}-\alpha^2\right)}.
\end{equation}

As we would expect, large frequency-dependent noise washes out all information from the protocol: $F_\mathrm{HOM}^\eta \to 0$ as $\eta_\epsilon \to \infty$.

\subsection{MZ2s}
For MZ2s, our noisy probabilities can be calculated as in the \ac{HOM} case but with an additional averaging over $\theta$. The resulting probabilities are
\begin{align}
	P_{1,\mathrm{MZ2s}}^\eta &= \frac{1}{8} \left( 2 + \frac{e^{-\frac{2 \delta^2 \sigma^2}{4 \eta_\epsilon^2 \sigma^2+1}}}{\sqrt{4 \eta_\epsilon^2 \sigma^2+1}}+\cos (\delta  \omegap) e^{-2 \eta_\theta^2-\frac{\eta_\epsilon^2 \omegap^2}{2}} \right. \nonumber \\*
	&\left.\qquad\qquad-\frac{4 \cos \left(\frac{\delta  \omegap}{2 \eta_\epsilon^2 \sigma^2+2}\right) e^{-\kappa}}{\sqrt{\eta_\epsilon^2 \sigma^2+1}}\right),
\label{eq_MZ2s_noisy_prob_1}
\\
	P_{2,\mathrm{MZ2s}}^\eta &= \frac{1}{8} \left( 2 + \frac{e^{-\frac{2 \delta^2 \sigma^2}{4 \eta_\epsilon^2 \sigma^2+1}}}{\sqrt{4 \eta_\epsilon^2 \sigma^2+1}}+\cos (\delta  \omegap) e^{-2 \eta_\theta^2-\frac{\eta_\epsilon^2 \omegap^2}{2}} \right. \nonumber \\*
&\left.\qquad\qquad+\frac{4 \cos \left(\frac{\delta  \omegap}{2 \eta_\epsilon^2 \sigma^2+2}\right) e^{-\kappa}}{\sqrt{\eta_\epsilon^2 \sigma^2+1}}\right),
\label{eq_MZ2s_noisy_prob_2}
\\
P_{c,\mathrm{MZ2s}}^\eta &= \frac{1}{4} \left(2-\frac{e^{-\frac{2 \delta^2 \sigma^2}{4 \eta_\epsilon^2 \sigma^2+1}}}{\sqrt{4 \eta_\epsilon^2 \sigma^2+1}}-\cos (\delta  \omegap) e^{-2 \eta_\theta^2-\frac{\eta_\epsilon^2 \omegap^2}{2}}\right),
\label{eq_MZ2s_noisy_prob_c}
\end{align}
with
\begin{equation}
\kappa = \frac{4 \left(\sigma^2 \left(\delta^2+\eta_\theta^2 \eta_\epsilon^2\right)+\eta_\theta^2\right)+\eta_\epsilon^2 \omegap^2}{8(1 + \eta_\epsilon^2 \sigma^2)}.
\label{eq_kappa}
\end{equation}

The Fisher information can then be calculated from the above probabilities, though the resulting expression is not particularly illuminating so we here omit it, see the Supplemental Material \cite{supp} for full expression.
Once again, information decays to zero as $\eta_\epsilon$ is increased.
However, intriguingly the same is not true when increasing $\eta_\theta$:
in this case the information instead decays until it resembles a ``HOM-like'' Fisher information curve, converging at around $\eta_\theta\approx3.5$.
This curious result will be discussed in more detail in Sec.~\ref{sec_results} and Appendix~\ref{app_cc}.

\begin{table*}[htbp]
\centering
\begin{tabular}{p{10cm}cccccccccccc}
	\toprule
	& \multicolumn{4}{c}{HOM} & \multicolumn{4}{c}{MZ2s} & \multicolumn{4}{c}{MZ2d} \\
	\cmidrule(r){2-5} \cmidrule(r){6-9} \cmidrule(r){10-13} 
	& \multicolumn{2}{c}{MC} & \multicolumn{2}{c}{MU} & \multicolumn{2}{c}{MC} & \multicolumn{2}{c}{MU} & \multicolumn{2}{c}{MC} & \multicolumn{2}{c}{MU} \\
	\cmidrule(r){2-3} \cmidrule(r){4-5} \cmidrule(r){6-7} \cmidrule(r){8-9} \cmidrule(r){10-11} \cmidrule(r){12-13}
	& FE & Ind. & FE & Ind. & FE & Ind. & FE & Ind. & FE & Ind. & FE & Ind. \\
	\midrule
	Residual oscillating information as $\delta\to\infty$ (Fig.~\ref{fig_fringes}) & & & & & \checkmark & & \checkmark & & \checkmark & & \checkmark & \\
	Visibility-dependence & \checkmark & \checkmark & \checkmark & \checkmark &  &  & $\dagger$ & $\dagger$ & \checkmark  & \checkmark & \checkmark & \checkmark \\
	HOM-like residual at high \( \eta_{\theta} \) independent of $\alpha$ (Fig.~\ref{fig_3d} (left)) & & & & & \checkmark & \checkmark & & & & & & \\
	HOM-like residual at high \( \eta_{\theta} \) \emph{inversely} proportional to $\alpha$ (Fig.~\ref{fig_3d} (right)) & & & & & & & & & \checkmark & \checkmark & & \\
	HOM-like residual at high \( \eta_{\theta} \) proportional to $\alpha$ & \checkmark & \checkmark & \checkmark & \checkmark & & & \checkmark & \checkmark & & & &  \\
	No HOM-like residual at high \( \eta_{\theta} \) & & & & & & & & & & & \checkmark & \checkmark \\
	\bottomrule
\end{tabular}
\caption{Comparing the FI behaviour of our different protocols and mode-correlated noise (MC), mode-uncorrelated noise (MU), frequency-entangled photons (FE), and independendent photons (Ind.) variants.
$\alpha$ is the visibility of our input photon pair.
$^\dagger$ For MZ2s with mode-uncorrelated noise, the visibility dependence is only present at nonzero noise.
If $\eta_\epsilon=\eta_\theta=0$, the information remains constant as visibility is varied.}
\label{table}
\end{table*}

\subsection{MZ2d}
We follow the same method to obtain the noisy probabilities for MZ2d:
\begin{align}
	P_{1,\mathrm{MZ2d}}^\eta &= P_{2,\mathrm{MZ2d}}^\eta = \frac{1}{8} \left(2-\frac{ (1-\alpha) e^{-\frac{2 \delta^2 \sigma^2}{4 \eta_\epsilon^2 \sigma^2+1}} }{ \sqrt{4 \eta_\epsilon^2 \sigma^2+1} } \right. \nonumber \\*
	&\mkern64mu\left. - (1+\alpha) \cos (\delta  \omegap) e^{-2 \eta_\theta^2-\frac{\eta_\epsilon^2 \omegap^2}{2}}\right),
\\
	P_{c,\mathrm{MZ2d}}^\eta &= \frac{1}{4} \left(2+\frac{(1-\alpha) e^{-\frac{2 \delta^2 \sigma^2}{4 \eta_\epsilon^2 \sigma^2+1}} }{ \sqrt{4 \eta_\epsilon^2 \sigma^2+1} } \right. \nonumber \\*
	&\mkern64mu\left.+ (1+\alpha) \cos (\delta \omegap) e^{-2 \eta_\theta^2-\frac{\eta_\epsilon^2 \omegap^2}{2}}\right).
\label{eq_MZ2d_noisy_prob_c}
\end{align}

Once again, we omit the unwieldy the Fisher information, the full expression is given in the Supplemental Material~\cite{supp}.
Like MZ2s, we see a full decay to zero at high $\eta_\epsilon$ but decay to a fixed ``HOM-like'' curve at high $\eta_\theta$.

\subsection{MZ1}
Finally, following the same method, we derive the noisy probabilities for MZ1:
\begin{align}
	P_{1,\mathrm{MZ1}}^\eta &= \frac{1}{2} \left( 1-\frac{\cos \left(\frac{\delta \omegap}{2 \eta_\epsilon^2 \sigma^2+2}\right) e^{-\kappa} }{ \sqrt{\eta_\epsilon^2 \sigma^2+1} } \right),
	\\
	P_{2,\mathrm{MZ1}}^\eta &= \frac{1}{2} \left( 1+\frac{\cos \left(\frac{\delta \omegap}{2 \eta_\epsilon^2 \sigma^2+2}\right) e^{-\kappa} }{ \sqrt{\eta_\epsilon^2 \sigma^2+1} } \right),
\end{align}
with $\kappa$ as defined in Eq.~(\ref{eq_kappa}).

The Fisher information is therefore
\begin{equation}
	F_\mathrm{MZ1}^\eta = 
	\frac{ \left(\omegap \sin \left(\frac{\delta  \omegap}{2 \eta_\epsilon^2 \sigma^2+2}\right)+2 \delta  \sigma^2 \cos \left(\frac{\delta  \omegap}{2 \eta_\epsilon^2 \sigma^2+2}\right)\right)^2 }{ 4 \left(\eta_\epsilon^2 \sigma^2+1\right)^2 \left(\left(\eta_\epsilon^2 \sigma^2+1\right) e^{2\kappa}-\cos^2\left(\frac{\delta  \omegap}{2 \eta_\epsilon^2 \sigma^2+2}\right)\right) }.
\end{equation}
Again, as $\eta_\epsilon$ increases the Fisher information tends to zero.
In this case, the same is true for frequency-independent noise via its dependence on $\kappa$ [Eq.~(\ref{eq_kappa})],
i.e.~$F_\mathrm{MZ1}^\eta\to0$ as $\eta_\theta \to \infty$.


\section{Two-photon model variations}\label{sec_alternatives}

For the purpose of probing and better understanding the origin of some of the subtleties we shall discuss in the next section, we introduce some slight variations for the noise-model, which primarily affect the two-photon MZ case. These are explained in the following; and the full expression for the probabilities and Fisher information, for every combination of protocol and model, are given in the Supplemental Material \cite{supp}.

\subsection{Mode-uncorrelated noise}
So far we have applied a common phase shift to both the \( \hat{c} \) and \( \hat{d} \) photonic modes, and so the noise affects the indistinguishable and distinguishable components in a correlated fashion.
We now consider the case where the two photon modes experience distinct phase shifts
such that the propagation transformations for MZ2s and MZ2d, previously given in Eq.~\eqref{eq:delay_encoding}, now take the form
\begin{equation}
\begin{gathered}
	\hat{c}_1^\dagger(\omega) \to e^{-i\omega(\delta_1+\epsilon_1)}e^{-i\theta_1}\hat{c}_1^\dagger(\omega), \\*
	\hat{c}_2^\dagger(\omega) \to e^{-i\omega\delta_2}\hat{c}_2^\dagger(\omega), \\*
	\hat{d}_1^\dagger(\omega) \to e^{-i\omega(\delta_1+\epsilon_2)}e^{-i\theta_2}\hat{d}_1^\dagger(\omega), \\*
	\hat{d}_2^\dagger(\omega) \to e^{-i\omega\delta_2}\hat{d}_2^\dagger(\omega).
\end{gathered}
\label{eq_trans_prop_modes}
\end{equation}
Now, $\epsilon_1$ and $\theta_1$ solely shift the $\hat{c}_1$ mode, with $\epsilon_2$ and $\theta_2$ being a shifts for the $\hat{d}_1$ mode.
The initial state and other transformations remain unchanged.
This leads to a subtly different output state, from which we can derive the probabilities as before.

For MZ1, this obviously need not be considered as only one photon---and thus one single mode---is considered.
For HOM, the orthogonal mode does not interfere at the beamsplitter, and any phase shifts applied drop out.
Hence, this only has a tangible effect on MZ2s and MZ2d, where a photon in a superposition of modes interferes \textit{with itself}.

A minor effect of working with mode-uncorrelated noise is that the resulting probabilities for MZ2s are now visibility dependent
as the uncorrelated noise means the two photonic modes see different noise and so only experience the same optics on average.
This dependence naturally drops out at zero noise.

\subsection{Independent photons}
We have previously assumed our photons to be frequency-entangled, such as a photon pair generated by \ac{SPDC}.
We can also compare to the case where the input photons are independent (but with frequencies peaked around the same value).

For \ac{HOM} and MZ2d, the initial state then looks like
\begin{align}
	\ket{\psi_\mathrm{HOM}^\mathrm{in}} = \ket{\psi_\mathrm{MZ2d}^\mathrm{in}} = &\int d\omega_1 d\omega_2 \> \phi(\omega_1) \phi(\omega_2) [\sqrt{\alpha}\>\hat{a}_2^\dagger(\omega_1) \nonumber \\*
	&+\sqrt{1-\alpha}\>\hat{b}_2^\dagger(\omega_1)]\hat{a}_1^\dagger(\omega_2)\ket{0},
\end{align}
where $\hat{a}$ and $\hat{b}$ can be relabelled $\hat{c}$ and $\hat{d}$ for the HOM case as we omit the first beamsplitter.
For MZ2s we get
\begin{align}
	\ket{\psi_\mathrm{MZ2s}^\mathrm{in}} = \frac{1}{\sqrt{1+\alpha}} &\int d\omega_1 d\omega_2 \> \phi(\omega_1)\phi(\omega_2) [\sqrt{\alpha}\>\hat{a}_1^\dagger(\omega_1)\nonumber \\*
&+\sqrt{1-\alpha}\>\hat{b}_1^\dagger(\omega_1)] \hat{a}_1^\dagger(\omega_2) \ket{0}.
\end{align}
MZ1 features only a single photon so is as Eq.~(\ref{eq_MZ1_input}).

In deriving the output states the same transformations as detailed in Sec.~\ref{sec_protocols} can be followed, optionally replacing the transformations in Eq.~\eqref{eq:delay_encoding} with those in Eq.~(\ref{eq_trans_prop_modes}) if we wish to model mode-uncorrelated noise.
These output states are of a common form, and the general detection probabilities for states of this form are given in Eqs.~(\ref{eq_general_prob_j_ind}, \ref{eq_general_prob_c_ind}).
From here, obtaining the noisy probabilities and the Fisher information for each protocol follows the same identical steps as for frequency-entangled photons. noise.

\section{Results}\label{sec_results}
The three two-photon protocols (HOM, MZ2s, and MZ2d), for frequency-entangled and separable photons, with photonic-mode correlated and uncorrelated noise are tabulated in Table~\ref{table} along with a characterisation of their behaviour in different regimes.
In the following subsections we discuss in more depth some of the more interesting observations.

For simplicity we choose not to specify a specific value for the pump frequency, and instead scale other quantities relative to an arbitrary $\omegap$.
The spectral width is then written as some fraction of the pump frequency; we fix $\sigma=\frac{\omegap}{100}$ in the following which is within the experimentally viable range of the ratios of Refs.~\cite{Olindo-2006,Lyons-2018}.

While both $\eta_{\theta}$ and $\eta_{\epsilon}$ are unbounded in principle, their action within the Mach-Zehnder is limited to the $2\pi$ periodicity, with \( \eta_{\theta} \sim 2\pi \) or \( \eta_{\epsilon} \sim 4\pi/\omegap \) approaching the regime where the effective phase is uniform in \( 2\pi \).

\begin{figure}[htb]
\centering
\tikzsetnextfilename{epsnoise}
\begin{tikzpicture}
	\begin{groupplot}[group style={
		group size = 1 by 3,
		horizontal sep=0pt,
		vertical sep=0pt,
		xticklabels at=edge bottom,
	},
	xmin=-100,xmax=100,
	xtick={-100,-50,0,50,100},
	width=0.95\linewidth,
	height=0.6\linewidth,
	ylabel style={at={(axis description cs:-0.125,.5)},rotate=-90},
	ylabel = $\frac{F}{\omegap^2}$,
	every axis title/.style={below left,at={(1,1)},fill=white,fill opacity=0.75,text opacity=1},
	every y tick scale label/.style={at={(0.05,0.9)},anchor=south west,inner sep=0pt}
	]
\nextgroupplot[
	title = {$\eta_\epsilon\omegap=0$},
	ytick={0,0.25,0.5,0.75,1},
	yticklabels={0,0.25,0.5,0.75,1},
	ymin=0,
	ymax=1,
]
	\addplot[HOMcolour, very thick] file {data/epsnoise1a.dat};
	\addplot[MZ2scolour, thick] file {data/epsnoise1b.dat};
	\addplot[MZ2dcolour] file {data/epsnoise1c.dat};

\nextgroupplot[
	title = {$\eta_\epsilon\omegap=5$},
	ytick={0,2/10000,4/10000,6/10000},
	ymin=0,
	ymax=7/10000,
]
	\addplot[MZ2dcolour, very thick] file {data/epsnoise2c.dat};
	\addplot[MZ2scolour, thick] file {data/epsnoise2b.dat};
	\addplot[HOMcolour, very thick] file {data/epsnoise2a.dat};

\nextgroupplot[
	title = {$\eta_\epsilon\omegap=10$},
	ytick={0,1/10000,2/10000},
	ymin=0,
	ymax=0.00022,
	xlabel = $\delta\>\omegap$,
	legend style = {
		draw = none,
		at={(0.5,-0.25)},
		anchor=north,
		legend columns = 3,
		/tikz/every even column/.append style={column sep=4mm}},
]
	\addplot[MZ2dcolour, very thick] file {data/epsnoise3c.dat};
	\addplot[MZ2scolour, very thick] file {data/epsnoise3b.dat};
	\addplot[HOMcolour, very thick] file {data/epsnoise3a.dat};
	\legend{MZ2d,MZ2s,HOM};
\end{groupplot}
\end{tikzpicture}
\caption{Comparison of the two-photon protocols, in the frequency-entangled, mode-correlated noise configuration with
$\sigma=\frac{\omegap}{100}$ and $\alpha = 0.9$.
As frequency-noise increases, all protocols see a marked reduction in the Fisher information.
MZ2d performs best at low noise, but is most sensitive to noise.
\ac{HOM} initially performs by far the worst, but is significantly more resilient to noise.
MZ2s lies in between in regards to both initial performance and resilience to noise.}
\label{fig_noise_comp_dep}
\end{figure}
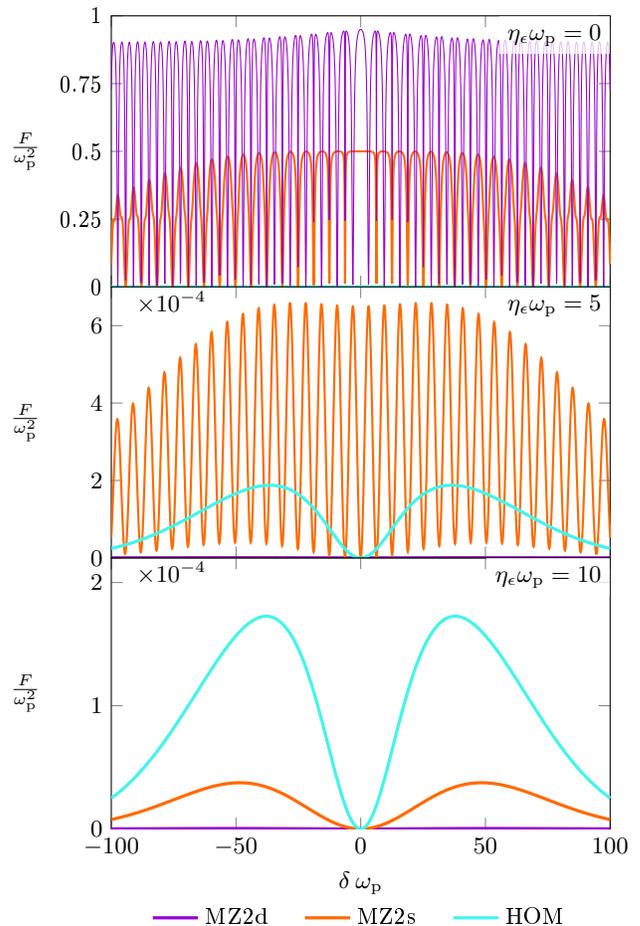

\subsection{Frequency-dependent noise}

Fig.~\ref{fig_noise_comp_dep} illustrates how increasing frequency-dependent noise affects our protocols.
Unsurprisingly, when noise is near zero, HOM is far outclassed by the two MZ protocols.
MZ2d performs best in this low-noise regime, but it drops rapidly as noise increases.
MZ2s is more resilient, suffering a sizeable performance hit as noise is increased but retaining its relative advantage for longer.
HOM performs worst but experiences only comparatively minor information loss.
For all protocols information decays as $\eta_\epsilon$ is increased, as the phase shifts due to noise grow larger and thus sensitivity to the true delay we wish to measure, $\delta$, is reduced.
In the limit $\eta_\epsilon\to\infty$ the phase shifts from noise become effectively wholly random and detection probabilities are thus constant at $P_1=P_2=\frac{P_c}{2}$.
The Fisher information is therefore everywhere equal to zero.

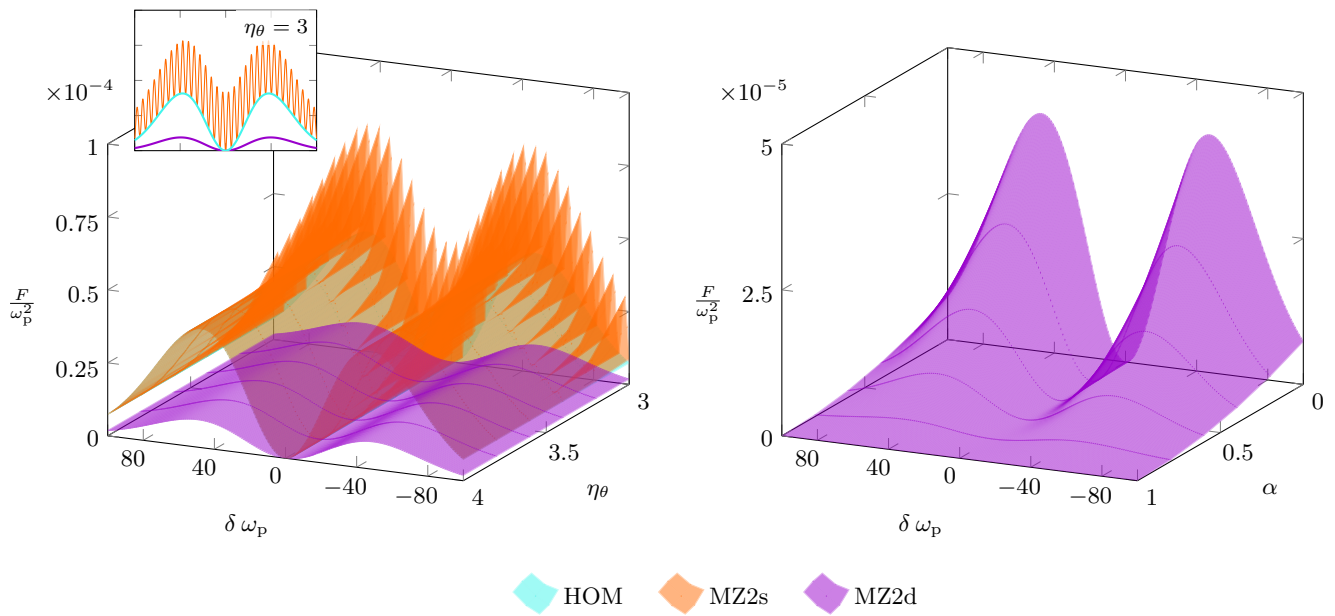
\begin{figure*}
\centering
\tikzsetnextfilename{3dplots}
\begin{tikzpicture}

\begin{axis}[
	width=0.475\linewidth,
	view={205}{20},
	legend style = {
		draw = none,
		at={(1.17,-0.2)},
		anchor=north,
		/tikz/every even column/.append style={column sep=4mm},
		legend columns = 5},
	xlabel = $\delta\>\omegap$,
	xtick={-80,-40,0,40,80},
	ylabel = $\eta_\theta$,
	ytick={3,3.5,4},
	zlabel style={rotate=-90},
	zlabel = $\frac{F}{\omegap^2}$,
	zmin = 0,
	zmax = 1/10000,
	ztick = {0,0.25/10000,0.5/10000,0.75/10000,1/10000}
	],
	\addplot3[surf,shader=faceted,mesh/ordering=y varies, mesh/rows=6,no marks,faceted color=HOMcolour,color=HOMcolour,opacity=0.3,fill opacity=0.5] file{data/thetanoisea.dat};
	\addplot3[surf,shader=faceted,mesh/ordering=y varies, mesh/rows=6,no marks,faceted color=MZ2scolour,color=MZ2scolour,opacity=0.3,fill opacity=0.5] file{data/thetanoiseb.dat};
	\addplot3[surf,shader=faceted,mesh/ordering=y varies, mesh/rows=6,no marks,faceted color=MZ2dcolour,color=MZ2dcolour,opacity=0.3,fill opacity=0.5] file{data/thetanoisec.dat};
	\legend{HOM, MZ2s, MZ2d};
\end{axis}
\begin{axis}[xshift=0.02\textwidth,yshift=0.245\textwidth,tiny,
	axis background/.style={fill=white},
	title style={below left,at={(1,1)},fill=white,fill opacity=0.75,text opacity=1},
	title = {$\eta_\theta=3$},
	xtick={-100,-50,0,50,100},
	xticklabels={},
	xmin=-100,
	xmax=100,
	ylabel style={rotate=-90},
	ymin=0,
	ytick={0,0.25/10000,0.5/10000,0.75/10000,1/10000},
	scaled y ticks = false,
	yticklabels={},
	yticklabel pos=right,
	ymax = 1/10000,
]
	\addplot[MZ2dcolour, thick] file {data/thetanoiseinsetc.dat};
	\addplot[MZ2scolour] file {data/thetanoiseinsetb.dat};
	\addplot[HOMcolour, thick] file {data/thetanoiseinseta.dat};
\end{axis}

\begin{axis}[
	width=0.475\linewidth,
	xshift=0.5\textwidth,
	view={205}{20},
	xlabel = $\delta\>\omegap$,
	xtick={-80,-40,0,40,80},
	ylabel = $\alpha$,
	ytick={0,0.5,1},
	zlabel style={rotate=-90},
	zlabel = $\frac{F}{\omegap^2}$,
	zmin = 0,
	zmax = 0.5/10000,
	ztick = {0,0.25/10000,0.5/10000}
	],
	\addplot3[surf,shader=faceted,mesh/ordering=y varies, mesh/rows=6,no marks,faceted color=MZ2dcolour,color=MZ2dcolour,opacity=0.3,fill opacity=0.5] file{data/thetanoisecalpha.dat};
\end{axis}

\end{tikzpicture}
\caption{We examine behaviour at and approaching the high frequency-independent noise limit. Left: Comparison of the two-photon protocols, in the frequency-entangled, mode-correlated noise configuration, at high frequency-independent noise with $\sigma=\frac{\omegap}{100}$ and $\alpha = 0.5$. 
Frequency-independent noise does not affect HOM.
Due to classical correlations (see Appendix~\ref{app_cc}) between the two photons, for our MZ protocols information is not entirely washed out.
Instead, as $\eta_\theta$ increases, both protocols tend towards a ``HOM-like'' information curve.
For MZ2d, this curve is visibility-dependent.
For MZ2s, which is always visibility-independent, it tends towards a curve that exactly matches the HOM curve at 50\% visibility.
The inset shows the three curves for the far face ($\eta_\theta=3$).
Right: The visibility dependence of MZ2d, in the high frequency-independent noise limit ($\eta_\theta = 4$) with $\sigma=\frac{\omegap}{100}$.
The information has an inverse dependence on the visibility, and at $\alpha=0$ the curve matches that of our HOM protocol with $\alpha=0.5$. See Appendix~\ref{app_cc} for further discussion.}
\label{fig_3d}
\end{figure*}

Similar plots, for a frequency-independent photon input, are given in Fig.~\ref{fig_noise_comp_ind} in Appendix~\ref{app_ind_noise}. Though the exact values differ, the same trends follow with increasing noise. This is also true for our mode-uncorrelated noise model.

\subsection{Frequency-independent noise}
Initially, increasing frequency-independent noise seems to affect the MZ protocols analagously to frequency-dependent noise, while HOM is the exception being entirely unaffected.
However there is an interesting high-noise limit, which we depict in the left plot of Fig.~\ref{fig_3d}.
Rather than information tending towards zero everywhere, increased $\eta_\theta$ merely washes out the fringes; and either side of $\delta=0$ two peaks remain: much like the familiar HOM information curve.
Indeed, for MZ2s this limit exactly matches a HOM curve with 50\% visibility.
MZ2d remains visibility-dependent even in this limit.
Curiously, this result suggests that for a low-visibility ($\alpha<0.5$) photon pair MZ2s is always preferable to HOM for maximising information.

In low-noise scenarios ($\eta_\theta\lesssim1.5$), MZ2d has an intuitive dependence on visibility: information is highest at $\alpha=1$ and lowest (though non-zero) at $\alpha=0$.
However when noise is larger ($\eta_\theta\gtrsim2.3$), and particularly in the high $\eta_\theta$ limit, information is inversely proportional to visibility, as seen in the right plot of Fig.~\ref{fig_3d}.
At $\alpha=1$, information vanishes, whereas at $\alpha=0$ the curve again matches the HOM curve for 50\% visibility.
In the transition region $1.5\lesssim\eta_\theta\lesssim2.3$ visbility dependence is more complicated, with some information peaks still proportional to visibility while others are inversely proportional. The visibility-proportional peaks gradually decay with larger $\eta_\theta$ until they become negligible at $\eta_\theta\approx2.3$.

Considering our mode-uncorrelated noise model, this limit remains for MZ2s: though it now has a direct visibility depedence (no information at $\alpha=0$, matches 50\% visibility HOM at $\alpha=1$).
However, the MZ2d information now tends to zero everywhere at high $\eta_\theta$.

If we instead take our initial photons to be frequency-independent we see the same general behaviour (though the precise value of the Fisher information varies) for both mode-correlated noise, and mode-uncorrelated noise.

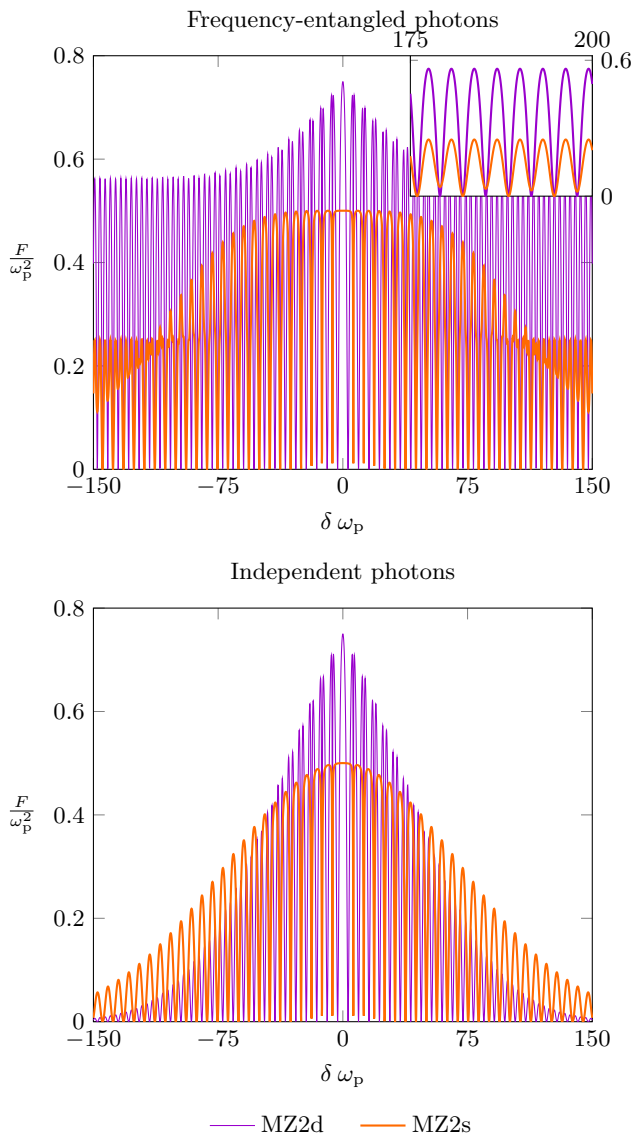
\begin{figure}[htb]
\centering
\tikzsetnextfilename{fringes}
\begin{tikzpicture}

\begin{axis}[
	width=0.95\linewidth,
	title = {Frequency-entangled photons},
	xlabel = $\delta\>\omegap$,
	xtick={-150,-75,0,75,150},
	xmin = -150,
	xmax = 150,
	ylabel style={rotate=-90},
	ylabel = $\frac{F}{\omegap^2}$,
	ytick={0,0.2,0.4,0.6,0.8},
	ymin=0,
	ymax=0.8,
]
	\addplot[MZ2dcolour] file {data/fringes1c.dat};
	\addplot[MZ2scolour, thick] file {data/fringes1b.dat};
	\coordinate (inset) at (rel axis cs:1.00,1.00);
\end{axis}

\begin{axis}[at = (inset),
	anchor = north east,
	tiny,
	every tick label/.append style={font=\small},
	xtick={175,200},
	xmin = 175, xmax = 200,
	xticklabel pos=top,
	yticklabel pos=right,
	ytick={0,0.6},
	ymin=0,
	ymax=0.62,
	every axis plot/.append style={thick},
	axis background/.style={fill=white},
]
	\addplot[MZ2dcolour] file {data/fringes1insetc.dat};
	\addplot[MZ2scolour] file {data/fringes1insetb.dat};
\end{axis}

\begin{axis}[
	yshift=-0.85\linewidth,
	width=0.95\linewidth,
	legend style = {
		draw = none,
		at={(0.5,-0.2)},
		anchor=north,
		legend columns = 3,
		/tikz/every even column/.append style={column sep=4mm}},
	title = {Independent photons},
	xlabel = $\delta\>\omegap$,
	xtick={-150,-75,0,75,150},
	xmin = -150,
	xmax = 150,
	ylabel style={rotate=-90},
	ylabel = $\frac{F}{\omegap^2}$,
	ytick={0,0.2,0.4,0.6,0.8},
	ymin=0,
	ymax=0.8,
]
	\addplot[MZ2dcolour] file {data/fringes2c.dat};
	\addplot[MZ2scolour, thick] file {data/fringes2b.dat};
	\legend{MZ2d,MZ2s};
\end{axis}

\end{tikzpicture}
\caption{Fisher information for frequency-entangled (top) vs independent (bottom) photons with $\sigma=\frac{\omegap}{100}$, $\alpha = 0.5$, and \( \eta_{\epsilon} = \eta_{\theta} = 0 \).
When the photons are frequency-entangled information plateaus to some regular oscillatory information at high $|\delta|$, either side of the central peak (a closeup is shown in inset to top plot).
These enduring fringes are not present when the input photons are frequency-independent, with information gradually decaying to zero at high $|\delta|$.}
\label{fig_fringes}
\end{figure}

This initially perplexing behaviour, which---given the \( \alpha \to 0 \) limit---may at first glance appear to suggest interference between wholly distinguishable photons, arises in fact from 
classical correlations between the paths of the two photons:
At large $\eta_\theta$ uncertainty in the value of the frequency-independent phase shift increases, and in the limit $\eta_\theta\to\infty$ the shift becomes wholly random.
However, though this $\theta$ shift is random, the same size shift is experienced by both photons.
This allows some \( \delta \)-dependence to remain even with distinguishable photons.
In the \( \eta_{\theta} \to \infty \) limit \( \delta \)-dependent interference at the second beam splitter only occurs when one photon is in each mode (when the photons are bunched they each exit the second beam splitter stochastically).
In MZ2d this component is surpressed, according to visibility, by HOM interference at the first beam splitter giving rise to the increasing Fisher information in spite of decreasing visibility.
A deeper analysis of these classical correlations, as well as a discussion of the extra subtleties in the mode-uncorrelated noise model, is given in Appendix~\ref{app_cc}.


\subsection{Oscillatory information}
For the MZ protocols, information peaks at $\delta=0$ which decays as the frequency components lose a common phase.
However, when our input photons are frequency-entangled $F$ does not decay to zero, but rather we see a constant regular oscillatory information.
This is not the case when the input photons are independent: there is still a central peak and some oscillation in the decay, but information ultimately drops to zero at large delays.
Both cases are shown in Fig.~\ref{fig_fringes}, in the scenario with zero noise.
\footnote{The central peaks (at $\delta\to0$) coincide at $\frac{1}{2}(4\sigma^2 +\omegap^2)$ for MZ2s regardless of whether the photons are frequency-entangled or independent (full expressions given in Supplemental Material~\cite{supp}).
For MZ2d, however, the frequency-entangled peak is $2 \sigma^2 + \frac{1}{2} (1+\alpha) \omegap^2 - 2\alpha\sigma^2$ whereas the frequency-independent peak drops the final term and is simply $2 \sigma^2 + \frac{1}{2} (1+\alpha) \omegap^2$.
This makes separable photons marginally favourable around $\delta \sim 0$.} 
When noise is introduced (not shown), these fringes decay faster than the central peaks, and no fringes remain in the high frequency-independent noise limit.

Such oscillations arise because the frequencies of down-converted photons, though themselves unknown, will always sum to the constant pump frequency $\omegap$. A quick inspection of the frequency-entangled coinicidence probabilities for MZ2s [Eq.~(\ref{eq_MZ2s_noisy_prob_c})] and MZ2d [Eq.~(\ref{eq_MZ2d_noisy_prob_c})] reveal these are indeed oscillations at the pump frequency.
These oscillations concur with results from previous two-photon MZ experiments where down-converted photons enter via the same port~\cite{Shih-1994} and via different ports~\cite{Rarity-1990}, and similar oscillatory behaviour is observed in the two-photon Franson interferometer~\cite{Franson-1989}.

By contrast, when the photons are independent, their frequencies no longer sum to some constant $\omegap$. As a result all probabilities tend to constants at high delays and information decays entirely.

Appendix~\ref{app_oscil} shows the relation between the detected signal (probabilities) and the resulting Fisher information in the oscillating region.

\subsection{Zero visibility and comparison with single-photon MZ}
We now want to compare our two-photon MZ protocols to MZ1.
MZ1 has no fringes, so it is natural to compare it more directly to MZ2s and MZ2d with indpendent photons.
We also choose $\alpha=0$ for MZ2d so as to prevent interference at the initial beamsplitter, which yields additional information.

\begin{figure}[htb]
\centering
\tikzsetnextfilename{zerovis}
\begin{tikzpicture}

\begin{axis}[
	yshift=-0.85\linewidth,
	width=0.95\linewidth,
	legend style = {
		draw = none,
		at={(0.5,-0.2)},
		anchor=north,
		legend columns = 3,
		/tikz/every even column/.append style={column sep=4mm}},
	title = {Independent photons},
	xlabel = $\delta\>\omegap$,
	xtick={-150,-75,0,75,150},
	xmin = -150,
	xmax = 150,
	ylabel style={rotate=-90},
	ylabel = $\frac{F}{\omegap^2}$,
	ytick={0,0.25,0.5},
]
	\addplot[MZ2scolour] file {data/MZ1compb.dat};
	\addplot[MZ2dcolour] file {data/MZ1compc.dat};
	\addplot[MZ1colour] file {data/MZ1compd.dat};
	\legend{MZ2s,MZ2d,MZ1};
\end{axis}

\begin{axis}[
	width=0.95\linewidth,
	legend style = {
		draw = none,
		at={(0.5,-0.2)},
		anchor=north,
		legend columns = 3,
		/tikz/every even column/.append style={column sep=4mm}},
	title = {Frequency-entangled photons},
	xlabel = $\delta\>\omegap$,
	xtick={-150,-75,0,75,150},
	xmin = -150,
	xmax = 150,
	ylabel style={rotate=-90},
	ylabel = $\frac{F}{\omegap^2}$,
	ytick={0,0.25,0.5},
]
	\addplot[MZ2scolour] file {data/FEzerovisb.dat};
	\addplot[MZ2dcolour] file {data/FEzerovisc.dat};
	\coordinate (inset) at (rel axis cs:1.00,1.00);
\end{axis}

\begin{axis}[at = (inset),
	anchor = north east,
	width=0.525\linewidth,
	every tick label/.append style={font=\small},
	xtick={200,215},
	xmin = 200, xmax = 215,
	xticklabel pos=top,
	yticklabel pos=right,
	ytick={0,0.25},
	ymin=0,
	ymax=0.26,
	every axis plot/.append style={thick},
	axis background/.style={fill=white},
]
	\addplot[MZ2dcolour, thin] file {data/FEzerovisinsetc.dat};
	\addplot[MZ2scolour,dashed,very thick] file {data/FEzerovisinsetb.dat};
\end{axis}

\end{tikzpicture}
\caption{Fisher information for frequency-entangled (top) vs independent (bottom) photons, with additional comparison to the single-photon MZ1 protocol in the bottom plot.
We choose zero visibility to minimise any two-photon interference effects, and plot $\sigma=\frac{\omegap}{100}$ and $\eta_{\epsilon} = \eta_{\theta} = 0$.
Though the central peaks of MZ2d and MZ2s coincide, MZ2d generally performs worse at zero visibility. This is true everywhere when the input photons are independent, but for the frequency-entangled case the two curves converge at the fringes (shown in inset to top plot).
When the photons are independent, MZ2s has exactly double the information compared to MZ1.}
\label{fig_zerovis}
\end{figure}
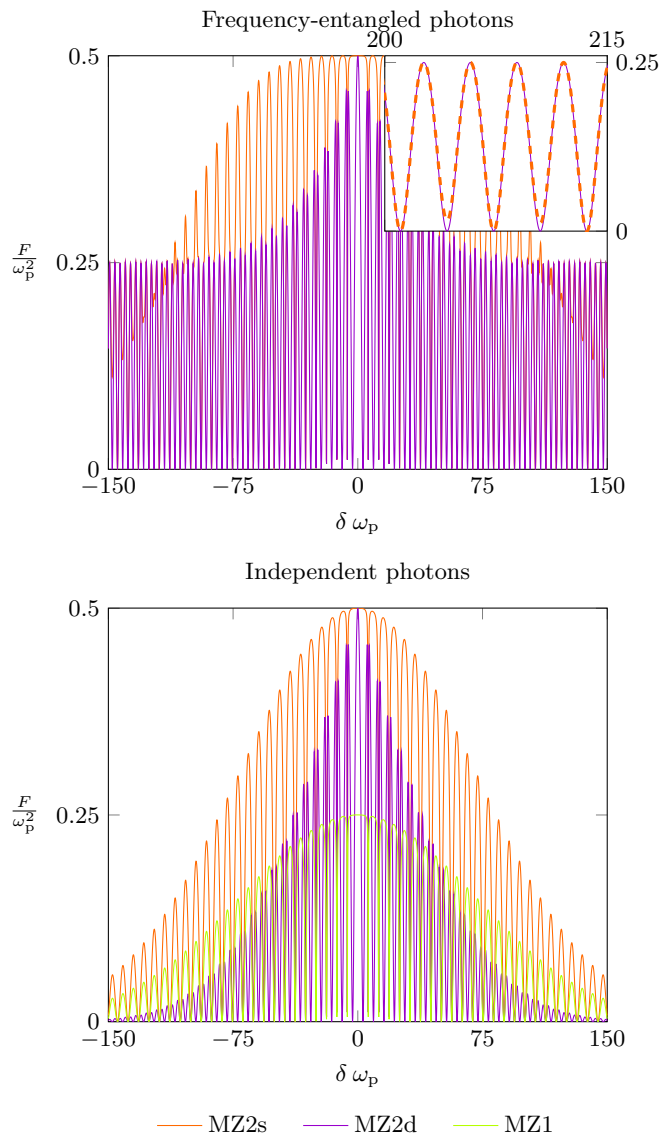

We might suppose that both MZ2s and MZ2d are now equivalent to twice MZ1.
However, the bottom plot of Fig.~\ref{fig_zerovis} shows this is only true for MZ2s which does indeed have double the information of MZ1.
Though MZ2d shares the same central peak, its overall shape is different to MZ2s (and thus, different to twice MZ1).
This is a limitation of the measurements which do not distinguish between the two photonic modes.
In MZ1s and MZ2s we can recognise whether the photon(s) are detected in the same spatial mode they started or the opposite, as the photon(s) always start in a common mode.
In MZ2d the same coincidence event is observed when each photon is detected in the original spatial mode as when they are detected in the opposite spatial mode due to this non-distinsuihing measurement.
This gives rise to an apparent loss in precision, albeit one which could be alleviated with a distinguishing measurement.

Returning to frequency-entangled photons, the top plot of Fig.~\ref{fig_zerovis} shows that we see similarly different Fisher information for both protocols despite $\alpha=0$, though at large delays the information coincides at the fringes.

\subsection{Maximal infomation}
\begin{figure}[htb]
\centering
\tikzsetnextfilename{maxinf}
\begin{tikzpicture}
\begin{axis}[
	width=0.9\linewidth,
	legend style = {
		draw = none,
		at={(0.5,-0.2)},
		anchor=north,
		/tikz/every even column/.append style={column sep=4mm},
		legend columns = 3},
	xlabel = $\eta$,
	xtick={0,1,2,3,4,5},
	xmin=0,
	xmax=5,
	ylabel style={rotate=-90},
	ylabel = $\frac{F}{\omegap^2}$,
	ytick={0,0.25,0.5,0.75,1},
	ymin=0,
	ymax=1,
	every axis plot/.append style={very thick},
]
	\addplot[MZ2scolour] file {data/maxinf1b.dat};
	\addplot[MZ2dcolour] file {data/maxinf1c.dat};
	\addplot[MZ1colour] file {data/maxinf1d.dat};
	\addplot[MZ2scolour, dashed] file {data/maxinf2b.dat};
	\addplot[MZ2dcolour, dashed] file {data/maxinf2c.dat};
	\addplot[MZ1colour, dashed] file {data/maxinf2d.dat};
	\legend{MZ2s ($\eta_\epsilon\omegap$),MZ2d ($\eta_\epsilon\omegap$),2$\times$MZ1 ($\eta_\epsilon\omegap$),MZ2s ($\eta_\theta$),MZ2d ($\eta_\theta$),2$\times$MZ1 ($\eta_\theta$)};
	\coordinate (inset) at (rel axis cs:1.00,1.00);
\end{axis}

\begin{axis}[at = (inset),
	anchor = north east,
	width=0.6\linewidth,
	every tick label/.append style={font=\small},
	xtick={0,40,80},
	xmin = 0, xmax = 100,
	xticklabel pos=bottom,
	yticklabel pos=right,
	scaled y ticks=false,
	ytick={0,0.0002},
	yticklabels={0,0.0002},
	ymin=0,
	ymax=0.0002,
	every axis plot/.append style={thick},
	axis background/.style={fill=white},
]
	\addplot[HOMcolour] file {data/maxinf1a.dat};
	\legend{HOM ($\eta_\epsilon\omegap$)};
\end{axis}
\end{tikzpicture}
\caption{Maximum Fisher information for the frequency-entangled two-photon Mach-Zehnder protocols against noise with $\sigma=\frac{\omegap}{100}$ and $\alpha=0.9$.
For comparison, we also plot twice the MZ1 maximum.
HOM performance is shown in the inset.
The solid lines show the information decrease as frequency-dependent noise (scaled with $\omegap$) grows, and the dashed lines show the decrease with increasing frequency-independent noise.
We saw previously that in the absence of noise two runs of MZ1 performs comparably to MZ2s, and MZ2d beats both provided $\alpha>0$.
We now note that MZ1 is in fact slightly more resilient to noise: while MZ2s emerges as superior to MZ2d as noise grows, it now dips below the performance of two MZ1 runs, except in the high $\eta_\theta$ limit where MZ2s retains some residual information [see Fig.~\ref{fig_3d} (left)] while MZ1 information decays fully to zero.
}
\label{fig_MZ2_v_MZ1_peaks}
\end{figure}
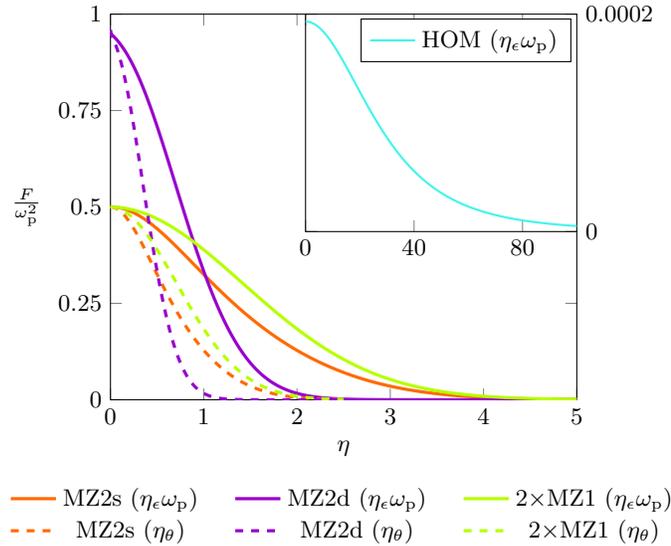

As a final comparison, Fig.~\ref{fig_MZ2_v_MZ1_peaks} explores the resilience of MZ1 to noise in comparison to its frequency-entangled two-photon counterparts.
Plotting only the maximum information, as both types of noise increase, we again see the superior performance of MZ2d at low noise (provided a sufficiently high visibility).
We also see, however, that MZ1 proves slightly more resilient to noise than MZ2s.
Thus, in scenarios where noise is sufficiently high so that MZ2d is no longer preferable, but before reaching such high values that HOM is preferable to all MZ protocols, the best choice is not MZ2s but rather two individual runs of MZ1.

We also see that we require higher values of $\eta_\epsilon\omegap$ to obtain a similar information decrease compared to lower $\eta_\theta$ values.
The disruptive effects of frequency-dependent noise naturally scale with the frequency. Specifically, recalling the central frequency is $\frac{\omegap}{2}$  we would expect performance with frequency-dependent noise $\eta_\epsilon$ to be roughly equivalent to performance with frequency-independent noise $\eta_\theta=\eta_\epsilon \frac{\omegap}{2}$.\footnote{For an immediate justification for why they cannot be exactly equal, recall that the Fisher information tends to different high noise limits for each type of noise.}
Thus, normalising as we do with respect to the pump frequency $\omegap$ we see both sets of maximal Fisher information curves in Fig.~\ref{fig_MZ2_v_MZ1_peaks} separated by a factor of $2$.

In perfect conditions (no noise, $\alpha=1$) MZ2d performs twice as well as MZ2s and four times as well as MZ1.
This can be associated with the monochromatic limits where the MZ2d probe state equates to a Heisenberg-scaling 2-photon N00N state, while the MZ1 and MZ2s probes are shot-noise limited Fock states~\cite{Demkowicz-Dobrzanski-2015}.

Additonally, Table~\ref{table_noise} gives the requisite noise values that reduce the peak information by \SI{50}{\percent} compared to the peak information in the absence of noise.

\begin{table}[htb]
\centering
\begin{tabular}{ccccc}
\toprule
   & MZ2d & MZ2s & MZ1  & HOM   \\ \midrule
$\eta_\epsilon\omegap$ & 0.8  & 1.32 & 1.66 & 27.36 \\
$\eta_\theta$ & 0.4  & 0.66 & 0.83 & N/A   \\ \bottomrule
\end{tabular}
\caption{Noise values where peak information is halved compared to the zero-noise peak. \ac{HOM} is impervious to frequency-independent noise.}
\label{table_noise}
\end{table}
%
\section{Conclusion}\label{sec_conclusion}

Our results confirm that under ideal, i.e.~noiseless, conditions Mach-Zehnder interferometry offers superior performance over Hong-Ou-Mandel interferometry for the purposes of delay estimation on a per photon basis. 
Once noise is introduced, we show that---in keeping with expectations---the HOM protocol proves remarkably resilient to frequency-dependent noise (equivalent to some unknown jitter in the actual delay), and is in fact wholly unaffected by frequency-independent noise (representing some random phase shift in one or both of the arms). 
In the following, we discuss the performance of MZ interferometry as the level of noise increases, with a focus on the difference between our three two-photon MZ protocols. 

Generally, we have found that two independent runs of a conventional single-photon MZ protocol (MZ1) is preferable to a simultaneous two-photon run where both enter via the \textit{same} input port of the \ac{MZI}, as in our MZ2s protocol, for \textit{any} finite amount of noise. Whilst both are matched at vanishing noise, the superiority of two independent input photons increases with noise severity. The ranking between the (\textit{different} input port) MZ2d protocol and $2 \times$MZ1 depends on the level of noise. The best protocol choice for a given noise regime is summarised in Table~\ref{table_optimals}.

Specifically, in low-noise scenarios, the MZ2d protocol offers optimal performance provided the visibility of the initial photon pair is sufficiently high.
However, beyond a certain noise threshold (the exact point varies depending on other parameters, but for $\alpha=0.9$ and $\sigma=\frac{\omegap}{100}$ the threshold is $\eta_\epsilon\omegap\approx0.9$ or $\eta_\theta\approx0.45$) the performance of MZ2d dips below both $2 \times$MZ1 and MZ2s, with $2 \times$MZ1 marginally preferable.
The experimental setup of Ref.~\cite{Lyons-2018} falls within this moderate noise regime: \SI{2}{fs} average delay drift and a spectral width $\sigma \approx \SI{4.6}{\ps^{-1}}$ corresponds to frequency-dependent noise of roughly $\eta_\epsilon\omegap\approx1.15$.

In high-noise scenarios ($\eta_\epsilon\omegap>5.6$ or $\eta_\theta>2.8$ for $\alpha=0.9$ and $\sigma=\frac{\omegap}{100}$) the performance of all MZ protocols drops substantially, and the best choice is generally to drop the initial beamsplitter and perform the delay estimation based on the \ac{HOM} approach.
The exception to this is if visibility is low ($\alpha<0.5$) and noise is largely frequency-independent, in this case MZ2s may remain the superior protocol owing to the classical correlations discussed in Appendix~\ref{app_cc}.

\begin{table}[htb]
\centering
\begin{tabular}{ccccc}
\toprule
\multicolumn{5}{c}{Noise regime}   \\ \midrule
Low & \hfill & Moderate & \hfill & High \\
$\eta_\epsilon\omegap<0.9$ & \hfill & $0.9<\eta_\epsilon\omegap<5.6$ & \hfill & $\eta_\epsilon\omegap>5.6$  \\
$\eta_\theta<0.45$ & \hfill & $0.45<\eta_\theta<2.8$ & \hfill & $\eta_\theta>2.8$  \\ \midrule
MZ2d & \hfill & $2\times$MZ1 & \hfill & HOM  \\ \bottomrule
\end{tabular}
\caption{The optimal protocol choice for varying noise regimes. Threshold values calculated for $\alpha=0.9$. At higher visibility both MZ2d and HOM perform better, tightening the range of noise values where $2\times$MZ1 (which is unaffected by visibility) is optimal. At lower visibilty the inverse is true: $2\times$MZ1 becomes optimal over a wider range of noise values.}
\label{table_optimals}
\end{table}
Throughout, we see broadly the same qualitative results whether our photons are frequency-entangled (as in an \ac{SPDC} pair) or frequency-independent.
The only remarkable difference being that the former produces regular oscillatory information at delays larger than the single-photon coherence time, whereas information decays rapidly at large delays if the photons are independent. 

The results for mode-correlated vs mode-uncorrelated noise are qualitatively similar, though notable differences arise in the interesting limit of high frequency-independent noise: where classical correlations between photon paths results in some residual information remaining when all other interference has been washed out.

In this paper we have not accounted for the possibility of photon loss, instead treating our detectors as $100\%$ efficient.
In practical scenarios, photon loss can result in some ambiguity as to the location of the second photon when only a single detector clicks. With conventional bucket detectors, which lack photon number-resolving capabilities, it is unclear whether this single click represents a true bunching event, or is simply a consequence of one photon being lost. This ambiguity can be overcome with the introduction of number-resolving detectors~\cite{Scott-2020}. In accounting for loss, the practical detection probabilities will differ slightly depending on the number-resolving capabilities of the detectors.
For either detector type, the Supplemental Material \cite{supp} contains the ability to generate loss-dependent expressions for any of the two-photon protocol configurations we have considered.

In summary, our results provide a rigorous investigation of the effects of noise in \ac{MZ} and \ac{HOM} interferometers traversed by photon pairs. We have confirmed that \ac{HOM} interferometry is indeed largely impervious to phase noise and thus remains the favoured choice for noisy scenarios.
By exploring three different two-photon \ac{MZ} protocols, we have quantitatively established their considerable resilience to phase noise and uncovered interesting subtleties and differences in performance depending on how the photons are fed in. Notably, the existence of `HOM'-like feature in the Fisher information arising from classical correlations and the fact that MZ2d performs up to twice as well in low-noise scenarios compared to MZ2s and $2 \times$MZ1 might be interesting for further exploration, and may become increasingly relevant as more phase-stable \acl{MZ} setups, e.g.~on integrated photonic chips, become more readily available.

\begin{acknowledgments}
This work was supported by UK EPSRC Grants EP/R030413/1 and EP/T00097X/1.
NW wishes to acknowledge support from the Royal Commission for the Exhibition of 1851.
\end{acknowledgments}

\appendix

\section{General detection probabilities}\label{app_general_probs}
\subsection{Frequency-entangled photons}
Our initial two-photon protocols involve a frequency-entangled photon pair (such as that generated by \ac{SPDC}).
Looking at the output states for our HOM [Eq.~\eqref{eq_HOM_output}], MZ2s [Eq.~\eqref{eq_MZ2s_output}], and MZ2d [Eq.~\eqref{eq_MZ2d_output}] protocols; these can be written in the general form
\begin{align}
	\ket{\psi^{\mathrm{out}}} = \int d\omega \sum_{i=1}^2 &\sum_{j=1}^2 \left[C_{e_i,e_j}(\omega) \> \hat{e}_i^\dagger(\omega_p-\omega)\hat{e}_j^\dagger(\omega) \right. \nonumber \\*
	 &\>\left. +\> C_{e_i,f_j}(\omega) \> \hat{e}_i^\dagger(\omega_p-\omega)\hat{f}_j^\dagger(\omega)\right],
\label{eq_general_output}
\end{align}
where the $C$ coefficient functions are now all that differ between protocols.

We can now define the \ac{POVM} elements associated with our detection events. For the case where both photons arrive at a single detector we have
\begin{align}
	\Pi_j = \int d\omega_1 d\omega_2 \> \left[\frac{1}{2}\hat{e}_j^\dagger(\omega_1)\hat{e}_j^\dagger(\omega_2)\right.&\ket{0}\bra{0}\hat{e}_j(\omega_2)\hat{e}_j(\omega_1) \nonumber \\*
+\hat{e}_j^\dagger(\omega_1)\hat{f}_j^\dagger(\omega_2)&\left.\ket{0}\bra{0}\hat{f}_j(\omega_2)\hat{e}_j(\omega_1)\right],
\label{eq_POVM_j}
\end{align}
where $j\in\{1,2\}$ indicates which detector we are considering.
The factor of $\frac{1}{2}$ in the first term is included to account for double counting. For the case of a coincidence at both detectors we have
\begin{align}
	\Pi_c = \int d\omega_1 d\omega_2 \> \left[\hat{e}_1^\dagger(\omega_1)\hat{e}_2^\dagger(\omega_2)\right.&\ket{0}\bra{0}\hat{e}_2(\omega_2)\hat{e}_1(\omega_1) \nonumber \\*
+\hat{e}_1^\dagger(\omega_1)\hat{f}_2^\dagger(\omega_2)&\ket{0}\bra{0}\hat{f}_2(\omega_2)\hat{e}_1(\omega_1)\nonumber \\*
+\hat{f}_1^\dagger(\omega_1)\hat{e}_2^\dagger(\omega_2)&\left.\ket{0}\bra{0}\hat{e}_2(\omega_2)\hat{f}_1(\omega_1)\right].
\label{eq_POVM_c}
\end{align}

These elements are straightforward sums of projectors onto orthogonal states, therefore their positivity is apparent.
We require one further element ${\mathbb{1}-(\Pi_1+\Pi_2+\Pi_c)}$ to strictly complete the \ac{POVM}, however as we are only considering states of the form of Eq.~(\ref{eq_general_output}) the events associated with that element all occur with probability zero and our sets of probabilities for each protocol sum to one.

We can now calculate the detection probabilities for our general output state.
We have
\begin{align}
	P_j &= \bra{\psi^{\mathrm{out}}}\Pi_j\ket{\psi^{\mathrm{out}}} \nonumber \\*
	&= \int d\omega \>\left[|C_{e_j,f_j}(\omega)|^2+\frac{1}{2}|C_{e_j,e_j}(\omega)+C_{e_j,e_j}(\omega_p-\omega)|^2\right]
\label{eq_general_prob_j}
\end{align}
for the probability of detection at detector $j$ and
\begin{align}
	P_c &= \bra{\psi^{\mathrm{out}}}\Pi_c\ket{\psi^{\mathrm{out}}} \nonumber \\*
	&=\int d\omega \> \left[|C_{e_1,f_2}(\omega)|^2+\>|C_{e_2,f_1}(\omega)|^2 \right.\nonumber \\*
	&\left.\qquad\qquad+|C_{e_1,e_2}(\omega)+C_{e_2,e_1}(\omega_p-\omega)|^2\right]
\label{eq_general_prob_c}
\end{align}
for the probability of coincidence at both detectors.

We can now insert the relevant coefficient functions for each output state to obtain the fixed shift probabilities presented in Sec.~\ref{sec_protocols}.
\subsection{Independent photons}
For the alternative model discussed in Sec.~\ref{sec_alternatives}, we now assume both input photons to have independent frequencies.
The general output state is now of the form
\begin{align}
	\ket{\psi^{\mathrm{out}}} = \int d\omega_1 d\omega_2 \sum_{i=1}^2 &\sum_{j=1}^2 \left[C_{e_i,e_j}(\omega_1,\omega_2) \> \hat{e}_i^\dagger(\omega_2)\hat{e}_j^\dagger(\omega_1) \right. \nonumber \\*
	 &\>\left. +\> C_{e_i,f_j}(\omega_1,\omega_2) \> \hat{e}_i^\dagger(\omega_2)\hat{f}_j^\dagger(\omega_1)\right] .
\label{eq_general_output_ind}
\end{align}

Our \ac{POVM} elements are the same as before [Eqs.~(\ref{eq_POVM_j}), (\ref{eq_POVM_c})] and we can calculate the detection probabilties as
\begin{align}
	P_j &= \bra{\psi^{\mathrm{out}}}\Pi_j\ket{\psi^{\mathrm{out}}} \nonumber \\*
	&= \int d\omega_1 d\omega_2 \>\left[|C_{e_j,f_j}(\omega_1,\omega_2)|^2\right. \nonumber \\*
	&\qquad\qquad \left.+\frac{1}{2}|C_{e_j,e_j}(\omega_1,\omega_2)+C_{e_j,e_j}(\omega_2,\omega_1)|^2\right]
\label{eq_general_prob_j_ind}
\end{align}
for the probability of detection at detector $j$ and
\begin{align}
	P_c &= \bra{\psi^{\mathrm{out}}}\Pi_c\ket{\psi^{\mathrm{out}}} \nonumber \\*
	&=\int d\omega_1 d\omega_2 \> \left[|C_{e_1,f_2}(\omega_2,\omega_1)|^2+\>|C_{e_2,f_1}(\omega_1,\omega_2)|^2 \right.\nonumber \\*
	&\left.\qquad\qquad+|C_{e_1,e_2}(\omega_2,\omega_1)+C_{e_2,e_1}(\omega_1,\omega_2)|^2\right]
\label{eq_general_prob_c_ind}
\end{align}
for the probability of coincidence at both detectors.

Once again, inserting the relevant coefficients for the output state of a specific protocol yields the (fixed $\delta$ / phase shift) detection probabilities for that protocol.

\vspace{5mm}
\section{Noise in two arms}\label{app_both_arms}
Throughout this paper we have opted to localise noise entirely within one arm of the interferometer. To motivate this choice and show that this neat simplification is sufficient (even if physically noise occurs in both arms),
we will explicitly model noise split between two arms for our MZ2s protocol in the folllowing.
This requires us to first introduce an additional pair of fixed phase shifts for the lower arm; we therefore modify the propogation transformations given in Eq.~\eqref{eq:delay_encoding} to now read
\begin{equation}
\begin{gathered}
	\hat{c}_1^\dagger(\omega) \to e^{-i\omega(\delta_1+\epsilon_1)}e^{-i\theta_1}\hat{c}_1^\dagger(\omega), \\*
	\hat{c}_2^\dagger(\omega) \to e^{-i\omega(\delta_2+\epsilon_2)}e^{-i\theta_2}\hat{c}_2^\dagger(\omega), \\*
	\hat{d}_1^\dagger(\omega) \to e^{-i\omega(\delta_1+\epsilon_1)}e^{-i\theta_1}\hat{d}_1^\dagger(\omega), \\*
	\hat{d}_2^\dagger(\omega) \to e^{-i\omega(\delta_2+\epsilon_2)}e^{-i\theta_2}\hat{d}_2^\dagger(\omega).
\end{gathered}
\end{equation}

The exact same method to derive the noisy probabilities now follows: obtain the output state, calculate the fixed shift probabilities, then average over the fixed shifts.
The only additional requirement is the introduction of a second pair of integrals over $\epsilon_2$ and $\theta_2$.
The resulting noisy probabilities are
\begin{widetext}
\begin{align}
	P_{1,\mathrm{MZ2s}}^\eta &= \frac{1}{8} \left(2+\frac{e^{-\frac{2 \delta^2 \sigma^2}{4 \etamag{\epsilon} \sigma^2+1}}}{\sqrt{4 \etamag{\epsilon} \sigma^2+1}}+\cos (\delta  \omegap) e^{-2 \etamag{\theta}-\frac{\etamag{\epsilon} \omegap^2}{2}} 
-\frac{4 \cos \left(\frac{\delta  \omegap}{2 \etamag{\epsilon} \sigma^2+2}\right) e^{-\frac{4 \delta^2 \sigma^2+4 \etamag{\theta} \left(\etamag{\epsilon} \sigma^2+1\right)+\etamag{\epsilon} \omegap^2}{8 \etamag{\epsilon} \sigma^2+8}}}{\sqrt{\etamag{\epsilon} \sigma^2+1}}\right),
\\
	P_{2,\mathrm{MZ2s}}^\eta &= \frac{1}{8} \left(2+\frac{e^{-\frac{2 \delta^2 \sigma^2}{4 \etamag{\epsilon} \sigma^2+1}}}{\sqrt{4 \etamag{\epsilon} \sigma^2+1}}+\cos (\delta  \omegap) e^{-2 \etamag{\theta}-\frac{\etamag{\epsilon} \omegap^2}{2}}
+\frac{4 \cos \left(\frac{\delta  \omegap}{2 \etamag{\epsilon} \sigma^2+2}\right) e^{-\frac{4 \delta^2 \sigma^2+4 \etamag{\theta} \left(\etamag{\epsilon} \sigma^2+1\right)+\etamag{\epsilon} \omegap^2}{8 \etamag{\epsilon} \sigma^2+8}}}{\sqrt{\etamag{\epsilon} \sigma^2+1}}\right),
\\
P_{c,\mathrm{MZ2s}}^\eta &= \frac{1}{4} \left(2-\frac{e^{-\frac{2 \delta^2 \sigma^2}{4 \etamag{\epsilon} \sigma^2+1}}}{\sqrt{4 \etamag{\epsilon} \sigma^2+1}}-\cos (\delta  \omegap) e^{-2 \etamag{\theta}-\frac{\etamag{\epsilon} \omegap^2}{2}}\right).
\end{align}
\end{widetext}
From the above we see that the frequency-dependent (frequency-independent) noise terms only enter through an effective total frequency-dependent (frequency-independent) noise term ${\etamag{\epsilon} = \sqrt{\eta_{\epsilon_1}^2+\eta_{\epsilon_2}^2}}$ (${\etamag{\theta} = \sqrt{\eta_{\theta_1}^2+\eta_{\theta_2}^2}}$).
The form matches Eqs.~(\ref{eq_MZ2s_noisy_prob_1}, \ref{eq_MZ2s_noisy_prob_2}, \ref{eq_MZ2s_noisy_prob_c}) which can be recovered by taking \( \etamag{\epsilon} = \eta_{\epsilon} \) and \( \etamag{\theta} = \eta_{\theta} \).

The same equivalence holds for \ac{HOM}, MZ1, and MZ2d; and also in the case of independent photons for each protocol.

\section{Classical correlations and the high frequency-independent noise limit}\label{app_cc}
In Sec.~\ref{sec_results}, we noted that at high frequency-independent noise some residual information remains unscathed for most of our model variants.
To justify our claim that this is a result of classical correlations, let us return to our single-photon MZ probabilities with fixed $\epsilon$ and $\theta$ shifts.
These are $P_{1,\mathrm{MZ1}}$, the probability that the photon is detected at detector 1, and $P_{2,\mathrm{MZ1}}$, the probability that the photon is detected at detector 2; and are given in Eq.~(\ref{eq_MZ1_P1}) and Eq.~(\ref{eq_MZ1_P2}), respectively.

We want to now consider what happens if we run MZ1 twice, but with the same fixed $\epsilon$ and $\theta$ shifts in both runs. This leads to three possible outcomes:
\begin{align}
	P_{1,\mathrm{CC}}&=(P_{1,\mathrm{MZ1}})^2, \\
	P_{2,\mathrm{CC}}&=(P_{2,\mathrm{MZ1}})^2, \\
	P_{c,\mathrm{CC}}&=2\times P_{1,\mathrm{MZ1}}\times P_{2,\mathrm{MZ1}},
\end{align}
the probabilites that both photons arrive at detector 1, both at detector 2, and a coincidence at both detectors.

We can now perform the same procedure in Sec.~\ref{sec_noise}, averaging over $\epsilon$ and $\theta$ with appropriate weighting to obtain noisy probabilities.
The full expressions for these are given in the Supplemental Material \cite{supp}.

If we now take the limit $\eta_\epsilon\to\infty$, our new probabilities tend to constants, just as in a single MZ1 run.
However, while the same holds for a single MZ1 run in the limit $\eta_\theta\to\infty$ what we instead see is that for two MZ1 runs, correlated with the same (albeit unknown) $\theta$, the probablities do still vary with delay. Specifically, we see
\begin{gather}
	\lim_{\eta_\theta\to\infty}P_{1,\mathrm{CC}}^\eta=\frac{1}{8} \left(2+\frac{e^{-\frac{\delta^2 \sigma^2}{2 \eta_\epsilon^2 \sigma^2+1}}}{\sqrt{2 \eta_\epsilon^2 \sigma^2+1}}\right),
	\\
	\lim_{\eta_\theta\to\infty}P_{2,\mathrm{CC}}^\eta=\frac{1}{8} \left(2+\frac{e^{-\frac{\delta^2 \sigma^2}{2 \eta_\epsilon^2 \sigma^2+1}}}{\sqrt{2 \eta_\epsilon^2 \sigma^2+1}}\right),
	\\
	\lim_{\eta_\theta\to\infty}P_{c,\mathrm{CC}}^\eta=\frac{1}{4} \left(2-\frac{e^{-\frac{\delta^2 \sigma^2}{2 \eta_\epsilon^2 \sigma^2+1}}}{\sqrt{2 \eta_\epsilon^2 \sigma^2+1}}\right).
\end{gather}
Because these probabilities were derived from two independent runs of MZ1, which could be taken some arbitrary time apart, with only the requirement that $\epsilon$ and $\theta$ remain constant (but unknown) between each run, these probabilities must be the result of classical correlations.

We can then note these are the same probabilities we get for HOM with independent photons and $\alpha=0.5$, which is the same $\eta_\theta\to\infty$ limit we see for independent photon MZ2s.
While the equivalent frequency-entangled probabilities differ slightly, this nevertheless suggests that the residual Fisher information seen in this limit is exactly the information that remains from these classical correlations.
\begin{figure}[htb]
\centering
\tikzsetnextfilename{indepsnoise}
\begin{tikzpicture}
	\begin{groupplot}[group style={
		group size = 1 by 3,
		horizontal sep=0pt,
		vertical sep=0pt,
		xticklabels at=edge bottom,
	},
	xmin=-100,xmax=100,
	xtick={-100,-50,0,50,100},
	width=0.95\linewidth,
	height=0.6\linewidth,
	ylabel style={at={(axis description cs:-0.125,.5)},rotate=-90},
	ylabel = $\frac{F}{\omegap^2}$,
	every axis title/.style={below left,at={(1,1)},fill=white,fill opacity=0.75,text opacity=1},
	every y tick scale label/.style={at={(0.05,0.9)},anchor=south west,inner sep=0pt}
	]
\nextgroupplot[
	title = {$\eta_\epsilon\omegap=0$},
	ytick={0,0.25,0.5,0.75,1},
	yticklabels={0,0.25,0.5,0.75,1},
	ymin=0,
	ymax=1,
]
	\addplot[HOMcolour, very thick] file {data/indepsnoise1a.dat};
	\addplot[MZ2scolour, thick] file {data/indepsnoise1b.dat};
	\addplot[MZ2dcolour] file {data/indepsnoise1c.dat};

\nextgroupplot[
	title = {$\eta_\epsilon\omegap=5$},
	ytick={0,2/10000,4/10000,6/10000},
	ymin=0,
	ymax=7/10000,
]
	\addplot[MZ2dcolour, very thick] file {data/indepsnoise2c.dat};
	\addplot[MZ2scolour, thick] file {data/indepsnoise2b.dat};
	\addplot[HOMcolour, very thick] file {data/indepsnoise2a.dat};

\nextgroupplot[
	title = {$\eta_\epsilon\omegap=10$},
	ytick={0,1/10000,2/10000},
	ymin=0,
	ymax=0.00022,
	xlabel = $\delta\>\omegap$,
	legend style = {
		draw = none,
		at={(0.5,-0.25)},
		anchor=north,
		legend columns = 3,
		/tikz/every even column/.append style={column sep=4mm}},
]
	\addplot[MZ2dcolour, very thick] file {data/indepsnoise3c.dat};
	\addplot[MZ2scolour, very thick] file {data/indepsnoise3b.dat};
	\addplot[HOMcolour, very thick] file {data/indepsnoise3a.dat};
	\legend{MZ2d,MZ2s,HOM};
\end{groupplot}
\end{tikzpicture}
\caption{Comparison of the two-photon protocols, in the frequency-independent, mode-correlated noise configuration with
$\sigma=\frac{\omegap}{100}$ and $\alpha = 0.9$.
We see behaviour qualitatively similar to the frequency-entangled case depicted in Fig.~\ref{fig_noise_comp_dep}. As also seen in the bottom plot of Fig.~\ref{fig_zerovis}, information decays at large $\delta$ even in the absence of noise, and we now note that once noise is introduced the Fisher information is generally slightly lower than the frequency-entangled equivalent. We also see the wider \ac{HOM} dip of an independent photon input, specifically the Fisher information is equivalent to that of frequency-entangled \ac{HOM} with the reduced spectral width $\sigma/\sqrt{2}$.}
\label{fig_noise_comp_ind}
\end{figure}
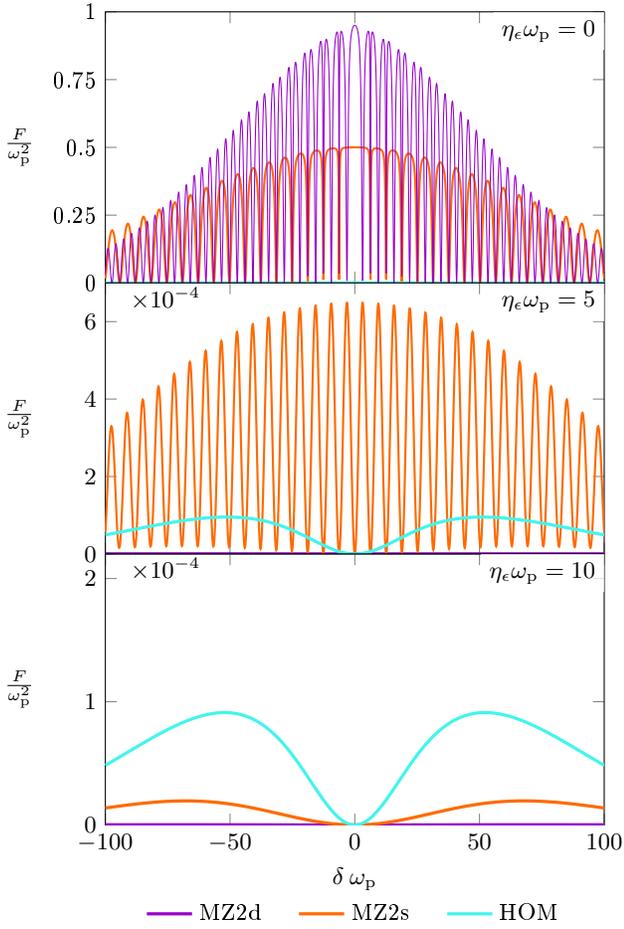

This is sufficient to explain the MZ2s case with mode-correlated noise.
However, MZ2d with mode-correlated noise sees a peculiar inverse visibility dependence.
At $\alpha=0$, the photon pair will not interfere at the initial beamsplitter.
The same logic now holds in terms of classical correlations: the two photons behave independently but experience the same $\theta$ shift.
As visibility increases, however, so does the degree of interference at the first beamsplitter.
When the two photons bunch the analogy to two correlated MZ1 runs breaks down.
Rather than each photon independently interfering with itself, the photon pair now acts as one.
At $\alpha=1$ it becomes wholly impossible to exploit the classical correlations as the two photons will be completely entangled after the first beamsplitter: hence the Fisher information drops to zero.

It is perhaps even more straightforward to understand the behaviour with mode-uncorrelated noise.
MZ2s retains the same residual information in the high $\eta_\theta$ limit, but it now decays at low visibility.
This naturally follows as low visibility means the second photon has a larger contribution from the orthogonal $\hat d$ mode, which experiences a different $\theta$ shift.
Because the shifts now differ between modes, classical correlations only exist between the paths of the $\hat c$ modes.
Letting $\alpha$ drop all the way to zero puts both photons in wholly distinguishable modes and thus the Fisher information vanishes as there are no correlations between the two.

MZ2d with mode-uncorrelated noise has no residual information at all in this limit.
This odd one out can be explained with a combination of the previous two cases: at high visibility the photons are more likely to bunch, which provides no classical correlations to exploit.
At low visibility the photons will behave independently, but both modes experience different $\theta$ shifts so no correlations exist.
Combining these two effects we will now always see zero information.

\section{Noise resilience of independent photons}\label{app_ind_noise}
For comparison to Fig.~\ref{fig_noise_comp_dep}, we plot in Fig.~\ref{fig_noise_comp_ind} the Fisher information at different frequency-dependent noise values when our input photons are now independent. The behaviour is qualitatively similar but the value of the resulting Fisher information is slightly reduced.

\section{Noise distributions}\label{app_dist}

In Eq.~\eqref{eq_dist_int} we derive our noisy probabilities by integrating our fixed shift probabilities with Gaussian weightings. 
While the choice of a Gaussian noise distribution is the most natural for frequency-dependent ($\epsilon$) noise, being equivalent to some jitter in the delay, for frequency-independent noise (where the $\theta$ shifts will always lie within an interval of width $2\pi$) a more conventional choice would be some circular distribution for $\theta$ shifts.

One such common circular distribution can be obtained by wrapping the standard Gaussian distribution around the circle. If we define an arbitrary Gaussian distribution
\begin{equation}
	G(x) = \frac{1}{\sqrt{2\pi}\sigma}e^{-\frac{(x-\mu)^2}{2\sigma^2}},
\end{equation}
then the wrapped Gaussian distribution is given by~\cite{Collett-1981}
\begin{equation}
	W(x) = \sum_{k=-\infty}^\infty G(x+2\pi k).
\end{equation}
Then, suppose some periodic function $f(x)$ such that $f(x+2\pi k) = f(x)$ for all integer $k$. We can then demonstrate that integrating this function with a Gaussian distribution over $(-\infty,\infty)$ is equivalent to integrating with the wrapped Gaussian over a $2\pi$ window:
\begin{align}
	\int_{-\infty}^\infty dx f(x)G(x) &=  \sum_{k=-\infty}^\infty \int_{2\pi k-\pi}^{2\pi k+\pi} dx f(x)G(x) \nonumber \\*
	&=\int_{-\pi}^\pi dx \sum_{k=-\infty}^\infty f(x+2\pi k) G(x+ 2\pi k) \nonumber \\*
	&=\int_{-\pi}^\pi dx f(x) \sum_{k=-\infty}^\infty G(x+ 2\pi k) \nonumber \\*
	&=\int_{-\pi}^\pi dx f(x)W(x).
\end{align}

We now note that all of our fixed shift probabilities given in Sec.~\ref{sec_protocols} have appropriate periodic dependence on $\theta$, i.e. $P(\theta + 2\pi k) = P(\theta)$ for all integer $k$.
Therefore the integral in Eq.~\eqref{eq_dist_int} will produce the same results as if we had used a wrapped Gaussian distribution.
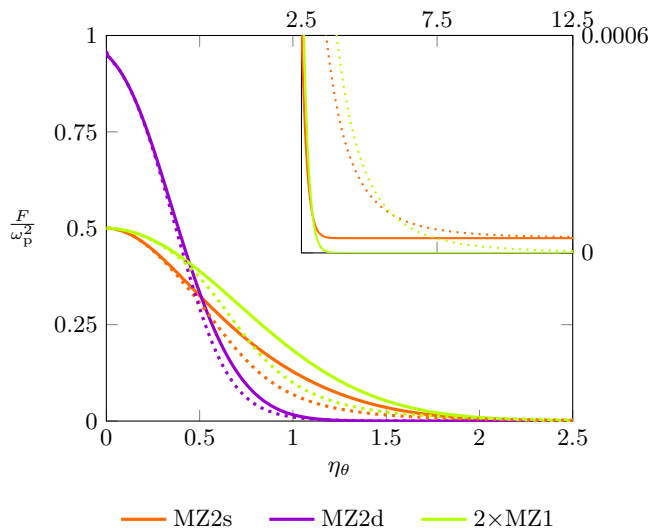
\begin{figure}[htb]
\centering
\tikzsetnextfilename{dists}
\begin{tikzpicture}
\begin{axis}[
	width=0.9\linewidth,
	legend style = {
		draw = none,
		at={(0.5,-0.2)},
		anchor=north,
		/tikz/every even column/.append style={column sep=4mm},
		legend columns = 3},
	xlabel = $\eta_\theta$,
	xtick={0,0.5,1,1.5,2,2.5},
	xmin=0,
	xmax=2.5,
	ylabel style={rotate=-90},
	ylabel = $\frac{F}{\omegap^2}$,
	ytick={0,0.25,0.5,0.75,1},
	ymin=0,
	ymax=1,
	every axis plot/.append style={very thick},
]
	\addplot[MZ2scolour] file {data/maxinfWGb.dat};
	\addplot[MZ2dcolour] file {data/maxinfWGc.dat};
	\addplot[MZ1colour] file {data/maxinfWGd.dat};
	\addplot[MZ2scolour, dotted] file {data/maxinfVMb.dat};
	\addplot[MZ2dcolour, dotted] file {data/maxinfVMc.dat};
	\addplot[MZ1colour, dotted] file {data/maxinfVMd.dat};
	\legend{MZ2s,MZ2d,2$\times$MZ1};
	\coordinate (inset) at (rel axis cs:1.00,1.00);
\end{axis}

\begin{axis}[at = (inset),
	anchor = north east,
	width=0.6\linewidth,
	every tick label/.append style={font=\small},
	xtick={2.5,7.5,12.5},
	xmin = 2.5, xmax = 12.5,
	xticklabel pos=top,
	yticklabel pos=right,
	scaled y ticks=false,
	ytick={0,0.0006},
	yticklabels={0,0.0006},
	ymin=0,
	ymax=0.0006,
	every axis plot/.append style={thick},
	axis background/.style={fill=white},
]
	\addplot[MZ2scolour] file {data/maxinfWGbinset.dat};
	\addplot[MZ1colour] file {data/maxinfWGdinset.dat};
	\addplot[MZ2scolour, dotted] file {data/maxinfVMbinset.dat};
	\addplot[MZ1colour, dotted] file {data/maxinfVMdinset.dat};
\end{axis}
\end{tikzpicture}
\caption{Comparison of the maximal information against frequency-independent noise modelled with a (wrapped) Gaussian noise distribution (solid) and a von Mises noise distribution (dotted).
We choose frequency-entangled photons and mode-correlated noise with $\sigma=\frac{\omegap}{100}$ and $\alpha = 0.9$.
At low noise, values are similar. As noise increases the curves diverge, the von Mises curves dropping faster. At higher noise the curves converge again.
In the inset, we see that for large noise values ($\eta_\theta>2.5$) the Gaussian curves, which initially sat above, have now dropped below the von Mises curves.
Information plateaus to the same values for both distributions (the residual ``HOM-like'' information from Fig.~\ref{fig_3d} (left) can be seen for MZ2s) but with the von Mises distribution this happens over a notably larger range of noise values.}
\label{fig_dists}
\end{figure}

An alternative circular distribution, the von Mises distribution, is given by~\cite{Collett-1981}
\begin{equation}
	\frac{e^{\kappa \cos(x-\mu)}}{2\pi I_0(\kappa)},
\end{equation}
with $I_0(\kappa)$ the modified Bessel function of the first kind. $\kappa$ is analogous to the Gaussian distribution's $1/\sigma^2$, so for our purposes we write the von Mises weighting in the form
\begin{equation}
	J_\theta(\vartheta) = \frac{e^{\frac{\cos(\vartheta)}{{\eta_\theta}^2}}}{2\pi I_0(\frac{1}{{\eta_\theta}^2})}.
\end{equation}
This can then replace the second expression in Eq.~\eqref{eq_dists}, and the $\vartheta$ integral from Eq.~\eqref{eq_dist_int} is now performed over the region $[-\pi,\pi]$.

Employing the von Mises distribution produces the same qualitative results, but with slightly different dependence on the noise parameter $\eta_\theta$. Considering only frequency-independent noise, we plot in Fig.~\ref{fig_dists} the maximal information for our protocols with both noise distributions. Values are most similar for high and low $\eta_\theta$, attaining the same limits at $\eta_\theta=0$ and $\eta_\theta \to \infty$, while the curves are furthest separated at moderate noise values. Also notable, from the inset, is that the von Mises model takes notably longer to decay to its high $\eta_\theta$ limits.

For the protocols in Fig.~\ref{fig_dists}, the Supplemental Material~\cite{supp} contains full probability and Fisher information expressions where a von Mises noise distribution was chosen for frequency-independent noise. Also included are all initial fixed shift probabilities from Sec.~\ref{sec_protocols}, plus those for the Sec.~\ref{sec_alternatives} model variations, so that all results can be easily reproduced for an alternative choice of noise distributions.

\section{Oscillations in the detected signal}\label{app_oscil}
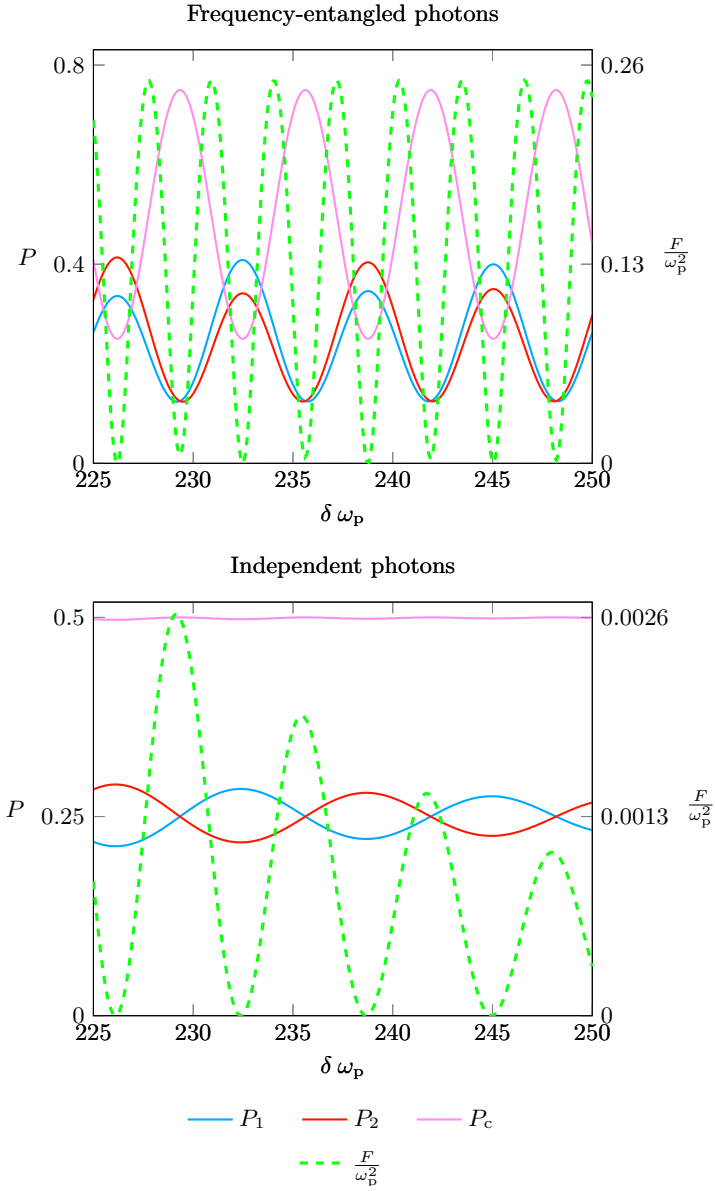
\begin{figure}[htb]
\centering
\tikzsetnextfilename{oscil}
\begin{tikzpicture}

\begin{axis}[
	yshift=-0.85\linewidth,
	width=0.95\linewidth,
	legend style = {
		draw = none,
		at={(0.5,-0.2)},
		anchor=north,
		legend columns = 3,
		/tikz/every even column/.append style={column sep=4mm}},
	title = {Independent photons},
	xlabel = $\delta\>\omegap$,
	xtick={225,230,235,240,245,250},
	xmin = 225,
	xmax = 250,
	ylabel style={rotate=-90},
	ylabel = $P$,
	ytick={0,0.25,0.5},
	ymin=0,
	ymax=0.51923,
]
	\addplot[P1colour, thick] file {data/oscil2P1.dat};
	\addplot[P2colour, thick] file {data/oscil2P2.dat};
	\addplot[Pccolour, thick] file {data/oscil2Pc.dat};
	\legend{$P_1$,$P_2$,$P_\mathrm{c}$};
\end{axis}

\begin{axis}[
	yshift=-0.85\linewidth,
	width=0.95\linewidth,
	legend style = {
		draw = none,
		at={(0.5,-0.3)},
		anchor=north,
		legend columns = 3,
		/tikz/every even column/.append style={column sep=4mm}},
	title = {Independent photons},
	xlabel = $\delta\>\omegap$,
	xtick={225,230,235,240,245,250},
	xmin = 225,
	xmax = 250,
	ylabel style={rotate=-90},
	ylabel = $\frac{F}{\omegap^2}$,
	yticklabel pos=right,
	ytick={0,0.0013,0.0026},
	yticklabels={0,0.0013,0.0026},
	scaled y ticks=false,
	ymin=0,
	ymax=0.0027,
]
	\addplot[FIcolour, very thick, dashed] file {data/oscil2FI.dat};
	\legend{$\frac{F}{\omegap^2}$};
\end{axis}

\begin{axis}[
	width=0.95\linewidth,
	legend style = {
		draw = none,
		at={(0.5,-0.2)},
		anchor=north,
		legend columns = 3,
		/tikz/every even column/.append style={column sep=4mm}},
	title = {Frequency-entangled photons},
	xlabel = $\delta\>\omegap$,
	xtick={225,230,235,240,245,250},
	xmin = 225,
	xmax = 250,
	ylabel style={rotate=-90},
	ylabel = $P$,
	ytick={0,0.4,0.8},
	ymin=0,
	ymax=0.830769,
]
	\addplot[P1colour, thick] file {data/oscil1P1.dat};
	\addplot[P2colour, thick] file {data/oscil1P2.dat};
	\addplot[Pccolour, thick] file {data/oscil1Pc.dat};
\end{axis}

\begin{axis}[
	width=0.95\linewidth,
	legend style = {
		draw = none,
		at={(0.5,-0.2)},
		anchor=north,
		legend columns = 3,
		/tikz/every even column/.append style={column sep=4mm}},
	title = {Frequency-entangled photons},
	xlabel = $\delta\>\omegap$,
	xtick={225,230,235,240,245,250},
	xmin = 225,
	xmax = 250,
	ylabel style={rotate=-90},
	ylabel = $\frac{F}{\omegap^2}$,
	yticklabel pos=right,
	ytick={0,0.13,0.26},
	ymin=0,
	ymax=0.27,
]
	\addplot[FIcolour, very thick, dashed] file {data/oscil1FI.dat};
\end{axis}

\end{tikzpicture}
\caption{Comparison of the detected signal to the resulting Fisher information for MZ2s in the absence of noise. Solid lines are the three probabilities (detection at detector 1, at detector 2, or a coincidence at both detectors), and the dashed line is the Fisher information.
We choose $\sigma=\frac{\omegap}{100}$ and examine behaviour at high $|\delta|$.
In the top plot (frequency-entangled input photons), we see regular oscillations in the detected signal produce a Fisher information that likewise oscillates but does not decay.
In the bottom plot (frequency-independent input photons) we see the detected probabilities, still slightly oscillating, are tending towards constants. Thus the Fisher information is decaying to zero.}
\label{fig_oscil}
\end{figure}

In Sec.~\ref{sec_results} we discussed how, at high $|\delta|$, the Fisher information oscillates without decay when the input photons are frequency-entangled. If the photons are independent, the Fisher information instead decays to zero. To further demonstrate this, Fig.~\ref{fig_oscil} plots the detection probabilities together with the Fisher information in a high $|\delta|$ region, for the MZ2s protocol.

Generally we expect a large Fisher information when probabilities change most quickly with respect to the parameter of interest, while it vanishes at the extrema of the probabilities as there is no local information at these points; Eq.~\eqref{eq_FI} shows that the Fisher information must vanish whenever the derivative of all probabilities is zero.
In practice, a degree of prior information (such as that obtained through some initial coarse calibration) enables one to tune the setup and operate in a region in which the Fisher information is high~\cite{Lyons-2018}.

The relation between detected signal and resulting Fisher information is similar for MZ2d.

\begin{thebibliography}{42}%
\makeatletter
\providecommand \@ifxundefined [1]{%
 \@ifx{#1\undefined}
}%
\providecommand \@ifnum [1]{%
 \ifnum #1\expandafter \@firstoftwo
 \else \expandafter \@secondoftwo
 \fi
}%
\providecommand \@ifx [1]{%
 \ifx #1\expandafter \@firstoftwo
 \else \expandafter \@secondoftwo
 \fi
}%
\providecommand \natexlab [1]{#1}%
\providecommand \enquote  [1]{``#1''}%
\providecommand \bibnamefont  [1]{#1}%
\providecommand \bibfnamefont [1]{#1}%
\providecommand \citenamefont [1]{#1}%
\providecommand \href@noop [0]{\@secondoftwo}%
\providecommand \href [0]{\begingroup \@sanitize@url \@href}%
\providecommand \@href[1]{\@@startlink{#1}\@@href}%
\providecommand \@@href[1]{\endgroup#1\@@endlink}%
\providecommand \@sanitize@url [0]{\catcode `\\12\catcode `\$12\catcode
  `\&12\catcode `\#12\catcode `\^12\catcode `\_12\catcode `\%12\relax}%
\providecommand \@@startlink[1]{}%
\providecommand \@@endlink[0]{}%
\providecommand \url  [0]{\begingroup\@sanitize@url \@url }%
\providecommand \@url [1]{\endgroup\@href {#1}{\urlprefix }}%
\providecommand \urlprefix  [0]{URL }%
\providecommand \Eprint [0]{\href }%
\providecommand \doibase [0]{https://doi.org/}%
\providecommand \selectlanguage [0]{\@gobble}%
\providecommand \bibinfo  [0]{\@secondoftwo}%
\providecommand \bibfield  [0]{\@secondoftwo}%
\providecommand \translation [1]{[#1]}%
\providecommand \BibitemOpen [0]{}%
\providecommand \bibitemStop [0]{}%
\providecommand \bibitemNoStop [0]{.\EOS\space}%
\providecommand \EOS [0]{\spacefactor3000\relax}%
\providecommand \BibitemShut  [1]{\csname bibitem#1\endcsname}%
\let\auto@bib@innerbib\@empty
\bibitem [{\citenamefont {Ley}\ and\ \citenamefont {Loudon}(1987)}]{Ley-1987}%
  \BibitemOpen
  \bibfield  {author} {\bibinfo {author} {\bibfnamefont {M.}~\bibnamefont
  {Ley}}\ and\ \bibinfo {author} {\bibfnamefont {R.}~\bibnamefont {Loudon}},\
  }\bibfield  {title} {\bibinfo {title} {Quantum {Theory} of {High}-resolution
  {Length} {Measurement} with a {Fabry}-{Perot} {Interferometer}},\ }\href
  {https://doi.org/10.1080/09500348714550251} {\bibfield  {journal} {\bibinfo
  {journal} {Journal of Modern Optics}\ }\textbf {\bibinfo {volume} {34}},\
  \bibinfo {pages} {227} (\bibinfo {year} {1987})}\BibitemShut {NoStop}%
\bibitem [{\citenamefont {Loudon}(2000)}]{Loudon-2000}%
  \BibitemOpen
  \bibfield  {author} {\bibinfo {author} {\bibfnamefont {R.}~\bibnamefont
  {Loudon}},\ }\href@noop {} {\emph {\bibinfo {title} {The Quantum Theory of
  Light}}},\ \bibinfo {edition} {3rd}\ ed.\ (\bibinfo  {publisher} {Oxford
  University Press},\ \bibinfo {year} {2000})\BibitemShut {NoStop}%
\bibitem [{\citenamefont {Demkowicz-Dobrza\'{n}ski}\ \emph
  {et~al.}(2015)\citenamefont {Demkowicz-Dobrza\'{n}ski}, \citenamefont
  {Jarzyna},\ and\ \citenamefont
  {Ko\l{}ody\'{n}ski}}]{Demkowicz-Dobrzanski-2015}%
  \BibitemOpen
  \bibfield  {author} {\bibinfo {author} {\bibfnamefont {R.}~\bibnamefont
  {Demkowicz-Dobrza\'{n}ski}}, \bibinfo {author} {\bibfnamefont
  {M.}~\bibnamefont {Jarzyna}},\ and\ \bibinfo {author} {\bibfnamefont
  {J.}~\bibnamefont {Ko\l{}ody\'{n}ski}},\ }\bibfield  {title} {\bibinfo
  {title} {Quantum limits in optical interferometry},\ }\href
  {https://doi.org/10.1016/bs.po.2015.02.003} {\bibfield  {journal} {\bibinfo
  {journal} {Progress in Optics}\ ,\ \bibinfo {pages} {345–435}} (\bibinfo
  {year} {2015})}\BibitemShut {NoStop}%
\bibitem [{\citenamefont {Bouchard}\ \emph {et~al.}(2021)\citenamefont
  {Bouchard}, \citenamefont {Sit}, \citenamefont {Zhang}, \citenamefont
  {Fickler}, \citenamefont {Miatto}, \citenamefont {Yao}, \citenamefont
  {Sciarrino},\ and\ \citenamefont {Karimi}}]{Bouchard-2021}%
  \BibitemOpen
  \bibfield  {author} {\bibinfo {author} {\bibfnamefont {F.}~\bibnamefont
  {Bouchard}}, \bibinfo {author} {\bibfnamefont {A.}~\bibnamefont {Sit}},
  \bibinfo {author} {\bibfnamefont {Y.}~\bibnamefont {Zhang}}, \bibinfo
  {author} {\bibfnamefont {R.}~\bibnamefont {Fickler}}, \bibinfo {author}
  {\bibfnamefont {F.~M.}\ \bibnamefont {Miatto}}, \bibinfo {author}
  {\bibfnamefont {Y.}~\bibnamefont {Yao}}, \bibinfo {author} {\bibfnamefont
  {F.}~\bibnamefont {Sciarrino}},\ and\ \bibinfo {author} {\bibfnamefont
  {E.}~\bibnamefont {Karimi}},\ }\bibfield  {title} {\bibinfo {title}
  {Two-photon interference: the {Hong}–{Ou}–{Mandel} effect},\ }\href
  {https://doi.org/10.1088/1361-6633/abcd7a} {\bibfield  {journal} {\bibinfo
  {journal} {Reports on Progress in Physics}\ }\textbf {\bibinfo {volume}
  {84}},\ \bibinfo {pages} {012402} (\bibinfo {year} {2021})}\BibitemShut
  {NoStop}%
\bibitem [{\citenamefont {Tse}\ \emph {et~al.}(2019)\citenamefont {Tse},
  \citenamefont {Yu}, \citenamefont {Kijbunchoo}, \citenamefont
  {Fernandez-Galiana}, \citenamefont {Dupej}, \citenamefont {Barsotti},
  \citenamefont {Blair}, \citenamefont {Brown}, \citenamefont {Dwyer},
  \citenamefont {Effler}, \citenamefont {Evans}, \citenamefont {Fritschel},
  \citenamefont {Frolov}, \citenamefont {Green}, \citenamefont {Mansell},
  \citenamefont {Matichard}, \citenamefont {Mavalvala}, \citenamefont
  {McClelland}, \citenamefont {McCuller}, \citenamefont {McRae}, \citenamefont
  {Miller}, \citenamefont {Mullavey}, \citenamefont {Oelker}, \citenamefont
  {Phinney}, \citenamefont {Sigg}, \citenamefont {Slagmolen}, \citenamefont
  {Vo}, \citenamefont {Ward}, \citenamefont {Whittle}, \citenamefont {Abbott},
  \citenamefont {Adams}, \citenamefont {Adhikari}, \citenamefont {Ananyeva},
  \citenamefont {Appert}, \citenamefont {Arai}, \citenamefont {Areeda},
  \citenamefont {Asali}, \citenamefont {Aston}, \citenamefont {Austin},
  \citenamefont {Baer}, \citenamefont {Ball}, \citenamefont {Ballmer},
  \citenamefont {Banagiri}, \citenamefont {Barker}, \citenamefont {Bartlett},
  \citenamefont {Berger}, \citenamefont {Betzwieser}, \citenamefont
  {Bhattacharjee}, \citenamefont {Billingsley}, \citenamefont {Biscans},
  \citenamefont {Blair}, \citenamefont {Bode}, \citenamefont {Booker},
  \citenamefont {Bork}, \citenamefont {Bramley}, \citenamefont {Brooks},
  \citenamefont {Buikema}, \citenamefont {Cahillane}, \citenamefont {Cannon},
  \citenamefont {Chen}, \citenamefont {Ciobanu}, \citenamefont {Clara},
  \citenamefont {Cooper}, \citenamefont {Corley}, \citenamefont {Countryman},
  \citenamefont {Covas}, \citenamefont {Coyne}, \citenamefont {Datrier},
  \citenamefont {Davis}, \citenamefont {Di~Fronzo}, \citenamefont {Driggers},
  \citenamefont {Etzel}, \citenamefont {Evans}, \citenamefont {Feicht},
  \citenamefont {Fulda}, \citenamefont {Fyffe}, \citenamefont {Giaime},
  \citenamefont {Giardina}, \citenamefont {Godwin}, \citenamefont {Goetz},
  \citenamefont {Gras}, \citenamefont {Gray}, \citenamefont {Gray},
  \citenamefont {Gupta}, \citenamefont {Gustafson}, \citenamefont {Gustafson},
  \citenamefont {Hanks}, \citenamefont {Hanson}, \citenamefont {Hardwick},
  \citenamefont {Hasskew}, \citenamefont {Heintze}, \citenamefont
  {Helmling-Cornell}, \citenamefont {Holland}, \citenamefont {Jones},
  \citenamefont {Kandhasamy}, \citenamefont {Karki}, \citenamefont {Kasprzack},
  \citenamefont {Kawabe}, \citenamefont {King}, \citenamefont {Kissel},
  \citenamefont {Kumar}, \citenamefont {Landry}, \citenamefont {Lane},
  \citenamefont {Lantz}, \citenamefont {Laxen}, \citenamefont {Lecoeuche},
  \citenamefont {Leviton}, \citenamefont {Liu}, \citenamefont {Lormand},
  \citenamefont {Lundgren}, \citenamefont {Macas}, \citenamefont {MacInnis},
  \citenamefont {Macleod}, \citenamefont {Márka}, \citenamefont {Márka},
  \citenamefont {Martynov}, \citenamefont {Mason}, \citenamefont {Massinger},
  \citenamefont {McCarthy}, \citenamefont {McCormick}, \citenamefont {McIver},
  \citenamefont {Mendell}, \citenamefont {Merfeld}, \citenamefont {Merilh},
  \citenamefont {Meylahn}, \citenamefont {Mistry}, \citenamefont {Mittleman},
  \citenamefont {Moreno}, \citenamefont {Mow-Lowry}, \citenamefont {Mozzon},
  \citenamefont {Nelson}, \citenamefont {Nguyen}, \citenamefont {Nuttall},
  \citenamefont {Oberling}, \citenamefont {Oram}, \citenamefont {O’Reilly},
  \citenamefont {Osthelder}, \citenamefont {Ottaway}, \citenamefont {Overmier},
  \citenamefont {Palamos}, \citenamefont {Parker}, \citenamefont {Payne},
  \citenamefont {Pele}, \citenamefont {Perez}, \citenamefont {Pirello},
  \citenamefont {Radkins}, \citenamefont {Ramirez}, \citenamefont {Richardson},
  \citenamefont {Riles}, \citenamefont {Robertson}, \citenamefont {Rollins},
  \citenamefont {Romel}, \citenamefont {Romie}, \citenamefont {Ross},
  \citenamefont {Ryan}, \citenamefont {Sadecki}, \citenamefont {Sanchez},
  \citenamefont {Sanchez}, \citenamefont {Saravanan}, \citenamefont {Savage},
  \citenamefont {Schaetzl}, \citenamefont {Schnabel}, \citenamefont
  {Schofield}, \citenamefont {Schwartz}, \citenamefont {Sellers}, \citenamefont
  {Shaffer}, \citenamefont {Smith}, \citenamefont {Soni}, \citenamefont
  {Sorazu}, \citenamefont {Spencer}, \citenamefont {Strain}, \citenamefont
  {Sun}, \citenamefont {Szczepańczyk}, \citenamefont {Thomas}, \citenamefont
  {Thomas}, \citenamefont {Thorne}, \citenamefont {Toland}, \citenamefont
  {Torrie}, \citenamefont {Traylor}, \citenamefont {Urban}, \citenamefont
  {Vajente}, \citenamefont {Valdes}, \citenamefont {Vander-Hyde}, \citenamefont
  {Veitch}, \citenamefont {Venkateswara}, \citenamefont {Venugopalan},
  \citenamefont {Viets}, \citenamefont {Vorvick}, \citenamefont {Wade},
  \citenamefont {Warner}, \citenamefont {Weaver}, \citenamefont {Weiss},
  \citenamefont {Willke}, \citenamefont {Wipf}, \citenamefont {Xiao},
  \citenamefont {Yamamoto}, \citenamefont {Yap}, \citenamefont {Yu},
  \citenamefont {Zhang}, \citenamefont {Zucker},\ and\ \citenamefont
  {Zweizig}}]{Tse-2019}%
  \BibitemOpen
  \bibfield  {author} {\bibinfo {author} {\bibfnamefont {M.}~\bibnamefont
  {Tse}}, \bibinfo {author} {\bibfnamefont {H.}~\bibnamefont {Yu}}, \bibinfo
  {author} {\bibfnamefont {N.}~\bibnamefont {Kijbunchoo}}, \bibinfo {author}
  {\bibfnamefont {A.}~\bibnamefont {Fernandez-Galiana}}, \bibinfo {author}
  {\bibfnamefont {P.}~\bibnamefont {Dupej}}, \bibinfo {author} {\bibfnamefont
  {L.}~\bibnamefont {Barsotti}}, \bibinfo {author} {\bibfnamefont
  {C.}~\bibnamefont {Blair}}, \bibinfo {author} {\bibfnamefont
  {D.}~\bibnamefont {Brown}}, \bibinfo {author} {\bibfnamefont
  {S.}~\bibnamefont {Dwyer}}, \bibinfo {author} {\bibfnamefont
  {A.}~\bibnamefont {Effler}}, \bibinfo {author} {\bibfnamefont
  {M.}~\bibnamefont {Evans}}, \bibinfo {author} {\bibfnamefont
  {P.}~\bibnamefont {Fritschel}}, \bibinfo {author} {\bibfnamefont
  {V.}~\bibnamefont {Frolov}}, \bibinfo {author} {\bibfnamefont
  {A.}~\bibnamefont {Green}}, \bibinfo {author} {\bibfnamefont
  {G.}~\bibnamefont {Mansell}}, \bibinfo {author} {\bibfnamefont
  {F.}~\bibnamefont {Matichard}}, \bibinfo {author} {\bibfnamefont
  {N.}~\bibnamefont {Mavalvala}}, \bibinfo {author} {\bibfnamefont
  {D.}~\bibnamefont {McClelland}}, \bibinfo {author} {\bibfnamefont
  {L.}~\bibnamefont {McCuller}}, \bibinfo {author} {\bibfnamefont
  {T.}~\bibnamefont {McRae}}, \bibinfo {author} {\bibfnamefont
  {J.}~\bibnamefont {Miller}}, \bibinfo {author} {\bibfnamefont
  {A.}~\bibnamefont {Mullavey}}, \bibinfo {author} {\bibfnamefont
  {E.}~\bibnamefont {Oelker}}, \bibinfo {author} {\bibfnamefont
  {I.}~\bibnamefont {Phinney}}, \bibinfo {author} {\bibfnamefont
  {D.}~\bibnamefont {Sigg}}, \bibinfo {author} {\bibfnamefont {B.}~\bibnamefont
  {Slagmolen}}, \bibinfo {author} {\bibfnamefont {T.}~\bibnamefont {Vo}},
  \bibinfo {author} {\bibfnamefont {R.}~\bibnamefont {Ward}}, \bibinfo {author}
  {\bibfnamefont {C.}~\bibnamefont {Whittle}}, \bibinfo {author} {\bibfnamefont
  {R.}~\bibnamefont {Abbott}}, \bibinfo {author} {\bibfnamefont
  {C.}~\bibnamefont {Adams}}, \bibinfo {author} {\bibfnamefont
  {R.}~\bibnamefont {Adhikari}}, \bibinfo {author} {\bibfnamefont
  {A.}~\bibnamefont {Ananyeva}}, \bibinfo {author} {\bibfnamefont
  {S.}~\bibnamefont {Appert}}, \bibinfo {author} {\bibfnamefont
  {K.}~\bibnamefont {Arai}}, \bibinfo {author} {\bibfnamefont {J.}~\bibnamefont
  {Areeda}}, \bibinfo {author} {\bibfnamefont {Y.}~\bibnamefont {Asali}},
  \bibinfo {author} {\bibfnamefont {S.}~\bibnamefont {Aston}}, \bibinfo
  {author} {\bibfnamefont {C.}~\bibnamefont {Austin}}, \bibinfo {author}
  {\bibfnamefont {A.}~\bibnamefont {Baer}}, \bibinfo {author} {\bibfnamefont
  {M.}~\bibnamefont {Ball}}, \bibinfo {author} {\bibfnamefont {S.}~\bibnamefont
  {Ballmer}}, \bibinfo {author} {\bibfnamefont {S.}~\bibnamefont {Banagiri}},
  \bibinfo {author} {\bibfnamefont {D.}~\bibnamefont {Barker}}, \bibinfo
  {author} {\bibfnamefont {J.}~\bibnamefont {Bartlett}}, \bibinfo {author}
  {\bibfnamefont {B.}~\bibnamefont {Berger}}, \bibinfo {author} {\bibfnamefont
  {J.}~\bibnamefont {Betzwieser}}, \bibinfo {author} {\bibfnamefont
  {D.}~\bibnamefont {Bhattacharjee}}, \bibinfo {author} {\bibfnamefont
  {G.}~\bibnamefont {Billingsley}}, \bibinfo {author} {\bibfnamefont
  {S.}~\bibnamefont {Biscans}}, \bibinfo {author} {\bibfnamefont
  {R.}~\bibnamefont {Blair}}, \bibinfo {author} {\bibfnamefont
  {N.}~\bibnamefont {Bode}}, \bibinfo {author} {\bibfnamefont {P.}~\bibnamefont
  {Booker}}, \bibinfo {author} {\bibfnamefont {R.}~\bibnamefont {Bork}},
  \bibinfo {author} {\bibfnamefont {A.}~\bibnamefont {Bramley}}, \bibinfo
  {author} {\bibfnamefont {A.}~\bibnamefont {Brooks}}, \bibinfo {author}
  {\bibfnamefont {A.}~\bibnamefont {Buikema}}, \bibinfo {author} {\bibfnamefont
  {C.}~\bibnamefont {Cahillane}}, \bibinfo {author} {\bibfnamefont
  {K.}~\bibnamefont {Cannon}}, \bibinfo {author} {\bibfnamefont
  {X.}~\bibnamefont {Chen}}, \bibinfo {author} {\bibfnamefont {A.}~\bibnamefont
  {Ciobanu}}, \bibinfo {author} {\bibfnamefont {F.}~\bibnamefont {Clara}},
  \bibinfo {author} {\bibfnamefont {S.}~\bibnamefont {Cooper}}, \bibinfo
  {author} {\bibfnamefont {K.}~\bibnamefont {Corley}}, \bibinfo {author}
  {\bibfnamefont {S.}~\bibnamefont {Countryman}}, \bibinfo {author}
  {\bibfnamefont {P.}~\bibnamefont {Covas}}, \bibinfo {author} {\bibfnamefont
  {D.}~\bibnamefont {Coyne}}, \bibinfo {author} {\bibfnamefont
  {L.}~\bibnamefont {Datrier}}, \bibinfo {author} {\bibfnamefont
  {D.}~\bibnamefont {Davis}}, \bibinfo {author} {\bibfnamefont
  {C.}~\bibnamefont {Di~Fronzo}}, \bibinfo {author} {\bibfnamefont
  {J.}~\bibnamefont {Driggers}}, \bibinfo {author} {\bibfnamefont
  {T.}~\bibnamefont {Etzel}}, \bibinfo {author} {\bibfnamefont
  {T.}~\bibnamefont {Evans}}, \bibinfo {author} {\bibfnamefont
  {J.}~\bibnamefont {Feicht}}, \bibinfo {author} {\bibfnamefont
  {P.}~\bibnamefont {Fulda}}, \bibinfo {author} {\bibfnamefont
  {M.}~\bibnamefont {Fyffe}}, \bibinfo {author} {\bibfnamefont
  {J.}~\bibnamefont {Giaime}}, \bibinfo {author} {\bibfnamefont
  {K.}~\bibnamefont {Giardina}}, \bibinfo {author} {\bibfnamefont
  {P.}~\bibnamefont {Godwin}}, \bibinfo {author} {\bibfnamefont
  {E.}~\bibnamefont {Goetz}}, \bibinfo {author} {\bibfnamefont
  {S.}~\bibnamefont {Gras}}, \bibinfo {author} {\bibfnamefont {C.}~\bibnamefont
  {Gray}}, \bibinfo {author} {\bibfnamefont {R.}~\bibnamefont {Gray}}, \bibinfo
  {author} {\bibfnamefont {A.}~\bibnamefont {Gupta}}, \bibinfo {author}
  {\bibfnamefont {E.}~\bibnamefont {Gustafson}}, \bibinfo {author}
  {\bibfnamefont {R.}~\bibnamefont {Gustafson}}, \bibinfo {author}
  {\bibfnamefont {J.}~\bibnamefont {Hanks}}, \bibinfo {author} {\bibfnamefont
  {J.}~\bibnamefont {Hanson}}, \bibinfo {author} {\bibfnamefont
  {T.}~\bibnamefont {Hardwick}}, \bibinfo {author} {\bibfnamefont
  {R.}~\bibnamefont {Hasskew}}, \bibinfo {author} {\bibfnamefont
  {M.}~\bibnamefont {Heintze}}, \bibinfo {author} {\bibfnamefont
  {A.}~\bibnamefont {Helmling-Cornell}}, \bibinfo {author} {\bibfnamefont
  {N.}~\bibnamefont {Holland}}, \bibinfo {author} {\bibfnamefont
  {J.}~\bibnamefont {Jones}}, \bibinfo {author} {\bibfnamefont
  {S.}~\bibnamefont {Kandhasamy}}, \bibinfo {author} {\bibfnamefont
  {S.}~\bibnamefont {Karki}}, \bibinfo {author} {\bibfnamefont
  {M.}~\bibnamefont {Kasprzack}}, \bibinfo {author} {\bibfnamefont
  {K.}~\bibnamefont {Kawabe}}, \bibinfo {author} {\bibfnamefont
  {P.}~\bibnamefont {King}}, \bibinfo {author} {\bibfnamefont {J.}~\bibnamefont
  {Kissel}}, \bibinfo {author} {\bibfnamefont {R.}~\bibnamefont {Kumar}},
  \bibinfo {author} {\bibfnamefont {M.}~\bibnamefont {Landry}}, \bibinfo
  {author} {\bibfnamefont {B.}~\bibnamefont {Lane}}, \bibinfo {author}
  {\bibfnamefont {B.}~\bibnamefont {Lantz}}, \bibinfo {author} {\bibfnamefont
  {M.}~\bibnamefont {Laxen}}, \bibinfo {author} {\bibfnamefont
  {Y.}~\bibnamefont {Lecoeuche}}, \bibinfo {author} {\bibfnamefont
  {J.}~\bibnamefont {Leviton}}, \bibinfo {author} {\bibfnamefont
  {J.}~\bibnamefont {Liu}}, \bibinfo {author} {\bibfnamefont {M.}~\bibnamefont
  {Lormand}}, \bibinfo {author} {\bibfnamefont {A.}~\bibnamefont {Lundgren}},
  \bibinfo {author} {\bibfnamefont {R.}~\bibnamefont {Macas}}, \bibinfo
  {author} {\bibfnamefont {M.}~\bibnamefont {MacInnis}}, \bibinfo {author}
  {\bibfnamefont {D.}~\bibnamefont {Macleod}}, \bibinfo {author} {\bibfnamefont
  {S.}~\bibnamefont {Márka}}, \bibinfo {author} {\bibfnamefont
  {Z.}~\bibnamefont {Márka}}, \bibinfo {author} {\bibfnamefont
  {D.}~\bibnamefont {Martynov}}, \bibinfo {author} {\bibfnamefont
  {K.}~\bibnamefont {Mason}}, \bibinfo {author} {\bibfnamefont
  {T.}~\bibnamefont {Massinger}}, \bibinfo {author} {\bibfnamefont
  {R.}~\bibnamefont {McCarthy}}, \bibinfo {author} {\bibfnamefont
  {S.}~\bibnamefont {McCormick}}, \bibinfo {author} {\bibfnamefont
  {J.}~\bibnamefont {McIver}}, \bibinfo {author} {\bibfnamefont
  {G.}~\bibnamefont {Mendell}}, \bibinfo {author} {\bibfnamefont
  {K.}~\bibnamefont {Merfeld}}, \bibinfo {author} {\bibfnamefont
  {E.}~\bibnamefont {Merilh}}, \bibinfo {author} {\bibfnamefont
  {F.}~\bibnamefont {Meylahn}}, \bibinfo {author} {\bibfnamefont
  {T.}~\bibnamefont {Mistry}}, \bibinfo {author} {\bibfnamefont
  {R.}~\bibnamefont {Mittleman}}, \bibinfo {author} {\bibfnamefont
  {G.}~\bibnamefont {Moreno}}, \bibinfo {author} {\bibfnamefont
  {C.}~\bibnamefont {Mow-Lowry}}, \bibinfo {author} {\bibfnamefont
  {S.}~\bibnamefont {Mozzon}}, \bibinfo {author} {\bibfnamefont
  {T.}~\bibnamefont {Nelson}}, \bibinfo {author} {\bibfnamefont
  {P.}~\bibnamefont {Nguyen}}, \bibinfo {author} {\bibfnamefont
  {L.}~\bibnamefont {Nuttall}}, \bibinfo {author} {\bibfnamefont
  {J.}~\bibnamefont {Oberling}}, \bibinfo {author} {\bibfnamefont
  {R.}~\bibnamefont {Oram}}, \bibinfo {author} {\bibfnamefont {B.}~\bibnamefont
  {O’Reilly}}, \bibinfo {author} {\bibfnamefont {C.}~\bibnamefont
  {Osthelder}}, \bibinfo {author} {\bibfnamefont {D.}~\bibnamefont {Ottaway}},
  \bibinfo {author} {\bibfnamefont {H.}~\bibnamefont {Overmier}}, \bibinfo
  {author} {\bibfnamefont {J.}~\bibnamefont {Palamos}}, \bibinfo {author}
  {\bibfnamefont {W.}~\bibnamefont {Parker}}, \bibinfo {author} {\bibfnamefont
  {E.}~\bibnamefont {Payne}}, \bibinfo {author} {\bibfnamefont
  {A.}~\bibnamefont {Pele}}, \bibinfo {author} {\bibfnamefont {C.}~\bibnamefont
  {Perez}}, \bibinfo {author} {\bibfnamefont {M.}~\bibnamefont {Pirello}},
  \bibinfo {author} {\bibfnamefont {H.}~\bibnamefont {Radkins}}, \bibinfo
  {author} {\bibfnamefont {K.}~\bibnamefont {Ramirez}}, \bibinfo {author}
  {\bibfnamefont {J.}~\bibnamefont {Richardson}}, \bibinfo {author}
  {\bibfnamefont {K.}~\bibnamefont {Riles}}, \bibinfo {author} {\bibfnamefont
  {N.}~\bibnamefont {Robertson}}, \bibinfo {author} {\bibfnamefont
  {J.}~\bibnamefont {Rollins}}, \bibinfo {author} {\bibfnamefont
  {C.}~\bibnamefont {Romel}}, \bibinfo {author} {\bibfnamefont
  {J.}~\bibnamefont {Romie}}, \bibinfo {author} {\bibfnamefont
  {M.}~\bibnamefont {Ross}}, \bibinfo {author} {\bibfnamefont {K.}~\bibnamefont
  {Ryan}}, \bibinfo {author} {\bibfnamefont {T.}~\bibnamefont {Sadecki}},
  \bibinfo {author} {\bibfnamefont {E.}~\bibnamefont {Sanchez}}, \bibinfo
  {author} {\bibfnamefont {L.}~\bibnamefont {Sanchez}}, \bibinfo {author}
  {\bibfnamefont {T.}~\bibnamefont {Saravanan}}, \bibinfo {author}
  {\bibfnamefont {R.}~\bibnamefont {Savage}}, \bibinfo {author} {\bibfnamefont
  {D.}~\bibnamefont {Schaetzl}}, \bibinfo {author} {\bibfnamefont
  {R.}~\bibnamefont {Schnabel}}, \bibinfo {author} {\bibfnamefont
  {R.}~\bibnamefont {Schofield}}, \bibinfo {author} {\bibfnamefont
  {E.}~\bibnamefont {Schwartz}}, \bibinfo {author} {\bibfnamefont
  {D.}~\bibnamefont {Sellers}}, \bibinfo {author} {\bibfnamefont
  {T.}~\bibnamefont {Shaffer}}, \bibinfo {author} {\bibfnamefont
  {J.}~\bibnamefont {Smith}}, \bibinfo {author} {\bibfnamefont
  {S.}~\bibnamefont {Soni}}, \bibinfo {author} {\bibfnamefont {B.}~\bibnamefont
  {Sorazu}}, \bibinfo {author} {\bibfnamefont {A.}~\bibnamefont {Spencer}},
  \bibinfo {author} {\bibfnamefont {K.}~\bibnamefont {Strain}}, \bibinfo
  {author} {\bibfnamefont {L.}~\bibnamefont {Sun}}, \bibinfo {author}
  {\bibfnamefont {M.}~\bibnamefont {Szczepańczyk}}, \bibinfo {author}
  {\bibfnamefont {M.}~\bibnamefont {Thomas}}, \bibinfo {author} {\bibfnamefont
  {P.}~\bibnamefont {Thomas}}, \bibinfo {author} {\bibfnamefont
  {K.}~\bibnamefont {Thorne}}, \bibinfo {author} {\bibfnamefont
  {K.}~\bibnamefont {Toland}}, \bibinfo {author} {\bibfnamefont
  {C.}~\bibnamefont {Torrie}}, \bibinfo {author} {\bibfnamefont
  {G.}~\bibnamefont {Traylor}}, \bibinfo {author} {\bibfnamefont
  {A.}~\bibnamefont {Urban}}, \bibinfo {author} {\bibfnamefont
  {G.}~\bibnamefont {Vajente}}, \bibinfo {author} {\bibfnamefont
  {G.}~\bibnamefont {Valdes}}, \bibinfo {author} {\bibfnamefont
  {D.}~\bibnamefont {Vander-Hyde}}, \bibinfo {author} {\bibfnamefont
  {P.}~\bibnamefont {Veitch}}, \bibinfo {author} {\bibfnamefont
  {K.}~\bibnamefont {Venkateswara}}, \bibinfo {author} {\bibfnamefont
  {G.}~\bibnamefont {Venugopalan}}, \bibinfo {author} {\bibfnamefont
  {A.}~\bibnamefont {Viets}}, \bibinfo {author} {\bibfnamefont
  {C.}~\bibnamefont {Vorvick}}, \bibinfo {author} {\bibfnamefont
  {M.}~\bibnamefont {Wade}}, \bibinfo {author} {\bibfnamefont {J.}~\bibnamefont
  {Warner}}, \bibinfo {author} {\bibfnamefont {B.}~\bibnamefont {Weaver}},
  \bibinfo {author} {\bibfnamefont {R.}~\bibnamefont {Weiss}}, \bibinfo
  {author} {\bibfnamefont {B.}~\bibnamefont {Willke}}, \bibinfo {author}
  {\bibfnamefont {C.}~\bibnamefont {Wipf}}, \bibinfo {author} {\bibfnamefont
  {L.}~\bibnamefont {Xiao}}, \bibinfo {author} {\bibfnamefont {H.}~\bibnamefont
  {Yamamoto}}, \bibinfo {author} {\bibfnamefont {M.}~\bibnamefont {Yap}},
  \bibinfo {author} {\bibfnamefont {H.}~\bibnamefont {Yu}}, \bibinfo {author}
  {\bibfnamefont {L.}~\bibnamefont {Zhang}}, \bibinfo {author} {\bibfnamefont
  {M.}~\bibnamefont {Zucker}},\ and\ \bibinfo {author} {\bibfnamefont
  {J.}~\bibnamefont {Zweizig}},\ }\bibfield  {title} {\bibinfo {title}
  {Quantum-{Enhanced} {Advanced} {LIGO} {Detectors} in the {Era} of
  {Gravitational}-{Wave} {Astronomy}},\ }\href
  {https://doi.org/10.1103/PhysRevLett.123.231107} {\bibfield  {journal}
  {\bibinfo  {journal} {Physical Review Letters}\ }\textbf {\bibinfo {volume}
  {123}},\ \bibinfo {pages} {231107} (\bibinfo {year} {2019})}\BibitemShut
  {NoStop}%
\bibitem [{\citenamefont {{Virgo Collaboration}}\ \emph
  {et~al.}(2019)\citenamefont {{Virgo Collaboration}}, \citenamefont
  {Vahlbruch}, \citenamefont {Mehmet}, \citenamefont {Lück},\ and\
  \citenamefont {Danzmann}}]{Virgo-2019}%
  \BibitemOpen
  \bibfield  {author} {\bibinfo {author} {\bibnamefont {{Virgo
  Collaboration}}}, \bibinfo {author} {\bibfnamefont {H.}~\bibnamefont
  {Vahlbruch}}, \bibinfo {author} {\bibfnamefont {M.}~\bibnamefont {Mehmet}},
  \bibinfo {author} {\bibfnamefont {H.}~\bibnamefont {Lück}},\ and\ \bibinfo
  {author} {\bibfnamefont {K.}~\bibnamefont {Danzmann}},\ }\bibfield  {title}
  {\bibinfo {title} {Increasing the {Astrophysical} {Reach} of the {Advanced}
  {Virgo} {Detector} via the {Application} of {Squeezed} {Vacuum} {States} of
  {Light}},\ }\href {https://doi.org/10.1103/PhysRevLett.123.231108} {\bibfield
   {journal} {\bibinfo  {journal} {Physical Review Letters}\ }\textbf {\bibinfo
  {volume} {123}},\ \bibinfo {pages} {231108} (\bibinfo {year}
  {2019})}\BibitemShut {NoStop}%
\bibitem [{\citenamefont {Hong}\ \emph {et~al.}(1987)\citenamefont {Hong},
  \citenamefont {Ou},\ and\ \citenamefont {Mandel}}]{Hong-1987}%
  \BibitemOpen
  \bibfield  {author} {\bibinfo {author} {\bibfnamefont {C.~K.}\ \bibnamefont
  {Hong}}, \bibinfo {author} {\bibfnamefont {Z.~Y.}\ \bibnamefont {Ou}},\ and\
  \bibinfo {author} {\bibfnamefont {L.}~\bibnamefont {Mandel}},\ }\bibfield
  {title} {\bibinfo {title} {Measurement of subpicosecond time intervals
  between two photons by interference},\ }\href
  {https://doi.org/10.1103/PhysRevLett.59.2044} {\bibfield  {journal} {\bibinfo
   {journal} {Phys. Rev. Lett.}\ }\textbf {\bibinfo {volume} {59}},\ \bibinfo
  {pages} {2044} (\bibinfo {year} {1987})}\BibitemShut {NoStop}%
\bibitem [{\citenamefont {Kok}\ \emph {et~al.}(2004)\citenamefont {Kok},
  \citenamefont {Braunstein},\ and\ \citenamefont {Dowling}}]{Kok-2004}%
  \BibitemOpen
  \bibfield  {author} {\bibinfo {author} {\bibfnamefont {P.}~\bibnamefont
  {Kok}}, \bibinfo {author} {\bibfnamefont {S.~L.}\ \bibnamefont
  {Braunstein}},\ and\ \bibinfo {author} {\bibfnamefont {J.~P.}\ \bibnamefont
  {Dowling}},\ }\bibfield  {title} {\bibinfo {title} {Quantum lithography,
  entanglement and heisenberg-limited parameter estimation},\ }\href
  {https://doi.org/10.1088/1464-4266/6/8/029} {\bibfield  {journal} {\bibinfo
  {journal} {Journal of Optics B: Quantum and Semiclassical Optics}\ }\textbf
  {\bibinfo {volume} {6}},\ \bibinfo {pages} {S811–S815} (\bibinfo {year}
  {2004})}\BibitemShut {NoStop}%
\bibitem [{\citenamefont {Ben-Aryeh}(2012)}]{Ben-Aryeh-2012}%
  \BibitemOpen
  \bibfield  {author} {\bibinfo {author} {\bibfnamefont {Y.}~\bibnamefont
  {Ben-Aryeh}},\ }\bibfield  {title} {\bibinfo {title} {Phase estimation by
  photon counting measurements in the output of a linear {Mach-Zehnder}
  interferometer},\ }\href {https://doi.org/10.1364/JOSAB.29.002754} {\bibfield
   {journal} {\bibinfo  {journal} {J. Opt. Soc. Am. B}\ }\textbf {\bibinfo
  {volume} {29}},\ \bibinfo {pages} {2754} (\bibinfo {year}
  {2012})}\BibitemShut {NoStop}%
\bibitem [{\citenamefont {Olindo}\ \emph {et~al.}(2006)\citenamefont {Olindo},
  \citenamefont {Sagioro}, \citenamefont {Monken}, \citenamefont {P\'adua},\
  and\ \citenamefont {Delgado}}]{Olindo-2006}%
  \BibitemOpen
  \bibfield  {author} {\bibinfo {author} {\bibfnamefont {C.}~\bibnamefont
  {Olindo}}, \bibinfo {author} {\bibfnamefont {M.~A.}\ \bibnamefont {Sagioro}},
  \bibinfo {author} {\bibfnamefont {C.~H.}\ \bibnamefont {Monken}}, \bibinfo
  {author} {\bibfnamefont {S.}~\bibnamefont {P\'adua}},\ and\ \bibinfo {author}
  {\bibfnamefont {A.}~\bibnamefont {Delgado}},\ }\bibfield  {title} {\bibinfo
  {title} {{Hong-Ou-Mandel} interferometer with cavities: Theory},\ }\href
  {https://doi.org/10.1103/PhysRevA.73.043806} {\bibfield  {journal} {\bibinfo
  {journal} {Phys. Rev. A}\ }\textbf {\bibinfo {volume} {73}},\ \bibinfo
  {pages} {043806} (\bibinfo {year} {2006})}\BibitemShut {NoStop}%
\bibitem [{\citenamefont {Lyons}\ \emph {et~al.}(2018)\citenamefont {Lyons},
  \citenamefont {Knee}, \citenamefont {Bolduc}, \citenamefont {Roger},
  \citenamefont {Leach}, \citenamefont {Gauger},\ and\ \citenamefont
  {Faccio}}]{Lyons-2018}%
  \BibitemOpen
  \bibfield  {author} {\bibinfo {author} {\bibfnamefont {A.}~\bibnamefont
  {Lyons}}, \bibinfo {author} {\bibfnamefont {G.~C.}\ \bibnamefont {Knee}},
  \bibinfo {author} {\bibfnamefont {E.}~\bibnamefont {Bolduc}}, \bibinfo
  {author} {\bibfnamefont {T.}~\bibnamefont {Roger}}, \bibinfo {author}
  {\bibfnamefont {J.}~\bibnamefont {Leach}}, \bibinfo {author} {\bibfnamefont
  {E.~M.}\ \bibnamefont {Gauger}},\ and\ \bibinfo {author} {\bibfnamefont
  {D.}~\bibnamefont {Faccio}},\ }\bibfield  {title} {\bibinfo {title}
  {Attosecond-resolution {Hong-Ou-Mandel} interferometry},\ }\href
  {https://doi.org/10.1126/sciadv.aap9416} {\bibfield  {journal} {\bibinfo
  {journal} {Science Advances}\ }\textbf {\bibinfo {volume} {4}},\ \bibinfo
  {pages} {eaap9416} (\bibinfo {year} {2018})}\BibitemShut {NoStop}%
\bibitem [{\citenamefont {Chen}\ \emph {et~al.}(2019)\citenamefont {Chen},
  \citenamefont {Fink}, \citenamefont {Steinlechner}, \citenamefont {Torres},\
  and\ \citenamefont {Ursin}}]{Chen-2019}%
  \BibitemOpen
  \bibfield  {author} {\bibinfo {author} {\bibfnamefont {Y.}~\bibnamefont
  {Chen}}, \bibinfo {author} {\bibfnamefont {M.}~\bibnamefont {Fink}}, \bibinfo
  {author} {\bibfnamefont {F.}~\bibnamefont {Steinlechner}}, \bibinfo {author}
  {\bibfnamefont {J.~P.}\ \bibnamefont {Torres}},\ and\ \bibinfo {author}
  {\bibfnamefont {R.}~\bibnamefont {Ursin}},\ }\bibfield  {title} {\bibinfo
  {title} {{Hong-Ou-Mandel} interferometry on a biphoton beat note},\ }\href
  {https://doi.org/10.1038/s41534-019-0161-z} {\bibfield  {journal} {\bibinfo
  {journal} {npj Quantum Information}\ }\textbf {\bibinfo {volume} {5}},\
  \bibinfo {pages} {43} (\bibinfo {year} {2019})}\BibitemShut {NoStop}%
\bibitem [{\citenamefont {Yang}\ \emph {et~al.}(2019)\citenamefont {Yang},
  \citenamefont {Xu},\ and\ \citenamefont {Giovannetti}}]{Yang-2019}%
  \BibitemOpen
  \bibfield  {author} {\bibinfo {author} {\bibfnamefont {Y.}~\bibnamefont
  {Yang}}, \bibinfo {author} {\bibfnamefont {L.}~\bibnamefont {Xu}},\ and\
  \bibinfo {author} {\bibfnamefont {V.}~\bibnamefont {Giovannetti}},\
  }\bibfield  {title} {\bibinfo {title} {Two-parameter {Hong-Ou-Mandel} dip},\
  }\href {https://doi.org/10.1038/s41598-019-47207-3} {\bibfield  {journal}
  {\bibinfo  {journal} {Scientific Reports}\ }\textbf {\bibinfo {volume} {9}},\
  \bibinfo {pages} {10821} (\bibinfo {year} {2019})}\BibitemShut {NoStop}%
\bibitem [{\citenamefont {Restuccia}\ \emph {et~al.}(2019)\citenamefont
  {Restuccia}, \citenamefont {Toro{\v{s}}}, \citenamefont {Gibson},
  \citenamefont {Ulbricht}, \citenamefont {Faccio},\ and\ \citenamefont
  {Padgett}}]{Restuccia-2019}%
  \BibitemOpen
  \bibfield  {author} {\bibinfo {author} {\bibfnamefont {S.}~\bibnamefont
  {Restuccia}}, \bibinfo {author} {\bibfnamefont {M.}~\bibnamefont
  {Toro{\v{s}}}}, \bibinfo {author} {\bibfnamefont {G.~M.}\ \bibnamefont
  {Gibson}}, \bibinfo {author} {\bibfnamefont {H.}~\bibnamefont {Ulbricht}},
  \bibinfo {author} {\bibfnamefont {D.}~\bibnamefont {Faccio}},\ and\ \bibinfo
  {author} {\bibfnamefont {M.~J.}\ \bibnamefont {Padgett}},\ }\bibfield
  {title} {\bibinfo {title} {Photon bunching in a rotating reference frame},\
  }\href {https://doi.org/10.1103/physrevlett.123.110401} {\bibfield  {journal}
  {\bibinfo  {journal} {Physical Review Letters}\ }\textbf {\bibinfo {volume}
  {123}},\ \bibinfo {pages} {110401} (\bibinfo {year} {2019})}\BibitemShut
  {NoStop}%
\bibitem [{\citenamefont {Scott}\ \emph {et~al.}(2020)\citenamefont {Scott},
  \citenamefont {Branford}, \citenamefont {Westerberg}, \citenamefont {Leach},\
  and\ \citenamefont {Gauger}}]{Scott-2020}%
  \BibitemOpen
  \bibfield  {author} {\bibinfo {author} {\bibfnamefont {H.}~\bibnamefont
  {Scott}}, \bibinfo {author} {\bibfnamefont {D.}~\bibnamefont {Branford}},
  \bibinfo {author} {\bibfnamefont {N.}~\bibnamefont {Westerberg}}, \bibinfo
  {author} {\bibfnamefont {J.}~\bibnamefont {Leach}},\ and\ \bibinfo {author}
  {\bibfnamefont {E.~M.}\ \bibnamefont {Gauger}},\ }\bibfield  {title}
  {\bibinfo {title} {Beyond coincidence in {Hong-Ou-Mandel} interferometry},\
  }\href {https://doi.org/10.1103/PhysRevA.102.033714} {\bibfield  {journal}
  {\bibinfo  {journal} {Phys. Rev. A}\ }\textbf {\bibinfo {volume} {102}},\
  \bibinfo {pages} {033714} (\bibinfo {year} {2020})}\BibitemShut {NoStop}%
\bibitem [{\citenamefont {Fabre}\ and\ \citenamefont
  {Felicetti}(2021)}]{Fabre-2021}%
  \BibitemOpen
  \bibfield  {author} {\bibinfo {author} {\bibfnamefont {N.}~\bibnamefont
  {Fabre}}\ and\ \bibinfo {author} {\bibfnamefont {S.}~\bibnamefont
  {Felicetti}},\ }\bibfield  {title} {\bibinfo {title} {Parameter estimation of
  time and frequency shifts with generalized {Hong-Ou-Mandel} interferometry},\
  }\href {https://doi.org/10.1103/PhysRevA.104.022208} {\bibfield  {journal}
  {\bibinfo  {journal} {Phys. Rev. A}\ }\textbf {\bibinfo {volume} {104}},\
  \bibinfo {pages} {022208} (\bibinfo {year} {2021})}\BibitemShut {NoStop}%
\bibitem [{\citenamefont {Shih}\ \emph {et~al.}(1994)\citenamefont {Shih},
  \citenamefont {Sergienko}, \citenamefont {Rubin}, \citenamefont {Kiess},\
  and\ \citenamefont {Alley}}]{Shih-1994}%
  \BibitemOpen
  \bibfield  {author} {\bibinfo {author} {\bibfnamefont {Y.~H.}\ \bibnamefont
  {Shih}}, \bibinfo {author} {\bibfnamefont {A.~V.}\ \bibnamefont {Sergienko}},
  \bibinfo {author} {\bibfnamefont {M.~H.}\ \bibnamefont {Rubin}}, \bibinfo
  {author} {\bibfnamefont {T.~E.}\ \bibnamefont {Kiess}},\ and\ \bibinfo
  {author} {\bibfnamefont {C.~O.}\ \bibnamefont {Alley}},\ }\bibfield  {title}
  {\bibinfo {title} {Two-photon interference in a standard {Mach-Zehnder}
  interferometer},\ }\href {https://doi.org/10.1103/PhysRevA.49.4243}
  {\bibfield  {journal} {\bibinfo  {journal} {Phys. Rev. A}\ }\textbf {\bibinfo
  {volume} {49}},\ \bibinfo {pages} {4243} (\bibinfo {year}
  {1994})}\BibitemShut {NoStop}%
\bibitem [{\citenamefont {Rarity}\ \emph {et~al.}(1990)\citenamefont {Rarity},
  \citenamefont {Tapster}, \citenamefont {Jakeman}, \citenamefont {Larchuk},
  \citenamefont {Campos}, \citenamefont {Teich},\ and\ \citenamefont
  {Saleh}}]{Rarity-1990}%
  \BibitemOpen
  \bibfield  {author} {\bibinfo {author} {\bibfnamefont {J.~G.}\ \bibnamefont
  {Rarity}}, \bibinfo {author} {\bibfnamefont {P.~R.}\ \bibnamefont {Tapster}},
  \bibinfo {author} {\bibfnamefont {E.}~\bibnamefont {Jakeman}}, \bibinfo
  {author} {\bibfnamefont {T.}~\bibnamefont {Larchuk}}, \bibinfo {author}
  {\bibfnamefont {R.~A.}\ \bibnamefont {Campos}}, \bibinfo {author}
  {\bibfnamefont {M.~C.}\ \bibnamefont {Teich}},\ and\ \bibinfo {author}
  {\bibfnamefont {B.~E.~A.}\ \bibnamefont {Saleh}},\ }\bibfield  {title}
  {\bibinfo {title} {Two-photon interference in a {Mach-Zehnder}
  interferometer},\ }\href {https://doi.org/10.1103/PhysRevLett.65.1348}
  {\bibfield  {journal} {\bibinfo  {journal} {Phys. Rev. Lett.}\ }\textbf
  {\bibinfo {volume} {65}},\ \bibinfo {pages} {1348} (\bibinfo {year}
  {1990})}\BibitemShut {NoStop}%
\bibitem [{\citenamefont {Kim}\ \emph {et~al.}(2003)\citenamefont {Kim},
  \citenamefont {Kim}, \citenamefont {Ko},\ and\ \citenamefont
  {Park}}]{Kim-2003}%
  \BibitemOpen
  \bibfield  {author} {\bibinfo {author} {\bibfnamefont {T.-S.}\ \bibnamefont
  {Kim}}, \bibinfo {author} {\bibfnamefont {H.-O.}\ \bibnamefont {Kim}},
  \bibinfo {author} {\bibfnamefont {J.-H.}\ \bibnamefont {Ko}},\ and\ \bibinfo
  {author} {\bibfnamefont {G.-D.}\ \bibnamefont {Park}},\ }\bibfield  {title}
  {\bibinfo {title} {Two-photon interference experiment in a {Mach-Zehnder}
  interferometer},\ }\href
  {http://www.osapublishing.org/josk/abstract.cfm?URI=josk-7-2-113} {\bibfield
  {journal} {\bibinfo  {journal} {J. Opt. Soc. Korea}\ }\textbf {\bibinfo
  {volume} {7}},\ \bibinfo {pages} {113} (\bibinfo {year} {2003})}\BibitemShut
  {NoStop}%
\bibitem [{\citenamefont {Lang}\ and\ \citenamefont {Caves}(2014)}]{Lang-2014}%
  \BibitemOpen
  \bibfield  {author} {\bibinfo {author} {\bibfnamefont {M.~D.}\ \bibnamefont
  {Lang}}\ and\ \bibinfo {author} {\bibfnamefont {C.~M.}\ \bibnamefont
  {Caves}},\ }\bibfield  {title} {\bibinfo {title} {Optimal quantum-enhanced
  interferometry},\ }\href {https://doi.org/10.1103/PhysRevA.90.025802}
  {\bibfield  {journal} {\bibinfo  {journal} {Physical Review A}\ }\textbf
  {\bibinfo {volume} {90}},\ \bibinfo {pages} {025802} (\bibinfo {year}
  {2014})}\BibitemShut {NoStop}%
\bibitem [{\citenamefont {Polino}\ \emph {et~al.}(2020)\citenamefont {Polino},
  \citenamefont {Valeri}, \citenamefont {Spagnolo},\ and\ \citenamefont
  {Sciarrino}}]{Polino-2020}%
  \BibitemOpen
  \bibfield  {author} {\bibinfo {author} {\bibfnamefont {E.}~\bibnamefont
  {Polino}}, \bibinfo {author} {\bibfnamefont {M.}~\bibnamefont {Valeri}},
  \bibinfo {author} {\bibfnamefont {N.}~\bibnamefont {Spagnolo}},\ and\
  \bibinfo {author} {\bibfnamefont {F.}~\bibnamefont {Sciarrino}},\ }\bibfield
  {title} {\bibinfo {title} {Photonic quantum metrology},\ }\href
  {https://doi.org/10.1116/5.0007577} {\bibfield  {journal} {\bibinfo
  {journal} {AVS Quantum Science}\ }\textbf {\bibinfo {volume} {2}},\ \bibinfo
  {pages} {024703} (\bibinfo {year} {2020})}\BibitemShut {NoStop}%
\bibitem [{\citenamefont {Wolfgramm}\ \emph {et~al.}(2013)\citenamefont
  {Wolfgramm}, \citenamefont {Vitelli}, \citenamefont {Beduini}, \citenamefont
  {Godbout},\ and\ \citenamefont {Mitchell}}]{Wolfgramm-2013}%
  \BibitemOpen
  \bibfield  {author} {\bibinfo {author} {\bibfnamefont {F.}~\bibnamefont
  {Wolfgramm}}, \bibinfo {author} {\bibfnamefont {C.}~\bibnamefont {Vitelli}},
  \bibinfo {author} {\bibfnamefont {F.~A.}\ \bibnamefont {Beduini}}, \bibinfo
  {author} {\bibfnamefont {N.}~\bibnamefont {Godbout}},\ and\ \bibinfo {author}
  {\bibfnamefont {M.~W.}\ \bibnamefont {Mitchell}},\ }\bibfield  {title}
  {\bibinfo {title} {Entanglement-enhanced probing of a delicate material
  system},\ }\href {https://doi.org/10.1038/nphoton.2012.300} {\bibfield
  {journal} {\bibinfo  {journal} {Nature Photonics}\ }\textbf {\bibinfo
  {volume} {7}},\ \bibinfo {pages} {28} (\bibinfo {year} {2013})}\BibitemShut
  {NoStop}%
\bibitem [{\citenamefont {Taylor}\ and\ \citenamefont
  {Bowen}(2016)}]{Taylor-2016}%
  \BibitemOpen
  \bibfield  {author} {\bibinfo {author} {\bibfnamefont {M.~A.}\ \bibnamefont
  {Taylor}}\ and\ \bibinfo {author} {\bibfnamefont {W.~P.}\ \bibnamefont
  {Bowen}},\ }\bibfield  {title} {\bibinfo {title} {Quantum metrology and its
  application in biology},\ }\href
  {https://doi.org/10.1016/j.physrep.2015.12.002} {\bibfield  {journal}
  {\bibinfo  {journal} {Physics Reports}\ }\textbf {\bibinfo {volume} {615}},\
  \bibinfo {pages} {1–59} (\bibinfo {year} {2016})}\BibitemShut {NoStop}%
\bibitem [{\citenamefont {Casacio}\ \emph {et~al.}(2021)\citenamefont
  {Casacio}, \citenamefont {Madsen}, \citenamefont {Terrasson}, \citenamefont
  {Waleed}, \citenamefont {Barnscheidt}, \citenamefont {Hage}, \citenamefont
  {Taylor},\ and\ \citenamefont {Bowen}}]{Casacio-2021}%
  \BibitemOpen
  \bibfield  {author} {\bibinfo {author} {\bibfnamefont {C.~A.}\ \bibnamefont
  {Casacio}}, \bibinfo {author} {\bibfnamefont {L.~S.}\ \bibnamefont {Madsen}},
  \bibinfo {author} {\bibfnamefont {A.}~\bibnamefont {Terrasson}}, \bibinfo
  {author} {\bibfnamefont {M.}~\bibnamefont {Waleed}}, \bibinfo {author}
  {\bibfnamefont {K.}~\bibnamefont {Barnscheidt}}, \bibinfo {author}
  {\bibfnamefont {B.}~\bibnamefont {Hage}}, \bibinfo {author} {\bibfnamefont
  {M.~A.}\ \bibnamefont {Taylor}},\ and\ \bibinfo {author} {\bibfnamefont
  {W.~P.}\ \bibnamefont {Bowen}},\ }\bibfield  {title} {\bibinfo {title}
  {Quantum-enhanced nonlinear microscopy},\ }\href
  {https://doi.org/10.1038/s41586-021-03528-w} {\bibfield  {journal} {\bibinfo
  {journal} {Nature}\ }\textbf {\bibinfo {volume} {594}},\ \bibinfo {pages}
  {201} (\bibinfo {year} {2021})}\BibitemShut {NoStop}%
\bibitem [{\citenamefont {Triginer~Garces}\ \emph {et~al.}(2020)\citenamefont
  {Triginer~Garces}, \citenamefont {Chrzanowski}, \citenamefont {Daryanoosh},
  \citenamefont {Thiel}, \citenamefont {Marchant}, \citenamefont {Patel},
  \citenamefont {Humphreys}, \citenamefont {Datta},\ and\ \citenamefont
  {Walmsley}}]{Triginer-Garces-2020}%
  \BibitemOpen
  \bibfield  {author} {\bibinfo {author} {\bibfnamefont {G.}~\bibnamefont
  {Triginer~Garces}}, \bibinfo {author} {\bibfnamefont {H.~M.}\ \bibnamefont
  {Chrzanowski}}, \bibinfo {author} {\bibfnamefont {S.}~\bibnamefont
  {Daryanoosh}}, \bibinfo {author} {\bibfnamefont {V.}~\bibnamefont {Thiel}},
  \bibinfo {author} {\bibfnamefont {A.~L.}\ \bibnamefont {Marchant}}, \bibinfo
  {author} {\bibfnamefont {R.~B.}\ \bibnamefont {Patel}}, \bibinfo {author}
  {\bibfnamefont {P.~C.}\ \bibnamefont {Humphreys}}, \bibinfo {author}
  {\bibfnamefont {A.}~\bibnamefont {Datta}},\ and\ \bibinfo {author}
  {\bibfnamefont {I.~A.}\ \bibnamefont {Walmsley}},\ }\bibfield  {title}
  {\bibinfo {title} {Quantum-enhanced stimulated emission detection for
  label-free microscopy},\ }\href {https://doi.org/10.1063/5.0009681}
  {\bibfield  {journal} {\bibinfo  {journal} {Applied Physics Letters}\
  }\textbf {\bibinfo {volume} {117}},\ \bibinfo {pages} {024002} (\bibinfo
  {year} {2020})}\BibitemShut {NoStop}%
\bibitem [{\citenamefont {Xavier}\ \emph {et~al.}(2021)\citenamefont {Xavier},
  \citenamefont {Yu}, \citenamefont {Jones}, \citenamefont {Zossimova},\ and\
  \citenamefont {Vollmer}}]{Xavier-2021}%
  \BibitemOpen
  \bibfield  {author} {\bibinfo {author} {\bibfnamefont {J.}~\bibnamefont
  {Xavier}}, \bibinfo {author} {\bibfnamefont {D.}~\bibnamefont {Yu}}, \bibinfo
  {author} {\bibfnamefont {C.}~\bibnamefont {Jones}}, \bibinfo {author}
  {\bibfnamefont {E.}~\bibnamefont {Zossimova}},\ and\ \bibinfo {author}
  {\bibfnamefont {F.}~\bibnamefont {Vollmer}},\ }\bibfield  {title} {\bibinfo
  {title} {Quantum nanophotonic and nanoplasmonic sensing: towards quantum
  optical bioscience laboratories on chip},\ }\href
  {https://doi.org/10.1515/nanoph-2020-0593} {\bibfield  {journal} {\bibinfo
  {journal} {Nanophotonics}\ }\textbf {\bibinfo {volume} {10}},\ \bibinfo
  {pages} {1387} (\bibinfo {year} {2021})}\BibitemShut {NoStop}%
\bibitem [{\citenamefont {Itoh}(1982)}]{Itoh-1982}%
  \BibitemOpen
  \bibfield  {author} {\bibinfo {author} {\bibfnamefont {K.}~\bibnamefont
  {Itoh}},\ }\bibfield  {title} {\bibinfo {title} {Analysis of the phase
  unwrapping algorithm},\ }\href {https://doi.org/10.1364/AO.21.002470}
  {\bibfield  {journal} {\bibinfo  {journal} {Appl. Opt.}\ }\textbf {\bibinfo
  {volume} {21}},\ \bibinfo {pages} {2470} (\bibinfo {year}
  {1982})}\BibitemShut {NoStop}%
\bibitem [{\citenamefont {von Toussaint}(2015)}]{von-Toussaint-2015}%
  \BibitemOpen
  \bibfield  {author} {\bibinfo {author} {\bibfnamefont {U.}~\bibnamefont {von
  Toussaint}},\ }\bibfield  {title} {\bibinfo {title} {Robust phase estimation
  for signals with a low signal-to-noise-ratio},\ }\href
  {https://doi.org/10.1063/1.4905985} {\bibfield  {journal} {\bibinfo
  {journal} {AIP Conference Proceedings}\ }\textbf {\bibinfo {volume} {1641}},\
  \bibinfo {pages} {246} (\bibinfo {year} {2015})}\BibitemShut {NoStop}%
\bibitem [{\citenamefont {Hayashi}\ \emph {et~al.}(2018)\citenamefont
  {Hayashi}, \citenamefont {Vinjanampathy},\ and\ \citenamefont
  {Kwek}}]{Hayashi-2018}%
  \BibitemOpen
  \bibfield  {author} {\bibinfo {author} {\bibfnamefont {M.}~\bibnamefont
  {Hayashi}}, \bibinfo {author} {\bibfnamefont {S.}~\bibnamefont
  {Vinjanampathy}},\ and\ \bibinfo {author} {\bibfnamefont {L.~C.}\
  \bibnamefont {Kwek}},\ }\bibfield  {title} {\bibinfo {title} {Resolving
  unattainable {Cramér}–{Rao} bounds for quantum sensors},\ }\href
  {https://doi.org/10.1088/1361-6455/aaf348} {\bibfield  {journal} {\bibinfo
  {journal} {Journal of Physics B: Atomic, Molecular and Optical Physics}\
  }\textbf {\bibinfo {volume} {52}},\ \bibinfo {pages} {015503} (\bibinfo
  {year} {2018})}\BibitemShut {NoStop}%
\bibitem [{\citenamefont {Genoni}\ \emph {et~al.}(2011)\citenamefont {Genoni},
  \citenamefont {Olivares},\ and\ \citenamefont {Paris}}]{Genoni-2011}%
  \BibitemOpen
  \bibfield  {author} {\bibinfo {author} {\bibfnamefont {M.~G.}\ \bibnamefont
  {Genoni}}, \bibinfo {author} {\bibfnamefont {S.}~\bibnamefont {Olivares}},\
  and\ \bibinfo {author} {\bibfnamefont {M.~G.~A.}\ \bibnamefont {Paris}},\
  }\bibfield  {title} {\bibinfo {title} {Optical phase estimation in the
  presence of phase diffusion},\ }\href
  {https://doi.org/10.1103/PhysRevLett.106.153603} {\bibfield  {journal}
  {\bibinfo  {journal} {Phys. Rev. Lett.}\ }\textbf {\bibinfo {volume} {106}},\
  \bibinfo {pages} {153603} (\bibinfo {year} {2011})}\BibitemShut {NoStop}%
\bibitem [{\citenamefont {Escher}\ \emph {et~al.}(2012)\citenamefont {Escher},
  \citenamefont {Davidovich}, \citenamefont {Zagury},\ and\ \citenamefont
  {de~Matos~Filho}}]{Escher-2012}%
  \BibitemOpen
  \bibfield  {author} {\bibinfo {author} {\bibfnamefont {B.~M.}\ \bibnamefont
  {Escher}}, \bibinfo {author} {\bibfnamefont {L.}~\bibnamefont {Davidovich}},
  \bibinfo {author} {\bibfnamefont {N.}~\bibnamefont {Zagury}},\ and\ \bibinfo
  {author} {\bibfnamefont {R.~L.}\ \bibnamefont {de~Matos~Filho}},\ }\bibfield
  {title} {\bibinfo {title} {Quantum metrological limits via a variational
  approach},\ }\href {https://doi.org/10.1103/PhysRevLett.109.190404}
  {\bibfield  {journal} {\bibinfo  {journal} {Phys. Rev. Lett.}\ }\textbf
  {\bibinfo {volume} {109}},\ \bibinfo {pages} {190404} (\bibinfo {year}
  {2012})}\BibitemShut {NoStop}%
\bibitem [{\citenamefont {Genoni}\ \emph {et~al.}(2012)\citenamefont {Genoni},
  \citenamefont {Olivares}, \citenamefont {Brivio}, \citenamefont {Cialdi},
  \citenamefont {Cipriani}, \citenamefont {Santamato}, \citenamefont
  {Vezzoli},\ and\ \citenamefont {Paris}}]{Genoni-2012}%
  \BibitemOpen
  \bibfield  {author} {\bibinfo {author} {\bibfnamefont {M.~G.}\ \bibnamefont
  {Genoni}}, \bibinfo {author} {\bibfnamefont {S.}~\bibnamefont {Olivares}},
  \bibinfo {author} {\bibfnamefont {D.}~\bibnamefont {Brivio}}, \bibinfo
  {author} {\bibfnamefont {S.}~\bibnamefont {Cialdi}}, \bibinfo {author}
  {\bibfnamefont {D.}~\bibnamefont {Cipriani}}, \bibinfo {author}
  {\bibfnamefont {A.}~\bibnamefont {Santamato}}, \bibinfo {author}
  {\bibfnamefont {S.}~\bibnamefont {Vezzoli}},\ and\ \bibinfo {author}
  {\bibfnamefont {M.~G.~A.}\ \bibnamefont {Paris}},\ }\bibfield  {title}
  {\bibinfo {title} {Optical interferometry in the presence of large phase
  diffusion},\ }\href {https://doi.org/10.1103/PhysRevA.85.043817} {\bibfield
  {journal} {\bibinfo  {journal} {Phys. Rev. A}\ }\textbf {\bibinfo {volume}
  {85}},\ \bibinfo {pages} {043817} (\bibinfo {year} {2012})}\BibitemShut
  {NoStop}%
\bibitem [{\citenamefont {Dorner}\ \emph {et~al.}(2009)\citenamefont {Dorner},
  \citenamefont {Demkowicz-Dobrza{\'{n}}ski}, \citenamefont {Smith},
  \citenamefont {Lundeen}, \citenamefont {Wasilewski}, \citenamefont
  {Banaszek},\ and\ \citenamefont {Walmsley}}]{Dorner-2009}%
  \BibitemOpen
  \bibfield  {author} {\bibinfo {author} {\bibfnamefont {U.}~\bibnamefont
  {Dorner}}, \bibinfo {author} {\bibfnamefont {R.}~\bibnamefont
  {Demkowicz-Dobrza{\'{n}}ski}}, \bibinfo {author} {\bibfnamefont {B.~J.}\
  \bibnamefont {Smith}}, \bibinfo {author} {\bibfnamefont {J.~S.}\ \bibnamefont
  {Lundeen}}, \bibinfo {author} {\bibfnamefont {W.}~\bibnamefont {Wasilewski}},
  \bibinfo {author} {\bibfnamefont {K.}~\bibnamefont {Banaszek}},\ and\
  \bibinfo {author} {\bibfnamefont {I.~A.}\ \bibnamefont {Walmsley}},\
  }\bibfield  {title} {\bibinfo {title} {Optimal {Quantum} {Phase}
  {Estimation}},\ }\href {https://doi.org/10.1103/PhysRevLett.102.040403}
  {\bibfield  {journal} {\bibinfo  {journal} {Physical Review Letters}\
  }\textbf {\bibinfo {volume} {102}},\ \bibinfo {pages} {040403} (\bibinfo
  {year} {2009})}\BibitemShut {NoStop}%
\bibitem [{\citenamefont {Lee}\ \emph {et~al.}(2009)\citenamefont {Lee},
  \citenamefont {Huver}, \citenamefont {Lee}, \citenamefont {Kaplan},
  \citenamefont {McCracken}, \citenamefont {Min}, \citenamefont {Uskov},
  \citenamefont {Wildfeuer}, \citenamefont {Veronis},\ and\ \citenamefont
  {Dowling}}]{Lee-2009}%
  \BibitemOpen
  \bibfield  {author} {\bibinfo {author} {\bibfnamefont {T.-W.}\ \bibnamefont
  {Lee}}, \bibinfo {author} {\bibfnamefont {S.~D.}\ \bibnamefont {Huver}},
  \bibinfo {author} {\bibfnamefont {H.}~\bibnamefont {Lee}}, \bibinfo {author}
  {\bibfnamefont {L.}~\bibnamefont {Kaplan}}, \bibinfo {author} {\bibfnamefont
  {S.~B.}\ \bibnamefont {McCracken}}, \bibinfo {author} {\bibfnamefont
  {C.}~\bibnamefont {Min}}, \bibinfo {author} {\bibfnamefont {D.~B.}\
  \bibnamefont {Uskov}}, \bibinfo {author} {\bibfnamefont {C.~F.}\ \bibnamefont
  {Wildfeuer}}, \bibinfo {author} {\bibfnamefont {G.}~\bibnamefont {Veronis}},\
  and\ \bibinfo {author} {\bibfnamefont {J.~P.}\ \bibnamefont {Dowling}},\
  }\bibfield  {title} {\bibinfo {title} {Optimization of quantum
  interferometric metrological sensors in the presence of photon loss},\ }\href
  {https://doi.org/10.1103/PhysRevA.80.063803} {\bibfield  {journal} {\bibinfo
  {journal} {Physical Review A}\ }\textbf {\bibinfo {volume} {80}},\ \bibinfo
  {pages} {063803} (\bibinfo {year} {2009})}\BibitemShut {NoStop}%
\bibitem [{\citenamefont {Datta}\ \emph {et~al.}(2011)\citenamefont {Datta},
  \citenamefont {Zhang}, \citenamefont {Thomas-Peter}, \citenamefont {Dorner},
  \citenamefont {Smith},\ and\ \citenamefont {Walmsley}}]{Datta-2011}%
  \BibitemOpen
  \bibfield  {author} {\bibinfo {author} {\bibfnamefont {A.}~\bibnamefont
  {Datta}}, \bibinfo {author} {\bibfnamefont {L.}~\bibnamefont {Zhang}},
  \bibinfo {author} {\bibfnamefont {N.}~\bibnamefont {Thomas-Peter}}, \bibinfo
  {author} {\bibfnamefont {U.}~\bibnamefont {Dorner}}, \bibinfo {author}
  {\bibfnamefont {B.~J.}\ \bibnamefont {Smith}},\ and\ \bibinfo {author}
  {\bibfnamefont {I.~A.}\ \bibnamefont {Walmsley}},\ }\bibfield  {title}
  {\bibinfo {title} {Quantum metrology with imperfect states and detectors},\
  }\href {https://doi.org/10.1103/PhysRevA.83.063836} {\bibfield  {journal}
  {\bibinfo  {journal} {Physical Review A}\ }\textbf {\bibinfo {volume} {83}},\
  \bibinfo {pages} {063836} (\bibinfo {year} {2011})}\BibitemShut {NoStop}%
\bibitem [{\citenamefont {Bra\'nczyk}(2017)}]{Branczyk-2017}%
  \BibitemOpen
  \bibfield  {author} {\bibinfo {author} {\bibfnamefont {A.~M.}\ \bibnamefont
  {Bra\'nczyk}},\ }\href@noop {} {\bibinfo {title} {{Hong-Ou-Mandel}
  interference}} (\bibinfo {year} {2017}),\ \Eprint
  {https://arxiv.org/abs/1711.00080} {arXiv:1711.00080 [quant-ph]} \BibitemShut
  {NoStop}%
\bibitem [{\citenamefont {Holland}\ and\ \citenamefont
  {Burnett}(1993)}]{Holland-1993}%
  \BibitemOpen
  \bibfield  {author} {\bibinfo {author} {\bibfnamefont {M.~J.}\ \bibnamefont
  {Holland}}\ and\ \bibinfo {author} {\bibfnamefont {K.}~\bibnamefont
  {Burnett}},\ }\bibfield  {title} {\bibinfo {title} {Interferometric detection
  of optical phase shifts at the {Heisenberg} limit},\ }\href
  {https://doi.org/10.1103/PhysRevLett.71.1355} {\bibfield  {journal} {\bibinfo
   {journal} {Physical Review Letters}\ }\textbf {\bibinfo {volume} {71}},\
  \bibinfo {pages} {1355} (\bibinfo {year} {1993})}\BibitemShut {NoStop}%
\bibitem [{\citenamefont {Jachura}\ \emph {et~al.}(2016)\citenamefont
  {Jachura}, \citenamefont {Chrapkiewicz}, \citenamefont
  {Demkowicz-Dobrza{\'{n}}ski}, \citenamefont {Wasilewski},\ and\ \citenamefont
  {Banaszek}}]{Jachura-2016}%
  \BibitemOpen
  \bibfield  {author} {\bibinfo {author} {\bibfnamefont {M.}~\bibnamefont
  {Jachura}}, \bibinfo {author} {\bibfnamefont {R.}~\bibnamefont
  {Chrapkiewicz}}, \bibinfo {author} {\bibfnamefont {R.}~\bibnamefont
  {Demkowicz-Dobrza{\'{n}}ski}}, \bibinfo {author} {\bibfnamefont
  {W.}~\bibnamefont {Wasilewski}},\ and\ \bibinfo {author} {\bibfnamefont
  {K.}~\bibnamefont {Banaszek}},\ }\bibfield  {title} {\bibinfo {title} {Mode
  engineering for realistic quantum-enhanced interferometry},\ }\href
  {https://doi.org/10.1038/ncomms11411} {\bibfield  {journal} {\bibinfo
  {journal} {Nature Communications}\ }\textbf {\bibinfo {volume} {7}},\
  \bibinfo {pages} {11411} (\bibinfo {year} {2016})}\BibitemShut {NoStop}%
\bibitem [{\citenamefont {Kay}(1993)}]{Kay-1993}%
  \BibitemOpen
  \bibfield  {author} {\bibinfo {author} {\bibfnamefont {S.}~\bibnamefont
  {Kay}},\ }\href@noop {} {\emph {\bibinfo {title} {Fundamentals of statistical
  signal processing}}}\ (\bibinfo  {publisher} {Prentice-Hall PTR},\ \bibinfo
  {address} {Englewood Cliffs, N.J},\ \bibinfo {year} {1993})\BibitemShut
  {NoStop}%
\bibitem [{sup()}]{supp}%
  \BibitemOpen
  \href@noop {} {\bibinfo {title} {See {arXiv} ancillary files for
  {Mathematica} notebook and associated {PDF} containing full detection
  probabilities and {Fisher} information expressions.}}\BibitemShut {Stop}%
\bibitem [{\citenamefont {Franson}(1989)}]{Franson-1989}%
  \BibitemOpen
  \bibfield  {author} {\bibinfo {author} {\bibfnamefont {J.~D.}\ \bibnamefont
  {Franson}},\ }\bibfield  {title} {\bibinfo {title} {Bell inequality for
  position and time},\ }\href {https://doi.org/10.1103/PhysRevLett.62.2205}
  {\bibfield  {journal} {\bibinfo  {journal} {Phys. Rev. Lett.}\ }\textbf
  {\bibinfo {volume} {62}},\ \bibinfo {pages} {2205} (\bibinfo {year}
  {1989})}\BibitemShut {NoStop}%
\bibitem [{\citenamefont {Collett}\ and\ \citenamefont
  {Lewis}(1981)}]{Collett-1981}%
  \BibitemOpen
  \bibfield  {author} {\bibinfo {author} {\bibfnamefont {D.}~\bibnamefont
  {Collett}}\ and\ \bibinfo {author} {\bibfnamefont {T.}~\bibnamefont
  {Lewis}},\ }\bibfield  {title} {\bibinfo {title} {Discriminating between the
  von mises and wrapped normal distributions},\ }\href
  {https://doi.org/https://doi.org/10.1111/j.1467-842X.1981.tb00763.x}
  {\bibfield  {journal} {\bibinfo  {journal} {Australian Journal of
  Statistics}\ }\textbf {\bibinfo {volume} {23}},\ \bibinfo {pages} {73}
  (\bibinfo {year} {1981})}\BibitemShut {NoStop}%
\end{thebibliography}
\end{document}